\documentclass[10pt,aps,prd,amsmath,amssymb,superscriptaddress,twocolumn,nofootinbib,showpacs,preprintnumbers,amsfonts,notitlepage]{revtex4-2}
\usepackage{style}

\newcommand{\compass}{\mbox{COMPASS}\xspace}
\newcommand{\ale}[1]{{\color{blue} \bf }}

\definecolor{jpac_blue}   {HTML}{1F77B4}
\definecolor{jpac_red}    {HTML}{D61D28}
\definecolor{jpac_green}  {HTML}{2CA02C}
\definecolor{jpac_orange} {HTML}{FF7F0E}
\definecolor{jpac_purple} {HTML}{9467BD}
\definecolor{jpac_brown}  {HTML}{8C564B}
\definecolor{jpac_pink}   {HTML}{E377C2}
\definecolor{jpac_gold}   {HTML}{BCBD22}
\definecolor{jpac_aqua}   {HTML}{17BECF}
\definecolor{jpac_grey}   {HTML}{7F7F7F}

\begin{document}

\title{Production Effects and Final-State Interactions in \texorpdfstring{$\pi_1\to 3\pi$}{pi1 -> 3pi}}

\newcommand{\icn}{Instituto de Ciencias Nucleares,
    Universidad Nacional Aut\'onoma de M\'exico, Ciudad de M\'exico 04510, Mexico}

\newcommand{\ub}{Departament de F\'isica Qu\`antica i Astrof\'isica, Universitat de Barcelona, E-08028 Barcelona, Spain}
\newcommand{\iccub}{Institut de Ci\`encies del Cosmos, Universitat de Barcelona, E-08028 Barcelona, Spain.}

\newcommand{\UBarcelona}{Departament de Física Quàntica i 
Astrofísica (FQA), Universitat de Barcelona (UB),  c. Martí i Franquès, 1, 08028 Barcelona, Spain}

\newcommand{\iccubBarcelona}{Institut de Ciències del Cosmos (ICCUB), Universitat de Barcelona (UB), c. Martí i Franquès, 1, 08028 Barcelona, Spain.}

\newcommand{\iu}{Indiana University, Bloomington, IN 47405, USA}

\newcommand{\ceem}{Center for  Exploration  of  Energy  and  Matter,  Indiana  University,  Bloomington,  IN  47403,  USA}

\newcommand{\jlab}{Theory Center, Thomas  Jefferson  National  Accelerator  Facility,  Newport  News,  VA  23606,  USA}

\newcommand{\uned}{Departamento de F\'isica Interdisciplinar, Universidad Nacional de Educaci\'on a Distancia (UNED), E-28040 Madrid, Spain}

\newcommand{\bochum}{Ruhr University Bochum, D-44801 Bochum, Germany}

\newcommand{\AGH}{AGH University of Krakow, Faculty of Physics and Applied Computer Science, PL-30-059 Krak\'ow, Poland}

\newcommand{\odu}{Department of Physics, Old Dominion University, Norfolk, VA 23529, USA}

\newcommand{\catania}{INFN Sezione di Catania, I-95123 Catania, Italy}
\newcommand{\messina}{Dipartimento di Scienze Matematiche e Informatiche, Scienze Fisiche e Scienze della Terra, Universit\`a degli Studi di Messina, I-98166 Messina, Italy}

\author{D.~Winney\orcidlink{0000-0002-8076-243X}}
\email{daniel.winney@nucleares.unam.mx}
\affiliation{\icn}
\author{S.~Gonz\`{a}lez-Sol\'is\orcidlink{0000-0003-1947-5420}}
\affiliation{\ub}
\affiliation{\iccub}

\author{M.~Mikhasenko\,\orcidlink{0000-0002-6969-2063}}
\affiliation{\bochum}

\author{{\L}.~Bibrzycki\orcidlink{0000-0002-6117-4894}}
\affiliation{\AGH}

\author{C.~\surname{Fern\'andez-Ram\'irez}\orcidlink{0000-0001-8979-5660}}
\affiliation{\uned}

\author{V.~\surname{Mathieu}\orcidlink{0000-0003-4955-3311}} 
\affiliation{\ub}
\affiliation{\iccub}

\author{G.~\surname{Monta\~na}\orcidlink{0000-0001-8093-6682}} 
\affiliation{\ub}
\affiliation{\iccub}

\author{A.~\surname{Pilloni}\orcidlink{0000-0003-4257-0928}}
\affiliation{\messina}
\affiliation{\catania}

\author{L.~\surname{Qiu}\orcidlink{0000-0002-2683-8851}}\affiliation{\jlab}\affiliation{\odu}

\author{A.~\surname{Rodas}\orcidlink{0000-0003-2702-5286}}\affiliation{\jlab}\affiliation{\odu}

\author{A.~P.~\surname{Szczepaniak}\orcidlink{0000-0002-4156-5492}}
\affiliation{\jlab}
\affiliation{\iu}
\affiliation{\ceem}

\collaboration{Joint Physics Analysis Center}
\preprint{JLAB-THY-26-4841}
\begin{abstract}
    We investigate the $\pi_1(1600)$ signal channel of $3\pi$ in the $J^{PC} = 1^{-+} \,[\pi\pi]_{1^{--}}\pi~P$-wave. This channel has recently been analyzed by the COMPASS collaboration using the ``freed-isobar" technique which provides direct experimental access to the lineshape modifications of the $P$-wave $\pi\pi$ sub-channel arising from both final state interactions and production processes such as the Deck mechanism. We motivate a unified formalism combining production effects and final state interactions using the Khuri-Treiman formalism, which is consistent with low-energy unitarity, analyticity, and crossing symmetry, seeded by an initial production amplitude. We demonstrate that the precise lineshape of the $\pi\pi$ mass spectrum can be used to determine the relative strength and complex phase of different production mechanisms contributing to the $3\pi$ final state. We conduct a global fit of the $\pi_1$ freed-isobar data set, yielding a multi-dimensional parameterization of the $\pi_1\to3\pi$ decay amplitude at COMPASS as a function of both decay and production variables. We identify the contributions from short-range production and the Deck mechanism and extract the total $3\pi$ invariant mass dependence which will be important for a future extraction of the $\pi_1$ pole position in the $\rho\pi$ channel.
\end{abstract}
\maketitle

\section{Introduction}

The observation of a resonant peak around 1400 \mev in the $\eta\pi$ $P$-wave mass spectra~\cite{E852:1997gvf,E852:1999xev,E862:2006cfp,CrystalBarrel:1998cfz} around the turn of the century marked a pivotal point in our understanding of the light hadron spectrum within quantum chromodynamics (QCD). 
This is because this signal carries exotic quantum numbers $J^{PC}(I^G) = 1^{-+}(1^-)$, which cannot be associated to a $q\bar{q}$ meson. Hence, if the signal corresponds to a resonance, it would thus point unambiguously to an exotic hadron. Similar spin-exotic signals were later observed also in the $\eta^\prime \pi$~\cite{E852:2001ikk} and $\rho\pi$ final states~\cite{Khokhlov:2000tk} at a higher mass of approximately 1600 \mev, pointing to the possibility of two exotic meson states in this sector.

The higher mass state was generally consistent with expectations for hybrid mesons, i.e., states with explicit valence-gluon degrees of freedom in addition to the usual $q\bar{q}$ structure, which had been postulated as early as the late `70s~\cite{Horn:1977rq,Isgur:1984bm,Chanowitz:1982qj,Barnes:1982tx}. More recent lattice QCD studies~\cite{Dudek:2013yja,Woss:2020ayi} indeed support the existence of a $1^{-+}$ hybrid meson in the $1.6$--$2$ GeV mass range, coupling predominantly to $b_1\pi$ but also to $\eta^{(\prime)}\pi$ and $\rho\pi$, among others. The presence of two such states, however, was unexpected and pointed to some unknown dynamics  complicated by the challenge of isolating exotic signatures and by limited statistics (for a review, see Refs.~\cite{Meyer:2010ku,Meyer:2015eta}).

In recent years, high-statistics extractions of exotic $\eta^{(\prime)}\pi$ partial waves by the \compass Collaboration~\cite{COMPASS:2014vkj} enabled a simultaneous coupled-channel analysis of both reactions, revealing that the two seemingly distinct peaks arose from a single pole in the complex plane~\cite{JPAC:2018zyd}. This supported the existence of a single hybrid, the $\pi_1(1600)$, consistent with lattice predictions. This result was later confirmed by analogous coupled-channel analyses from the Crystal Barrel Collaboration~\cite{CrystalBarrel:2019zqh,Kopf:2020yoa}. The case of the two apparent $\pi_1$ peaks clearly illustrates that investigating multiple decay channels, ideally also multiple production modes, is necessary for a robust understanding of resonance properties.

So far, the only extractions of the $\pi_1$ pole position have come from the ``golden modes" of $\eta^{(\prime)}\pi$, even though these are predicted to account for only about $2\%$ of the total decay width~\cite{Woss:2020ayi}. The primary reason is that these are two-body final states and thus well suited to rigorous, unitary partial-wave analyses to search for poles on unphysical Riemann sheets. In contrast, the vast majority of the remaining $\pi_1$ width is expected to couple to $b_1\pi$ and $f_1\pi$, both of which involve resonances decaying into final states of five or more pions, e.g.\ in the decay chain $\pi_1 \to b_1\pi \to \omega\pi\pi \to 5\pi$. While the large branching fractions can be exploited to look for exotic signals in previously unobserved production modes, e.g.\ in photoproduction searches at GlueX~\cite{GlueX:2024erj}, a robust pole-search analysis beyond simple bump-hunts is functionally impossible in such a multi-body final state.
This leaves the $\pi_1\to3\pi$ decay mode as the largest predicted branching fraction (possibly more than twice that of the combined $\eta^{(\prime)}\pi$), allowing a rigorous exploration of the $\pi_1$ properties.

The $3\pi$ final state has long been compelling as the simplest three-body hadronic system and thus an ideal laboratory for exploring multi-body resonances. Interest has surged more recently due to the possibility of \textit{ab initio} calculations of $3\pi\to3\pi$ scattering on the lattice, e.g.\ in Refs.~\cite{Mai:2018djl,Briceno:2025yuq,Hansen:2020otl,Blanton:2019vdk,Fischer:2020jzp,Dawid:2025doq,Yan:2024gwp,Yan:2025mdm,Mai:2021nul,Culver:2019vvu,Mai:2018djl}. While constructing unitary $3\to3$ amplitudes is dramatically more complex than their $2\to2$ analogues, recent years have seen significant formal developments in this regard~(see e.g.\  Refs.~\cite{Jackura:2022gib,Jackura:2019bmu} and references therein).

Unlike on the lattice, three-particle scattering is not directly accessible at experimental facilities, so most data-driven explorations in this sector focus on the production of resonances that then decay into $3\pi$. Of particular note is the \compass experiment, which has collected one of the largest $3\pi$ data sets to date in its spectroscopy flagship channel of $\pi^-p\to\pi^-\pi^-\pi^+p$ (an order of magnitude more than in the analogous $\eta^{(\prime)}\pi^-$ production)~\cite{Ketzer:2019wmd}. Such a large data set has enabled the development of the novel ``freed-isobar" analysis technique, which aims to extract not only $3\pi$ partial waves but also the intensity distributions of $\pi\pi$ subchannels projected onto specific quantum numbers in a minimally model-dependent way~\cite{COMPASS:2015gxz,Krinner:2017dba,Krinner:2018bwg}. Applied to the spin-exotic wave in Ref.~\cite{COMPASS:2021ogp}, this technique provides an unprecedented look at $3\pi$ dynamics in the $[\pi\pi]_{1^{--}}\pi~P$-wave by capturing possible modifications to the $\rho$ lineshape induced by final-state interactions in the mass range of the $\pi_1$. Despite its high potential physics impact, the freed-isobar data set has received relatively little attention because modeling both the high-energy resonance production and the subsequent three-body decay is intricate, making rigorous analysis difficult.

So far, the most successful theoretical formalism for handling $3\pi$ decay reactions was proposed by Khuri and Treiman (henceforth KT)~\cite{Khuri:1960zz}, in which the three-body decay is assumed to proceed through pairwise interactions. This assumption allows the decay amplitude to be decomposed into crossing-symmetric combinations of isobars, onto which two-body unitarity can be imposed. It has been applied to the decay of a number of (narrow) mesons~\cite{Guo:2016wsi,Danilkin:2014cra,Niecknig:2012sj,Garcia-Lorenzo:2025uzc,Bernard:2024ioq,JPAC:2023nhq,JPAC:2020umo,Colangelo:2018jxw,Cao:2025ncx,Akdag:2023pwx}. Despite relying on two-body unitarity relations, the KT construction has been argued to be compatible with three-body unitarity, at least to the extent that the isobar approximation holds~\cite{Aitchison:1966lpz,Aitchison:1976nk}. The KT formalism can therefore, in principle, be extended to study the three-body mass dependence in systems with more complicated dynamics, such as the $\pi_1$~\cite{Pasquier:1968zz,Pasquier:1969dt,Mikhasenko:2019vhk}. This is especially true when applied to production reactions, where subchannel unitarity is easier to implement than the analogous three-body unitarity~\cite{Aitchison:1966lpz}.

An initial exploration of the $\pi_1 \to 3\pi$ decay using the KT formalism was presented in Ref.~\cite{Stamen:2022eda}. There, the authors focused on the overall importance of the rescattering corrections \textit{per se}. Instead of comparing with existing data, they identified the experimental precision required to distinguish the unitarized lineshape (in the form of a once-subtracted solution to KT) from a simpler parameterization that does not incorporate rescattering.

In this work, we take a different approach. Specifically, the presence of a wide $\pi_1$ resonance provides sufficient evidence of nontrivial three-body dynamics that should not be ignored. We therefore assume that subchannel lineshape modifications arise from rescattering, as a consequence of three-body unitarity implemented by KT. The natural question then becomes: what can be learned about the $3\pi$ mass distribution, and therefore about the $\pi_1$, from the rescattering effects present in the $\pi\pi$ isobars extracted by COMPASS? 

Instead of comparing isobars with and without rescattering effects, we compare the $\rho$ lineshapes that arise from the dynamics of different initial production mechanisms contributing to the final Dalitz plot. This approach allows us to include and identify the key physics at play in the \compass data as a function of all relevant kinematic variables, while remaining consistent with low-energy unitarity and analyticity and without arbitrary subtraction polynomials. This work is thus intended as a first step toward extracting the $\pi_1$ pole position in the $\rho\pi$ channel. 

The rest of the paper is organized as follows: in \cref{sec:isobar_model}, we describe the decomposition of the $[3\pi]_{1^{-+}}$ Dalitz plot within an isobar model. We also motivate production amplitudes consistent with the isobar picture and relevant for \compass kinematicsN. I. MuskhelishviliN. I. MuskhelishviliN. I. Muskhelishvili. In \cref{sec:KT_with_production}, we outline how to combine these to implement unitarity using the KT equations, with the production amplitudes as driving terms. The unitarized amplitudes are then compared with data from the freed-isobar analysis in \cref{sec:fits}, with a summary and outlook toward future work in \cref{sec:conclusions}.

\section{Isobar decomposition and Production Amplitudes}
\label{sec:isobar_model}

\begin{figure}[h]
    \centering
    \begin{tikzpicture}[scale=4]
        \begin{feynman}
        \vertex (b)   at (-0.5, +0.35) {$\pi$};
        \vertex (et) [dot] at (0,  0.25) {};
        \vertex (eb) [dot] at (0, -0.15) {};
        \vertex (pom)   at (0, -0.55);
        \vertex (r)  [dot] at (+0.25, 0.25) {};
        \vertex (p1) at (+0.5, +0.35) {$\pi$};
        \vertex (p2) at (+0.5,  0.1) {$\pi$};
        \vertex (p3) at (+0.5, -0.15) {$\pi$};
        \vertex (b1) at (-0.5, -0.55) {$p$};
        \vertex (b2) at (+0.5, -0.55) {$p$};

        \vertex (bb1) at (-0.35, -0.57) {};
        \vertex (bb2) at (+0.35, -0.57) {};
        \vertex (tt1) at (-0.2, +0.31) {};
        \vertex (tt2) at (+0.16, 0.27) {};
        
        \diagram*
        {
            (b)  -- [plain]  (et),
            (et) -- [plain,edge label'={$\pi$}] (eb),
            (pom)  -- [zigzag, edge label'={$\;\,\mathbb{P}$}] (eb),
            (eb) -- [plain] (p3),
            (et) -- [double, edge label'={$\rho$}] (r),
            (r)  -- [plain] (p1),
            (r)  -- [plain] (p2),
            (b1) -- [fermion] (pom) -- [fermion] (b2)
        };

        \draw [<->,jpac_red, thick] (b)  to [bend right=30,edge label'=$s$] (b1);
        \draw [<->,jpac_red, thick] (p1) to [bend left=10,edge label=$\sigma$] (p2);
        \draw [<->,jpac_red, thick] (p1) to [bend left=65,edge label=$m_{3\pi}^2$] (p3);
        \draw [<->,jpac_red, thick] (p3) to [bend left=40,edge label=$s_{\pi p}$] (b2);
        \draw [<->,jpac_red, thick] (bb1) to [bend right=30,edge label'=$t$] (bb2);
        \draw [<->,jpac_red, thick] (tt1) to [bend left=30,edge label=$\tau$] (tt2);
        \end{feynman}
    \end{tikzpicture}
    \caption{Diagrammatic representation of the Deck production mechanism at \compass and relevant invariant variables. In all plots, $\pi$ is denoted with single lines, $[2\pi]_{1^{--}}$ by double lines, and $[3\pi]_{1^{-+}}$ by triple lines. The pomeron is denoted by a zigzag and protons by fermion lines.}
    \label{fig:deck_diagram}
\end{figure}
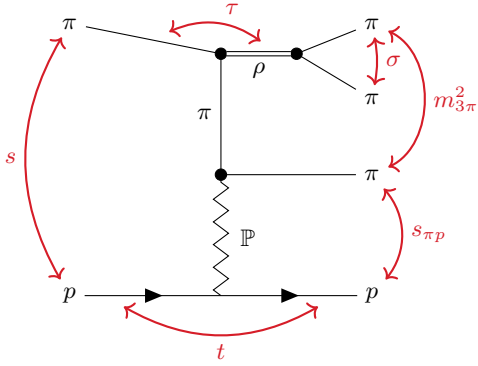

At the heart of the KT formalism is the isobar assumption that allows the $3\pi$ amplitude to be written as a crossing symmetric combination of analytic functions with only the unitarity cut. As previously mentioned, this formalism is typically applied to narrow resonance decays where the total invariant mass can be considered a fixed value and thus the only relevant dynamics enters only through functions of the subchannel invariant mass. Because of its larger width, a robust exploration of the $\pi_1$ requires an amplitude with explicit total $3\pi$ mass dependence. One must also account for production effects, such as the Deck mechanism (i.e., $\rho\pi$ production via a pion exchange with a spectator, as in \cref{fig:deck_diagram}), which will generate logarithmic singularities when projected onto the $P$-wave and may complicate the identification of the $\pi_1$ signal~\cite{Deck:1964hm,Ascoli:1973htr,Ascoli:1974hi,Aaron:1977wa,Basdevant:1977ya}.

In this section, we motivate the isobar decomposition for the $[\pi\pi]_{1^{--}}\pi\,P$-wave and how to include arbitrary model dependence external to the decay. Such a generalization will allow us to include  resonant $\pi_1$ lineshapes or the Deck mechanism, while staying consistent with low energy analyticity and crossing symmetry and will be the starting point for unitarization via the KT equations in the following section.

Although the full \compass process of interest is $2\to4$, to ease the connection with the usual KT formalism and the notion of isobars, we begin by motivating the description of the Dalitz plot for the simpler $1\to 3$ reaction:
\begin{equation}
    \label{eq:decay_process}
    \pi_1(q) \to \pi (q_a) \, \pi (q_b) \, \pi \, (q_c) ~,
\end{equation}
where \mbox{$q^2 = \mtp^2$}, and \mbox{$q_i^2 = \mu^2$} denotes the mass of the final-state pions.\footnote {We work in the isospin limit with $\mu^2 = m^2_{\pi^+}$ and ignore any complications from isospin breaking effects.} 
The initial state can be viewed as a shorthand for the projected $[3\pi]_{1^{-+}}$ system at a fixed $\mtp$ and not necessarily a resonance.
We can define the usual invariant-mass variables, which we denote:
\begin{align}
    \label{eq:sigmas_def}
    \sigma_i = (q - q_i)^2 \quad \text{with} \quad  i \in \{a,b,c\}~,
\end{align}
satisfying $\sum_i \sigma_i = 3\,\mu^2 + \mtp^2 \equiv \Sigma$. 

The decay amplitude for the process in \cref{eq:decay_process} will, in principle, depend on the helicity projection of the $\pi_1$ system. However, because the $1^{-+}$ wave has natural parity, we only need to consider a single scalar amplitude corresponding to a transverse $\pi_1$. The decay amplitude can be written in the factorized form~\cite{Albaladejo:2019huw,Stamen:2022eda}:
\begin{align}
    \label{eq:reduced_amplitude}
    \mathcal{F}(\sigma_a,\sigma_b) = 2\, \left[ \epsilon_{\mu\alpha\beta\gamma} \,\pi_1^\mu \, q_a^\alpha \, q_b^\beta \, q_c^\gamma\right]\, &\hat{\mathcal{F}}(\sigma_a, \sigma_b) \nonumber \\
    = \sqrt{\phi} \,  &\hat{\mathcal{F}}(\sigma_a, \sigma_b)  ~,
\end{align}
where $\pi_1^\mu$ is the polarization vector of the initial state system.\footnote{This prefactor is identical to those appearing in $V\to3\pi$ (see e.g.\ \cite{Niecknig:2012sj,JPAC:2020umo,JPAC:2023nhq}) which also have natural-parity initial states of spin-1.}
We include a factor of two in the normalization to remove overall numerical factors when evaluated in terms of the Kibble function~\cite{Kibble:1960zz}:
\begin{equation}
    \label{eq:kibble}
    \phi = \sigma_a \, \sigma_b \, \sigma_c - \mu^2 \, (\mtp^2 - \mu^2)^2 ~.
\end{equation}
We note that \cref{eq:reduced_amplitude} has implicit $\mtp$ dependence, e.g., through the phase space and in \cref{eq:kibble}. In anticipation of extensions below, we will denote functional dependence on $\mtp$ (or other kinematic variables) only if it enters through an explicit feature of the model.

The remaining reduced decay amplitude $\hat{\mathcal{F}}$ is related to the scattering process $\pi_1 \, \pi \to \pi \, \pi$ by crossing symmetry and can thus be decomposed into an expansion in partial-wave amplitudes of definite spin and isospin. In an isobar model, the dominance of two-body subsystems with the lowest spin is used to write a truncated, crossing-symmetric summation (a sketch of this derivation can be found in \cref{app:isobar_derivation} or a fully detailed discussion in Ref.~\cite{Albaladejo:2019huw}):
\begin{equation}
    \label{eq:isobar_decomposition}
    \mathcal{F}(\sigma_a, \sigma_b) = \sqrt{\phi} \times \left[F^{1}_1(\sigma_a) - F^1_1(\sigma_b) \right] ~.
\end{equation}
Here, $F^I_j(\sigma)$ is called an isobar and parameterizes the dynamics of a specific $\pi\pi$ subchannel. For \cref{eq:isobar_decomposition} (i.e., $I= j=1$), each term describes $\pi\pi$ scattering in $P$-wave with the third pion acting as a spectator. Since this partial wave is dominated by the $\rho$ resonance each isobar can be thought of as the propagation and subsequent decay of the $\rho$ in a two-body subsystem, e.g., in the decay chain $\pi_1 \to \pi[\rho\to\pi\pi]$. The two resonant channels appear with a relative minus sign to ensure the entire amplitude is manifestly crossing symmetric~\cite{Albaladejo:2019huw}.\footnote{For a hypothetical $\pi^0$ beam, the $\pi^+\pi^-\pi^0$ Dalitz plot will thus have a $\rho^+$ and a $\rho^-$ interfering. For the COMPASS $\pi^-$ beam, two final states are allowed: $\pi^0\pi^0\pi^-$, with two $\rho^-$ interfering, and $\pi^+\pi^-\pi^-$, with two $\rho^0$ interfering. Regardless, the interference is always destructive.}

Isobar functions differ crucially from partial-wave amplitudes in three ways: first, the anti-symmetry with respect to $\sigma_a \leftrightarrow \sigma_b$ induces the so-called ``isobar ambiguity" where the full amplitude, and therefore the Dalitz intensity, is invariant under the translation by an arbitrary constant (from here on, we drop the indices on the isobars)
    \begin{equation}    
        \label{eq:ambiguity}
        F(\sigma) \to \alpha  + F(\sigma) ~.
    \end{equation}
This invariance reflects an essential ambiguity in isobar decompositions wherein any given isobar (e.g. fixed to data) represents only one of an infinite number of solutions which yield the same total amplitude in Eq.~\eqref{eq:isobar_decomposition}. In the present case of the $J^{PC} = 1^{-+}$ quantum numbers, such ambiguity will prevent us from determining the absolute normalization of $F$ from the data.

Second, they are typically assumed to contain only a right-hand cut constrained by (two-body) unitarity and thus to obey  $m$-subtracted dispersion relations of the form:
\begin{subequations}
    \label{eq:subtracted_dispersion}
\begin{equation}
    F(\sigma) = P_{m-1}(\sigma)  + \frac{\sigma^m}{\pi}\int_{4\mu^2}^\infty \, \frac{d\sigma^\prime}{(\sigma^\prime)^m} \frac{\Disc F(\sigma^\prime)}{\sigma^\prime - \sigma} ~,
\end{equation}
with
    \begin{equation}
        \Disc F(\sigma) = \frac{1}{2i} \left[F(\sigma + i\epsilon) - F(\sigma-i\epsilon) \right] ~.
    \end{equation}
\end{subequations}
Due to analyticity, the isobar is determined by its discontinuity, modulo an \textit{a priori} unknown polynomial $P_{m-1}(\sigma)$ that parameterizes effects not fixed by unitarity and must be determined on a process-by-process basis.

Finally, isobars forego the analyticity in angular momentum expected of physical partial-wave amplitudes~\cite{Collins:1977jy,Gribov:2003nw}. This means the full amplitude will exhibit unphysical (i.e., non-Regge) behavior when continued to any of the scattering kinematics at high energies. 
While some of these shortcomings can be overcome with more complex isobar models~\cite{Stamen:2024gfz}, the simpler dispersive form of \cref{eq:subtracted_dispersion} is still expected to be reliable for the decay process in which the $\pi_1$ mass limits the physical region of interest well below the model's high-energy breakdown, and in which the unknown subtraction coefficients can be determined by fits to Dalitz plot distributions.

\Cref{eq:isobar_decomposition,eq:subtracted_dispersion} are the standard starting point for a traditional KT analysis where one absorbs any other effect not determined by unitarity into the unknown subtractions without explicitly modeling them. 
The isobar assumption itself, i.e., that the process proceeds entirely through pairwise interactions, can be used to include the core features of arbitrary production amplitudes in a straightforward way. Specifically, we consider production and decay to occur in two steps, allowing the replacement~\cite{Aitchison:1966lpz}:
\begin{equation}
    \label{eq:production_replacement}
    F(\sigma) \to F_B(t,\mtp^2;\sigma) = B(t,\mtp^2; \sigma) \, F(\sigma)~,
\end{equation}
where, a function $B$ parameterizes the production of quasi-stable $\rho \pi$ and depends on both the running mass of the $[\pi\pi]_{1^{--}}$ subsystem and $\mtp$ and $t$ are the relevant variables of the production process (e.g. in \cref{fig:deck_diagram}).~\footnote{Being a $2\to4$ process it will generally depend on 5 additional kinematic variables: the total energy of the reaction, $s$, and solid angles describing the $[3\pi] \to [\pi\pi] \pi$ and $[\pi\pi]\to \pi\pi$ decays. Since we always consider both these systems projected onto definite quantum numbers, this angular dependence is already integrated over. Further as \compass measures only a single value of $s$, we consider it a fixed parameter and do not show functional dependence.} The original isobar $F$ then weights the $\sigma$ distribution with the $\rho \to \pi\pi$ decay lineshape. When inserted into \cref{eq:isobar_decomposition}, this construction includes production in every possible $\pi\pi$ channel and thus preserves crossing symmetry.

If crossed-channel rescattering can be ignored (e.g., in some systems with large decay mass~\cite{Stamen:2022eda}), then \cref{eq:isobar_decomposition,eq:subtracted_dispersion} with \cref{eq:production_replacement} are sufficient to construct a $3\pi$ production amplitude consistent with low-energy analyticity, unitarity, and crossing symmetry. In this approximation, only the resonant $\pi\pi$ final state scattering is relevant and can be incorporated by imposing ``homogeneous" $\pi\pi$ unitarity:
    \begin{align}
        \label{eq:homogeneous}
        \text{Disc}_\sigma\, &F_B^{(0)}(t,\mtp^2;\sigma) = \text{Disc}_{\sigma} \, \left[B(t,\mtp^2;\sigma)\, F^{(0)}(\sigma)\right] \nonumber
         \\
        &= \sin\delta(\sigma) \, e^{-i\delta(\sigma)} \,B(t,\mtp^2;\sigma) \, F^{(0)}(\sigma)  ~, 
    \end{align}
with respect to the $P$-wave $\pi\pi$ phase shift, $\delta(\sigma)$. We use the superscript to denote the homogeneous solution which will be Born term, i.e. the ``zero-th" iteration, of the full unitarity corrections to be considered in \cref{sec:KT_with_production}.

As, in principle, $B$ represents a $3\pi$ production amplitude, it can contain the three-body cut starting at $m_{3\pi}^2 \geq (3\mu)^2$ (e.g. as in Ref.~\cite{Hoferichter:2014vra}). It, however, does not contain the $2\pi$ cut and thus $B$ should contain at most a left-hand cut for $\sigma \leq 0$.\footnote{In principle, finite cuts in the right-half plane are also possible (see, e.g., Refs.~\cite{Jackura:2018xnx,Dawid:2023jrj}) but not relevant for the current discussion as they do not overlap the physical region.} This means that above unitarity cut in $\sigma$, $B$ will factor from \cref{eq:homogeneous} such that the solution can be written in closed form as the Omn\`{e}s function~\cite{Omnes:1958hv,Muskhelishvili:1958}:
\begin{subequations}
    \begin{equation}
        \label{eq:fb_homogenous}
        F^{(0)}_B(t,\mtp^2;\sigma)  = B(t,\mtp^2;\sigma) \, \Omega(\sigma) ~,
    \end{equation}
with 
    \begin{equation}
        \label{eq:omnes}
        F^{(0)}(\sigma) =  \Omega(\sigma) = \exp\left(\frac{\sigma}{\pi}\int_{4\mu^2}^\infty \frac{d\sigma^\prime}{\sigma^\prime}\, \frac{\delta(\sigma^\prime)}{\sigma^\prime - \sigma }\right)~.
    \end{equation}
\end{subequations}
Here we omit the subtraction polynomial appearing in \cref{eq:subtracted_dispersion} which can be absorbed into the production function. 

This polynomial is typically used to parameterize ignorance of effects unconstrained by two-body unitarity and in some cases has been matched to fixed-order $\chi$PT expansions, as in so-called ``reconstruction theorems" (e.g., in Ref.~\cite{Stern:1993rg}). Insofar as the isobar and elastic rescattering approximations are valid, however, these can only arise from the left-hand cuts induced by production. The typical KT construction therefore assumes these effects are smooth and can be parameterized by a polynomial:
    \begin{equation}
        \label{eq:B_subs}
        B_{m\text{-subs}}(\sigma) = \sum_{i=0}^{m-1} N_i \, \sigma^i~.
    \end{equation}
More physics-motivated (non-polynomial) functions can be used if the production dynamics is reasonably known (see, \eg, Ref.~\cite{Albaladejo:2017hhj}). In the following, we therefore motivate specific Born terms relevant for the hadroproduction of the $\pi_1$ at \compass rooted in meson exchange and Regge phenomenology.

\subsection{Short-range production}
\label{subsec:contact}
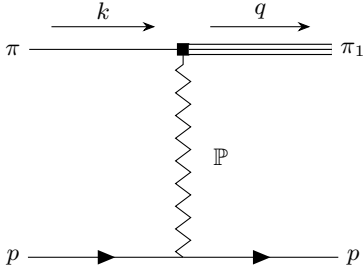
\begin{figure}[h]
    \centering
    \begin{tikzpicture}[scale=5]
    \begin{feynman}
    \vertex (b)   at (-0.45, 0.25) {$\pi$};
    \vertex (et) [square dot] at (0,  0.25) {};
    \vertex (r)    at (+0.45, 0.25) {$\pi_1$};
    \vertex (pom)  at (0,    -0.3);
    \vertex (pr1)  at (-0.45, -0.3) {$p$};
    \vertex (pr2)  at (+0.45, -0.3) {$p$};
    \vertex (t)  at (+0.1, -0.035) {$\mathbb{P}$};
    \diagram*
    {
        (b)    -- [plain, momentum=$k$]  (et),
        (pom)  -- [zigzag] (et),
        (et)   -- [triple,momentum=$q$] (r),
        (pr1)  -- [fermion] (pom) -- [fermion] (pr2)
    };
    \end{feynman}
\end{tikzpicture}
    \caption{Diagrammatic representation of contact, resonance-like hadro-production off a proton target.}
    \label{fig:contact_diagram}
\end{figure}

We first consider the simplest production mechanism for the $[3\pi]_{1^{-+}}$ system, which is direct, local production as in \cref{fig:contact_diagram}. Unlike \cref{fig:deck_diagram}, which we will consider below, this mechanism produces the $3\pi$ final state coherently and thus mimics the $2\to2$ production of a resonance via Pomeron exchange --- without microscopic dependence on any of the decay variables. As such, this production process will serve as a baseline with which to identify the general kinematic structure of the reaction and build more complicated production models.

Because this diagram is mediated by a Regge exchange, it is expected to factorize and thus be readily modeled with effective Lagrangians.
The most relevant piece for our purposes is the local $\pi\pi_1\pom$ interaction at the ``top" vertex for which we write the Lagrangian corresponding to the production of the $M^\epsilon = 1^+$ wave, ignoring numerical factors:
\begin{align}
    \label{eq:L_contact}
    \mathcal{L}^\text{Contact}_{\pi_1\pi\pom} & = \epsilon_{\mu\nu\alpha\beta} \, \pi_1^\mu \, \partial^\nu \pi \, \partial^\alpha \, \pom^\beta ~.
\end{align}
As we consider \cref{fig:contact_diagram} in the high-energy, near-forward kinematics, we may consider the Pomeron trajectory near $\alpha_\pom(t=0) \approx 1$ and approximate the Pomeron as a vector field~\cite{Close:1999bi,Lesniak:2003gf,JPAC:2017dbi,JPAC:2018zyd,Winney:2025tla}. As in \cref{eq:reduced_amplitude}, the $\pi_1$ field acts as a projector for the $3\pi$ system of mass $\mtp$, ensuring that the final-state pions are in the $1^{-+}$ wave without explicitly integrating over decay angles. Due to parity, we have only this single Lorentz structure. 

With this vertex and a generic, helicity-dependent nucleon vector $h^\mu_{\lambda\lambda^\prime}$ parameterizing the ``bottom" vertex (its explicit form will not be needed), the amplitude for \cref{fig:contact_diagram} can be written as:
\begin{align}
    \label{eq:amp_contact}
    B_\text{Contact}(s,t,\mtp^2) & =  \\
    N_c(\mtp^2) \,                  s^{\alpha_\pom(t)} \, & e^{c_0t} \left[\epsilon_{\mu\nu\alpha\beta} \, \pi_1^\mu k^\nu q^\alpha \, h^\beta_{\lambda\lambda^\prime} \right] ~, \nonumber
\end{align}
in terms of an arbitrary complex $\mtp$-dependent normalization,\footnote{Because the experimental intensity is not absolutely normalized, the mass dimension of $N_c$ will be irrelevant. Still, factors of $s_0 = 1\,\text{ GeV}^{2}$ are assumed wherever appropriate.} as well as the typical Regge factors associated with the Pomeron trajectory $\alpha_\pom(t)$ and the expected exponential fall-off for diffractive reactions. Note that, because the contact interaction does not depend microscopically on the $\rho\pi$, no arbitrary polynomials of $\sigma$ can arise, even with more complicated Lagrangians. 

The remaining term in the square brackets consists solely of kinematical factors that arise from the helicity structure, e.g., the so-called half-angle factors. In the Regge limit, relevant for \compass kinematics, this entire factor reduces to a power of $t$ that depends on the net helicity at each vertex~\cite{Collins:1977jy,Winney:2025tla}. Specifically, in this limit, we may replace:

\begin{equation}
    \label{eq:t_factors}
    \left[\epsilon_{\mu\nu\alpha\beta} \, \pi_1^\mu k^\nu q^\alpha \, h^\beta_{\lambda\lambda^\prime} \right] \longrightarrow \sqrt{-t}^{1 + |\lambda-\lambda^\prime|} = \sqrt{-t} ~,
\end{equation}
where we assume the dominance of $s$-channel helicity-conserving interactions between the target and the recoil proton (i.e., $\lambda = \lambda^\prime$). This assumption has been verified in a number of high-energy diffractive production processes involving the Pomeron (e.g., in Refs.~\cite{Mathieu:2018xyc,H1:2005dtp,ZEUS:1995bfs}). 
\Cref{eq:t_factors} indicates that Regge theory requires the $t$-distribution to vanish at forward angles, consistent with the observed event distributions in the initial (not freed-isobar) analysis at \compass~\cite{COMPASS:2018uzl}. 

Finally, because the freed-isobar data is only available for a fixed value of $s$, we will not be sensitive to the total invariant mass dependence. We may thus absorb the dependence of the Regge factor into the undetermined parameters by defining \hbox{$c = c_0 + \alpha^\prime_\mathbb{P} \log s$}, where we approximate the Pomeron trajectory as linear with slope $\alpha^\prime_\mathbb{P}$. This reduces \cref{eq:amp_contact} to:

\begin{equation}
    \label{eq:B_contact}
    B_\text{Contact}(t, \mtp^2) = N_c(\mtp^2) \, \sqrt{-t}\,e^{ct} ~.
\end{equation}
In principle, if \cref{eq:B_contact} represents the full production, the pole position of the $\pi_1$ can be extracted from knowledge of $N_c$ using explicit three-body unitarity. Additionally, because the short-range production treats the $[3\pi]_{1^{-+}}$ system coherently and contains no $\sigma$ dependence, it will be functionally equivalent to a constant from the perspective of the KT equations, which are insensitive to overall factors.

\subsection{Deck mechanism}
\label{subsec:deck}
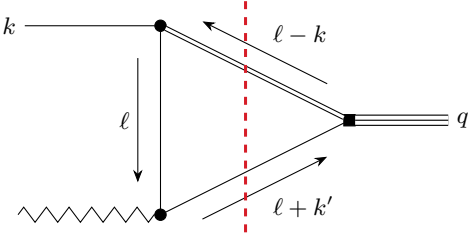
\begin{figure}[h]
    \centering
        \begin{tikzpicture}[scale=5]
        \begin{feynman}
        \vertex (b)   at (-0.4, +0.25) {$k$};
        \vertex (et) [dot] at (0,  0.25) {};
        \vertex (eb) [dot] at (0, -0.25) {};
        \vertex (pom)   at (-0.4, -0.25) {};
        \vertex (p3) [square dot] at (+0.5, 0.) {};
        \vertex (pi1) at (0.8, 0) {$q$};
        \diagram*
        {
            (b)  -- [plain]  (et),
            (et) -- [plain, momentum'={$\ell$}] (eb),
            (pom)  -- [zigzag] (eb),
            (eb) -- [plain, momentum'={$\ell +k^\prime$}] (p3) -- [triple] (pi1),
            (et) -- [double, rmomentum={$\ell-k$}] (p3)
        };
        \draw [dashed,color={jpac_red},line width = 0.4mm] ($(0.225,0.315)$) -- ($(0.225,-0.315)$);
        \end{feynman}
    \end{tikzpicture} 
    \caption{The Feynman loop whose discontinuity corresponds to the Deck projected onto the $1^{-+}\rho\pi~P$-wave. Particle labels are the same as \cref{fig:deck_diagram,fig:contact_diagram}. $k^\prime = q-k$ and we omit the nucleon vertex which is irrelevant. As we are interested on the amplitude for on-shell $\rho\pi$, we cut the diagram along the red line. The subenergy $\sigma$ is taken as the $\rho$ mass squared.}
    \label{fig:deck_triangle}
\end{figure}

For Deck production, we must consider the microscopic dependence on the $\rho\pi$ system being produced via pion exchange. This can, in principle, be readily calculated by considering the diagram in \cref{fig:deck_diagram} and projecting the $\rho\pi$ system onto the $P$-wave, such that:
    \begin{equation}
        \label{eq:deck_pwp}
        B_\text{Deck}(t,\mtp^2;\sigma) \propto \frac{1}{2} \int_{-1}^1 \, dz \,  \frac{1-z^2}{\mu^2 - \tau(z)} ~,
    \end{equation}
in terms of the momentum transfer $\tau(z)$, which also implicitly depends on $t,\,\mtp^2,$ and $\sigma$ (cf. \cref{eq:tau}). To draw an analogy to the structure of the contact term of the previous subsection, we will calculate the partial wave using effective Lagrangians, as this will allow a clear identification of the kinematic structure. To this end, we consider the $\pi_1$ field as a projector so that \cref{eq:deck_pwp} is equivalent to considering the discontinuity of the triangle diagram in \cref{fig:deck_triangle}, taking $\sigma$ as the $\rho$ mass squared. As we require that the $\rho\pi$ intermediate state be on-shell, the discontinuity of the loop must be taken across the $\mtp$ cut.\footnote{Cutting the loop diagram will also introduce a $P$-wave $\rho\pi$ phase space factor, which needs to be removed to not double count the kinematical factors already given by the Kibble function in \cref{eq:reduced_amplitude}.}

\begin{figure}
        \centering
        \includegraphics[width=\linewidth]{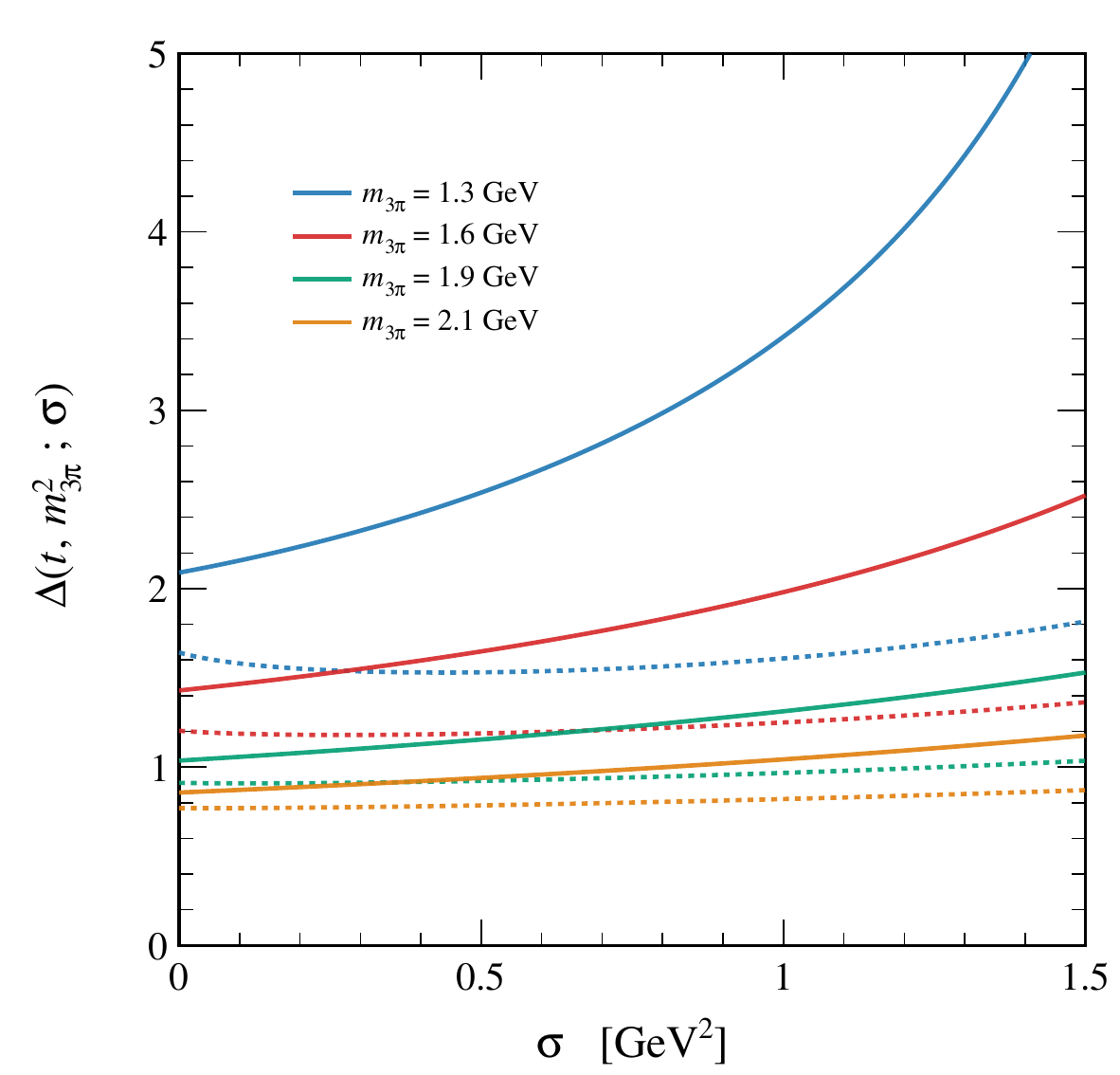}
        \caption{Dependence of $\Delta$ on the various production variables. Solid curves denote fixed $t= t_0 = -0.12$, which is the smallest $t$-bin considered by the \compass analysis, while dashed curves fix $t=t_3 = -0.66$, which is the largest. }
        \label{fig:delta}
    \end{figure}

Examining \cref{fig:deck_triangle}, the required vertices are (we again ignore couplings and other numerical factors):
\begin{subequations}
    \label{eq:deck_lagrangians}
    \begin{align}
        \mathcal{L}_{\rho\pi\pi} = i\,\rho^\mu \, \pi \overset{\leftrightarrow}{\partial}_\mu \pi
        ~, \quad
        \mathcal{L}_{\mathbb{P}\pi\pi} = i\, \mathbb{P}^\mu \, \pi \overset{\leftrightarrow}{\partial}_\mu  \pi ~,
    \end{align}
    and
    \begin{align}
        \mathcal{L}_{\pi_1\rho\pi} & = \epsilon_{\mu\nu\alpha\beta} \, \pi_1^\mu \, \partial^\nu\pi \, \partial^\alpha  \rho^\beta ~, 
    \end{align}
\end{subequations}
noting that the pion-exchange subprocess is of the form $V\pi\to V\pi$. With \cref{eq:deck_lagrangians}, one may write the loop in \cref{fig:deck_triangle} as
\begin{subequations}
    \label{eq:loop_integral}
    \begin{align}
        \label{eq:T_loop}
        \mathcal{T} & (t,\mtp^2;\sigma) =                                                                                                    \\
                    & \frac{i}{\pi^2}\int \frac{\mathcal{N}(\ell,q,k) \, d^4\ell}{[\ell^2-\mu^2] \, [(\ell-k)^2 - \sigma] \, [(\ell+q-k)^2 - \mu^2]}~,
        \nonumber
    \end{align}
    where the numerator is given by:
    \begin{align}
        \label{eq:Nhat}
        \mathcal{N}(\ell,q,k) = \ell^\mu \ell^\nu \left[\pom_\mu \, \epsilon_{\nu\alpha\beta\gamma} \pi_1^\alpha\, k^\beta q^\gamma\right] \equiv \ell^\mu \ell^\nu \, \hat{\mathcal{N}}_{\mu\nu}~.
    \end{align}
\end{subequations}
We observe that the reduced tensor $\hat{\mathcal{N}}_{\mu\nu}$ depends only on external momenta and obeys the orthogonality relations:
\begin{equation}
    \label{eq:N_orthog}
    (q-k)^\mu \, \hat{\mathcal{N}}_{\mu\nu} = q^\nu \, \hat{\mathcal{N}}_{\mu\nu} = k^\nu \, \hat{\mathcal{N}}_{\mu\nu} = 0 ~.
\end{equation}
As a result, performing a Passarino-Veltman decomposition~\cite{Passarino:1978jh} of the remaining tensor integral yields only a non-zero contribution proportional to $g^{\mu\nu}$. We can therefore write \cref{eq:loop_integral} as:
\begin{equation}
    \label{eq:mT}
    \mathcal{T}(t,\mtp^2; \sigma) =  \left[ \epsilon_{\mu\nu\alpha\beta}\, \pi_1^\mu\, k^\nu q^\alpha \,\pom^\beta\right] \times \hat{\mathcal{T}}(t,\mtp^2;\sigma) ~,
\end{equation}
where the prefactor is precisely that which arises from \cref{eq:L_contact} and $\hat{\mathcal{T}}$ is a scalar, logarithmically-divergent three-point integral.\footnote{In the common notation for one-loop integrals, e.g.\ used in \texttt{LoopTools}~\cite{Hahn:1998yk}, we have $\hat{\mathcal{T}}(t,\mtp^2;\sigma)=(4/\pi) \, C_{00}(\mu^2,t,\mtp^2,\mu^2,\mu^2,\sigma)$.} 
Cutting this loop, i.e. putting the intermediate state on-shell, we have
    \begin{align}
         \frac{1}{2i}[\hat{\mathcal{T}}(t,\mtp^2 +i\epsilon;\sigma) -  &\hat{\mathcal{T}}(t,\mtp^2-i\epsilon;\sigma)]  \nonumber \\
           &= \rho_{\pi\rho P} \, \Delta(t,\mtp^2;\sigma) ~,
    \end{align}
where $\rho_{\pi\rho P}$ is the quasi-two body phase space of the $[\pi\pi]_{1^{--}}\pi$ intermediate state in $P$-wave. The remaining function is the $P$-wave projection of the pion exchange (i.e., on the left-hand side of the cut in \cref{fig:deck_triangle}):
    \begin{align}
        \Delta(t,\mtp^2;\sigma)                                & =  \frac{1}{|\bm{q}| |\bm{k}|} \left[(1-\zeta^2)\,Q_0(\zeta)+\zeta\right] ~,
        \label{eq:Delta}
    \end{align}
which is consistent with the $\phantom{}^3P_1\to\phantom{}^3P_1$ projection of the one-particle exchange (OPE) as calculated in Ref.~\cite{Jackura:2023qtp}. We relegate kinematic definitions and discussion of these expressions to \cref{app:continuation_of_triangle} but comment that $\Delta$ will depend on both $t$ and $\mtp^2$ as well as $\sigma$, as can be seen in \cref{fig:delta}.

Matching the structure of \cref{eq:mT} to \cref{eq:L_contact}, with $\hat{\mathcal{T}} \to \Delta$ (i.e. cutting the loop and removing the phase space), the Deck induces an effective ``coupling" relative to the contact term, given by:
    \begin{equation}
        \label{eq:L_Deck}
        \mathcal{L}_{\pi_1\pi\pom}^\text{Deck} = \mathcal{L}^\text{Contact}_{\pi_1\pi\pom} \times \Delta(t,\mtp^2;\sigma) ~,
    \end{equation}
and thus the driving term can be written in complete analogy to \cref{eq:B_contact}:
\begin{equation}
    \label{eq:B_deck}
    B_\text{Deck}(t,\mtp^2;\sigma) = N_d(\mtp^2) \, \sqrt{-t}\, e^{dt} \, \Delta(t,\mtp^2;\sigma) ~,
\end{equation}
in terms of the function $\Delta$ and, in general, different normalization and $t$-slope, $N_d$ and $d$ respectively.

We briefly comment on three technical aspects implicit in \cref{eq:B_deck}. First, due to the spin factors involved, the loop integral in {\cref{eq:mT}} is divergent. Taken at face value, a counter-term should be introduced, as in an EFT, to remove this divergence and absorb the indeterminacy associated with the UV behavior of the (off-shell) particles propagating inside the loop. The loop, however, is only a tool to more effectively project the pion exchange onto the $1^{-+}\,[\pi\pi]_{1^{--}}\pi~P$-wave. After cutting the loop, the physical, on-shell amplitude of interest has no associated UV divergences or indeterminacies.

Second, similar to \cref{eq:amp_contact}, the Regge factor associated with the Pomeron appears as $s^{\alpha_\pom(t)}$ (and is subsequently absorbed). One might expect the factor to be expressed in terms of the invariant mass of the lower exchange, i.e., $s_{\pi p}$ (cf. \cref{fig:deck_diagram}). In the present kinematics, however, we are firmly in the single Regge regime, with $s \gg \mtp^2 > |t|$ as the only large mass scale. Thus, we may safely assume $s_{\pi p}\approx s$. This is especially true given that we will be insensitive to details of the recoil proton distribution in the Dalitz plot data.

Finally, in the loop integrand, we assume a simple propagator for the exchanged pion. In similar studies, a phenomenological form factor may be required to parameterize effects arising from the finite size of the exchanged particle, which lead to a steeper momentum-transfer distribution. The \compass analysis in Ref.~\cite{COMPASS:2021ogp}, for instance, adds an additional exponential factor to adequately describe the angular distribution within their Deck model. We can mimic this effect by modifying the loop integrand with a customary monopole form factor:
\begin{equation}
    \label{eq:monopole}
    \frac{1}{\ell^2 - \mu^2} \longrightarrow \frac{1}{\ell^2-\mu^2} \left(\frac{\mu^2-\Lambda^2}{\ell^2-\Lambda^2}\right)~,
\end{equation}

In terms of a cutoff momentum scale $\Lambda > \mu$, which is expected to yield an interaction range $R \approx \Lambda^{-1} = \mathcal{O}(1~\text{fm})$. Inserting this into \cref{eq:T_loop}, we can use

\begin{equation}
    \frac{\mu^2-\Lambda^2}{(\ell^2-\mu^2)\,(\ell^2-\Lambda^2)} = \frac{1}{\ell^2-\mu^2} - \frac{1}{\ell^2-\Lambda^2} ~,
\end{equation}
to observe that the effect of a form factor is functionally equivalent to the inclusion of additional production terms involving heavier exchanged particles. Reggeization effects for the exchanges pion similarly can be incorporated as a $t$-dependent form factor (since $s$ is fixed) and thus can also be included similarly. As discussed in \cref{app:not_considered}, diagrams with heavier exchanges can be effectively absorbed in the contact term and are not included explicitly. 

\subsection{Isobar ambiguities in the presence of left-hand cuts}
\label{subsec:ambiguities}
Here we very briefly comment on the issue of the polynomial ambiguity of \cref{eq:ambiguity}. In the present formalism, the $B$-terms incorporate additional dynamical features through their dependence on production variables, such as three-body cuts and exchange-particle poles in the crossed channel. This differentiates them from other terms in their analytic structure and gives them an intrinsic physical interpretation. As a result, any redefinition of the form \cref{eq:ambiguity}, which must not depend on $\sigma$, can only be absorbed by the contact term. This ambiguity would therefore complicate the resolution of the $\pi_1$ lineshape if only the contact term is considered, because arbitrary $\mtp$ dependence can be added to $N_c(\mtp^2)$ while still yielding the same Dalitz plot.

If, on the other hand, we were to expect (or the data suggest) a purely Deck-like production, i.e., one with no need for a contact term, then the ambiguity in \cref{eq:ambiguity} is irrelevant, since no contribution constant in $\sigma$ will exhibit the left-hand cut and logarithmic singularity of the $P$-wave projected OPE. When both are considered together, the ambiguity in the contact term is trivially absorbed into the overall normalization of the full amplitude, since only the relative sizes and phases of the terms in \cref{eq:F_basis} are of physical interest. From this perspective, any process which resolves a microscopic $\rho\pi$ interaction (in this case the Deck mechanism) is, in fact, the key to a meaningful extraction of the $\mtp$ lineshape, since the additional structure resolves the ambiguity of the isobar decomposition. 

\section{Inhomogenous Rescattering Corrections}
\label{sec:KT_with_production}

    \begin{figure*}
        \centering
        \begin{tikzpicture}[scale=4]
        \begin{feynman}
        \vertex (b1)  at (-0.5, -0.55) {$p$};
        \vertex (b2)  at (+0.35, -0.55) {$p$};
        \vertex (b)   at (-0.5, +0.35) {$\pi$};
        \vertex (et) [dot] at (0,  0.25) {};
        \vertex (eb) [dot] at (0, -0.15) {};
        \vertex (pom) at (-0.15, -0.35);
        \vertex (r)   at (+0.25, 0.25);
        \vertex (p1)  at (+0.85, +0.25);
        \vertex (p2)  at (+0.7,  -0.15);
        \vertex (p3)  at (+0.4, -0.15);
        \vertex (p4)  at (+0.95, +0.25);
        \vertex (p5)  at (+0.95, -0.15);
        \vertex (t1)  at (+1.1, 0.05) {$\dots$};
        \vertex (p6)  at (+1.25, 0.25);
        \vertex (p7)  at (+1.35, 0.25);
        \vertex (p8)  at (+1.25, -0.15);
        \vertex (p9)  at (+1.50, -0.15);
        \vertex (p10) at (+1.60, 0.25);
        \vertex (p11) at (+1.75, -0.15){$\pi$};
        \vertex (p12) at (+1.75, 0.05){$\pi$};
        \vertex (p13) at (+1.75, 0.35){$\pi$};
        \vertex (r1)  at (+1.85, 0.35);
        \vertex (r2)  at (+1.85, 0.05);
        \vertex (r3)  at (+2.05, 0.20) {\color{jpac_red} $L=1$};
        \vertex (r4)  at (+2.05, -.15);
        \vertex (r5)  at (+2.30, -.02){\color{jpac_red} $L=1$};

        \diagram*
        {
            (b)  -- [plain]  (et),
            (et) -- [plain] (eb),
            (pom)  -- [zigzag] (eb),
            (eb) -- [plain] (p3),
            (et) -- [double] (r),
            (r)  -- [plain] (p1),
            (r)  -- [plain] (p3),
            (p3) -- [double] (p2),
            (p2) -- [plain] (p1),
            (p1) -- [double] (p4),
            (p2) -- [plain] (p5),
            (p6) -- [plain] (p7),
            (p8) -- [double] (p9),
            (p7) -- [plain] (p9),
            (p7) -- [double] (p10),
            (p9) -- [plain] (p11),
            (p10) -- [plain] (p12),
            (p10) -- [plain] (p13),
            (b1) -- [fermion] (pom) -- [fermion] (b2)
        };
        \draw [<->,jpac_red, thick] (r1)  to [bend left=40] (r2);
        \draw [<->,jpac_red, thick] (r3)  to [bend left=40] (r4);
        
        \end{feynman}
    \end{tikzpicture}
        \caption{Diagrammatic representation of the solution to the KT integral equation \cref{eq:B_solution} with the Deck mechanism as a Born term. The $3\pi$ system is at all times projected onto the $1^{-+}[\pi\pi]_{1^{--}}\pi$ $P$-wave such that there is no decay angle dependence (until symmetrized in \cref{eq:isobar_decomposition}). }
        \label{fig:ladder}
    \end{figure*}
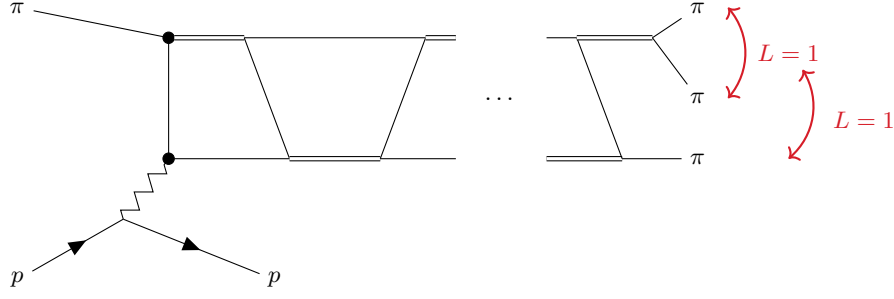

With the various forms of the $B$-functions for production mechanisms of interest, we return to the important constraint of unitarity. In a typical KT analysis, the titular ``KT equations" arise from requiring that the $P$-wave projection of \cref{eq:isobar_decomposition} satisfies two-body unitarity~\cite{Albaladejo:2019huw}. This yields a system of integral equations that encode how the isobars-spectator system must rescatter, as required by both unitarity and crossing symmetry. As argued in early work by Aitchison and Pasquier~\cite{Aitchison:1966lpz}, two-body subchannel unitarity implemented by KT is sufficient to satisfy three-body unitarity. Specifically, the KT formalism provides a model realization of $3\pi$ unitarity in which the isobars interact with the spectator pion through iterations of the OPE, without a genuine three-body force (see Ref.~\cite{Mikhasenko:2019vhk} and references therein). 

Restricting to elastic $\pi\pi$ and allowing isobar-spectator rescattering results in the ``inhomogenous" unitarity relation (cf., \cref{eq:homogeneous})
\begin{subequations}
    \label{eq:normal_KT}
    \begin{align}
        \label{eq:kt_disc}
        \Disc & F(\sigma) = \sin\delta(\sigma) \, e^{-i\delta(\sigma)} \, \bigg[ F(\sigma) + \tilde{F}(\sigma) \bigg]~,
    \end{align}
with the inhomogeneity
    \begin{equation}
        \label{eq:Ftilde}
        \tilde{F}(\sigma) = \int d\sigma^\prime \, K(\sigma, \sigma^\prime) \, F(\sigma^\prime) ~,
    \end{equation}
given in terms of an angular kernel,\footnote{For a technical discussion of solution strategies, including contour integration over $\sigma^\prime$ and handling the kinematic singularities that arise, we refer to Appendix D of Ref.~\cite{Albaladejo:2019huw} or the various appendices of Ref.~\cite{Akdag:2023oob}.} $K$, which depends only on the kinematic structure of the crossing matrix~\cite{Albaladejo:2019huw}. It is most easily expressed in a kinematic-singularity-free form by multiplying by factors of $\kappa \equiv \kappa(\sigma)$ (cf. \cref{eq:z_and_kappa}):
    \begin{align}
        \kappa^3 \, K(\sigma,\sigma^\prime) = -\frac{3}{2} \left[\kappa^2 - (2\sigma^\prime + \sigma - \Sigma)^2 \right]~.
    \end{align}
\end{subequations}

In the usual KT treatment, solutions to \cref{eq:normal_KT} can once again be obtained in terms of the Omn\`{e}s function in \cref{eq:omnes}. Specifically, one writes a dispersion relation for the ratio $F(\sigma)/\Omega(\sigma)$ such that the solution to \cref{eq:normal_KT} may be written as
\begin{align}
    \label{eq:normal_solution}
    F(\sigma & ) = \Omega(\sigma) \, \times
    \\
             & \left[\sum_{i=0}^{m-1} N_i\,\sigma^i +  \frac{\sigma^m}{\pi}\int_{4\mu^2}^\infty \, \frac{d\sigma^\prime}{(\sigma^\prime)^m} \frac{\sin\delta(\sigma^\prime) \, \tilde{F}(\sigma^\prime)}{|\Omega(\sigma^\prime)| \, (\sigma^\prime - \sigma)} \right]
    \nonumber  ~,
\end{align}
where the first term is the homogeneous solution to \cref{eq:homogeneous} with \cref{eq:B_subs}. \Cref{eq:normal_solution} typically cannot be solved analytically but is particularly amenable to numerical iteration or matrix inversion methods.

Guided by the generalization of the homogeneous case to arbitrary production functions, one might expect that the analogous extension for \cref{eq:normal_solution} simply replaces the polynomial term in \cref{eq:normal_solution} by the function $B$. Indeed, one can easily see that the modified isobar function $F_B$ introduced in \cref{eq:production_replacement} must satisfy the same unitarity relation as \cref{eq:kt_disc} at fixed values of production variables and thus admits similar solutions as \cref{eq:normal_solution}. Imposing the generalized ``boundary condition", that this recovers \cref{eq:fb_homogenous} when the inhomogeneity is zero, we arrive at
\begin{align}
    \label{eq:B_solution}
    F_B & (t,\mtp^2;\sigma) = F_B^{(0)}(t,\mtp^2;\sigma)
    \\
        &+ \Omega(\sigma)\, \frac{\sigma^m}{\pi}\int_{4\mu^2}^\infty  \frac{d\sigma^\prime}{(\sigma^\prime)^m} \frac{\sin\delta(\sigma^\prime) \, \tilde{F}_B(t,\mtp^2;\sigma^\prime)}{|\Omega(\sigma^\prime)| \, (\sigma^\prime - \sigma)}
    \nonumber  ~,
\end{align}
where $\tilde{F}_B$ follows from \cref{eq:Ftilde} except with $F\to F_B$.  

As an integral equation, \cref{eq:B_solution} iterates $\pi\pi$ rescattering in all subchannels to all orders, seeded by the initial production function, as shown diagrammatically in \cref{fig:ladder}. The $\sigma$ dependence introduced by $B$ will therefore determine the specific pattern of modifications which are induced to the $\rho$ lineshape by final-state interactions.
As a result it is important that the production amplitude can be analytically continued into the complex $\sigma$-plane for fixed values of the production variables (e.g., as discussed in
\cref{app:continuation_of_triangle}).

Formally, comparing the KT amplitudes with the $N/D$ decomposition in Ref.~\cite{Aitchison:1965zz} shows that the left-hand cut contribution introduced by the Born terms are closely related to the numerator amplitude. As a result, from the perspective of $3\pi$ unitarity, it is generally not possible to distinguish between resonant and non-resonant production (or, alternatively, signal and background) as is typically done in experimental analyses. A $\rho\pi$ system produced by the Deck process, for instance, can still resonate into the $\pi_1$ through final-state interactions.\footnote{To illustrate this, we can briefly consider a stable $\rho$. In the usual $N/D$ formalism, a unitary $\pom\pi\to\rho\pi$ amplitude (analogous to the amplitude in Ref.~\cite{JPAC:2018zyd}) with two production terms, say $F_\text{Contact} + F_\text{Deck}$, can be decomposed as $(N_\text{Contact}+N_\text{Deck})/D_{\rho\pi\to\rho\pi}$, where the $\pi_1$ would emerge as a complex root of the (common) denominator.} 

\subsection{Basis functions}
\label{subsec:basis}
Using the Born amplitudes of \cref{sec:isobar_model}, we may write a general isobar as \cref{eq:B_solution}, with $B$ given by the sum of \cref{eq:B_contact,eq:B_deck}. We remind that the KT kernel is agnostic to any $\sigma$-independent factors, which are functionally constant when solving \cref{eq:B_solution}. This admits a linear decomposition into two basis functions, in which the unknown normalizations and $t$-slopes factor out:
\begin{align}
    \label{eq:F_basis}
    F(t,\mtp^2;\sigma) =\, &N_c(\mtp^2) \, \sqrt{-t}\, e^{c t}  F_1(\sigma)  \nonumber            \\
    +\, &N_d(\mtp^2) \, \sqrt{-t} \, e^{dt} \,                  F_\Delta(t,\mtp^2;\sigma)
    ~. 
\end{align}
On the right-hand side, the two basis functions are given by \cref{eq:B_solution} with $B =1$ and $\Delta$ in \cref{eq:Delta} respectively. \Cref{eq:F_basis} can be symmetrized as in \cref{eq:isobar_decomposition} to obtain the final amplitude $\hat{\mathcal{F}}(t,\mtp^2;\sigma_a, \sigma_b)$ and to calculate the four-dimensional intensity.
As the basis functions contain no free parameters, they can be precomputed on an arbitrary grid of kinematic variables. \Cref{eq:F_basis} is therefore amenable to comparison with data to fix the normalizations and $t$-slopes, as we will see in \cref{sec:fits}.

Before then, we first examine these functions in isolation to compare the lineshape modifications resulting from each production mechanism. For each Born term and set of production variables, we calculate the basis functions by iterating the KT equations until convergence starting from the homogeneous solution \cref{eq:fb_homogenous}. As $|\Delta(t,\mtp^2;\sigma \to \infty)| \propto \sigma^{-1}$, we fix $m=1$ in \cref{eq:B_solution} to match the leading asymptotic behavior of the contact term.\footnote{More detailed discussion of this asymptotic behavior in this limit is relegated to \cref{app:asymptotic}.} In all cases, convergence is achieved in $10$ or fewer iterations. 

\begin{figure}
    \centering
    \includegraphics[width=\linewidth]{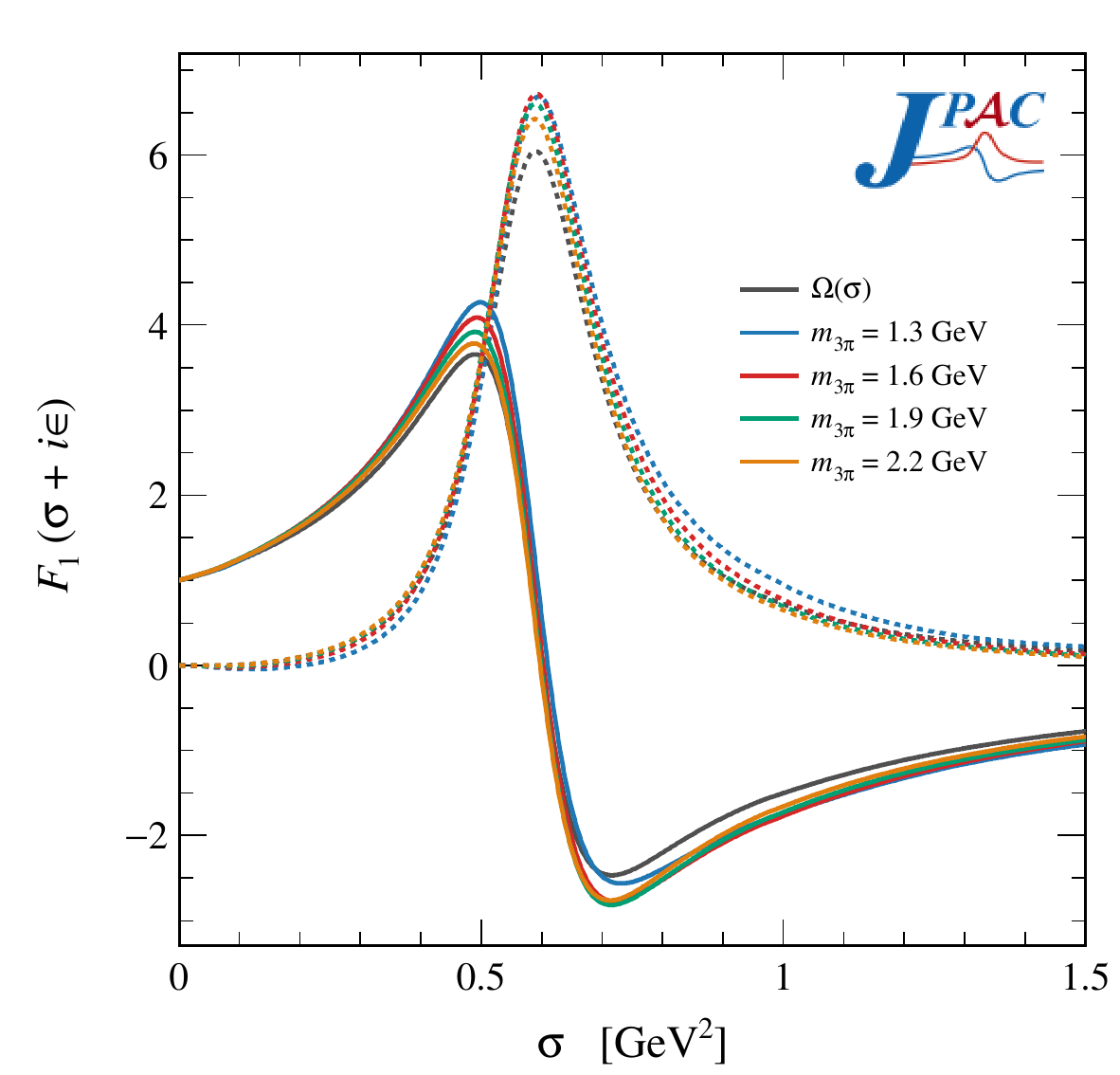}
    \caption{Real (solid) and imaginary (dashed) parts of the isobar unitarized $F_1(\sigma)$ at various values of $\mtp$. }
    \label{fig:fc_comparison}
\end{figure}

The simplest isobar is, of course, $F_1(\sigma)$, which is identical to the one that appears in a typical (once-subtracted) KT analysis, e.g. as in Ref.~\cite{Stamen:2022eda}. In this case, the $\mtp$ dependence appears only through the phase space and thus gives us a glimpse of the minimal three-body mass dependence required by unitarization. The lineshape is compared to the unmodified Omn\`{e}s function in \cref{fig:fc_comparison} for various values of $\mtp$ near the $\pi_1$ mass. Here we see that the effects of rescattering corrections are very mild, with deviations in the amplitude squared of a few percent at most.

\begin{figure}
    \centering
    \includegraphics[width=\linewidth]{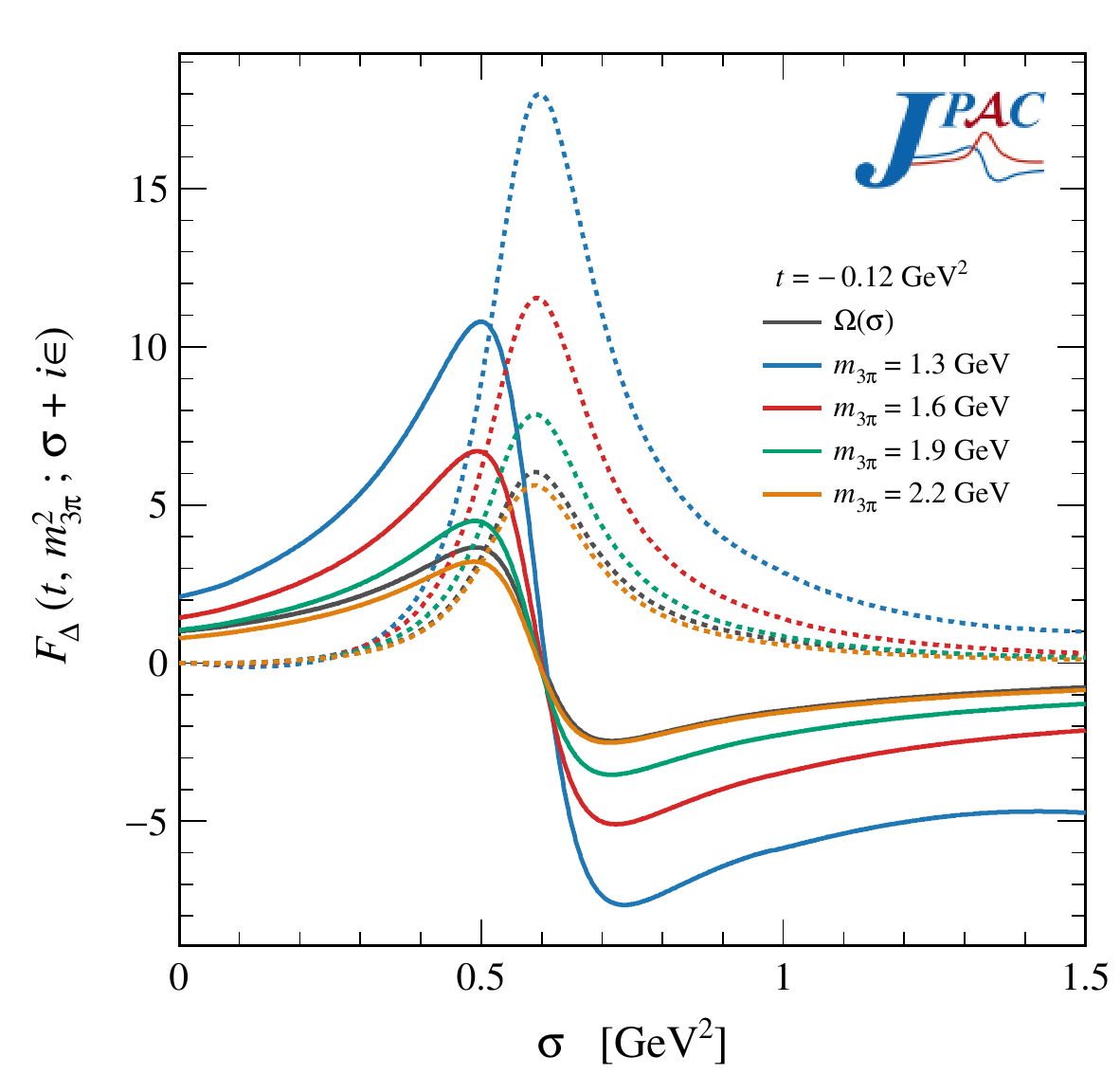}
    \caption{Analogous to \cref{fig:fc_comparison} but for $F_\Delta$. The $t$-dependence is mild compared to that on $\mtp$ in the kinematic region of interest and thus we fix $-t = 0.12$ GeV$^2$.}
    \label{fig:deck_comparison}
\end{figure}

More interesting is the $\mtp$ dependence of the isobar driven by Deck production, $F_\Delta$, which we plot in \cref{fig:deck_comparison}. 
Here, the additional $\sigma$ dependence introduced by the Born term, $\Delta$, enhances the rescattering corrections and produces a more pronounced $\rho$ peak that becomes shallower as $\mtp$ increases. As argued in \cref{sec:isobar_model}, such a dependence cannot arise from short-range production, even with a more complicated form of the driving term. The $\Delta$ function incorporates the effect of the logarithmic singularity of the $P$-wave-projected pion exchange and thus contributes to all derivatives of the isobar, unlike a typical subtraction polynomial, which determines only the first few terms. Moreover, this ``all-orders" contribution to the lineshape is not arbitrary, since, for fixed kinematics, the function $\Delta$ is fully determined and only the overall normalization is a free parameter. 

The size of the modification to the $\rho$ peak resulting from the unitarized Deck diagram compared to that of the contact production suggests that, to the extent that we the isobar approximation holds and we can ignore inelastic effects, the precise lineshape in the $[\pi\pi]_{1^{--}}$ mass spectrum at a given $\mtp$ is diagnostic of the relative strength and phase of the two terms and thus can tell us of the production dynamics as a function of production variables.

\section{Analysis of Freed-isobar Data}
\label{sec:fits}

As argued above, knowledge of the $[\pi\pi]_{1^{--}}$ lineshape can be used to constrain the dependence on other production variables. In this section, we do just that by comparing to Dalitz plots generated from the freed-isobar data by \compass~\cite{COMPASS:2021ogp}. Here, an identical decomposition to \cref{eq:isobar_decomposition} is used, except that, instead of assuming a fixed functional form for the isobars, the ``freed-isobar" technique assigns each $\pi\pi$ mass bin a floating, complex coefficient using
    \begin{align}
        \label{eq:freed_isobar}
        F(\sigma) = 
        &\sum_{k}(F_k^\text{(Re)} + i\, F_k^\text{(Im)}) \,\theta(\Delta\sigma_k - 2\,|\sigma-\sigma_k|) ~,
    \end{align}
where $\sigma_k$ and $\Delta\sigma_k$ denote the center and width of the \mbox{$k$-th} bin. The real and imaginary parts $F_k^\text{(Re/Im)}$ vary with the production variables and are determined by fit~\cite{Krinner:2018bwg,COMPASS:2021ogp}. The overall complex phase is defined relative to the reference channel, $1^{++}\ [\pi\pi]_{1^{--}}\pi\,S$-wave, but it will not be relevant for this analysis.

By not presupposing a dynamical model, the resulting freed-isobars aim to capture all relevant physics information in the event distributions, such as possible rescattering effects, while minimizing model bias. This procedure is repeated for 38 bins in $\mtp$ spanning $[0.98,2.48]$ GeV and for 4 bins of $-t$ spanning $[0.1,0.66]$ GeV$^2$, yielding the extracted freed-isobar as a function of $\sigma$ on a 2D grid of 152 bins in the production variables.\footnote{The data is in bins of $t^\prime = t_\text{min} - t$. However, as $s \gg \mtp^2$ in the \compass kinematics, we have $|t_\text{min}| \leq 10^{-3}$ GeV$^2$ and will simply use $-t$ and $t^\prime$ interchangeably.}

Because the extracted isobar in \cref{eq:freed_isobar} contains information on both modulus and phase, it should, in principle, be directly comparable to the KT isobar in \cref{eq:F_basis} at the amplitude level. However, because the experimental analysis uses the equivalent of \cref{eq:isobar_decomposition}, it is affected by the ambiguity discussed in \cref{subsec:ambiguities}. A considerable effort is made to resolve this ambiguity in the freed-isobar analysis to visualize the $\pi\pi$-subchannel intensity~\cite{Krinner:2017dba}, but ultimately the resulting freed-isobar is just one of infinitely many possible solutions. Because of this, we convert the one-dimensional isobar data back into two-dimensional amplitudes spanning the entire decay region for each value of $(\mtp^2,t)$:
    \begin{equation}
        \label{eq:freed-amplitude}
        \mathcal{F}^\text{ex}(t,\mtp^2; \sigma_a,\sigma_b) = N_\text{ex} \sqrt{\phi} \times \left(\frac{F(\sigma_a)}{Z(\sigma_a)}-\frac{F(\sigma_b)}{Z(\sigma_b)} \right) ~.
    \end{equation}
Here, $Z(\sigma)$ is the so-called zero-mode function, extracted by interpolating the Monte Carlo integral matrix from the experimental analysis~\cite{COMPASS:2021ogp}. The normalization $N_\text{ex}$ is chosen such that the experimental integrated intensity:
    \begin{equation}
        \Gamma^\text{ex}(t, \mtp^2) = \sum_{i,j} \,   |\mathcal{F}^\text{ex}(t,\mtp^2; \sigma_i, \sigma_j)|^2 \, \Delta\sigma_i \, \Delta\sigma_j ~,
    \end{equation}
matches the total number of events quoted in Ref.\cite{COMPASS:2021ogp} for each $(t,\mtp^2)$ bin. The amplitude of \cref{eq:freed_isobar} is free of ambiguity as it represents the physical intensity distribution and thus provides a more straight-forward comparison with the KT models.

It is important to note that because of the ambiguity, the covariance matrix of the $F^\text{(Re/Im)}_k$'s is nearly degenerate. The nearly degenerate eigenvectors corresponding to the values of $Z(\sigma_k) = Z_k$ are divided out in \cref{eq:freed-amplitude} to remove the near-singularity. For each bin of the four-dimensional intensity, the uncertainty $\delta|\mathcal{F}^\text{ex}|$ is computed as the standard deviation of $10^3$ pseudo-data sets resampled within the uncertainties of $F_k^{\text{(Re/Im)}}$ using the aforementioned correlations (and assuming that $Z(\sigma)$ has negligible uncertainties).

It is important to note that the uncertainties quoted in the COMPASS data are purely statistical and do not account for possible systematic effects which can arise from the freed-isobar analysis. As a result, the data points have generally small and almost certainly underestimated uncertainties. As the systematics of the COMPASS analysis are difficult to estimate, we do not attempt to rescale the original errors but instead consider a re-scaled bootstrapping procedure to attempt a more realistic estimation of own parameter uncertainties (see \cref{app:bootstap}). 

\subsection{Single bin fits}
\label{subsec:single_fits}
Before jumping into a simultaneous fit of 152 Dalitz plots, we begin with exploratory comparisons of single $(\mtp^2, t)$ bins, which are easier to visualize.
To this end, we will fix $t = t_0 \equiv -0.12$ GeV$^2$, which is the most forward bin in $t$, and consider a few $\mtp$ bins around the nominal $\pi_1$ mass. At this stage, we treat each bin independently. All $t$ and $\mtp$ dependence may be absorbed into the normalizations, and thus we may simplify \cref{eq:F_basis} to:
\begin{align}
    \label{eq:F_basis_simple}
    F & (\sigma) = N_c \, F_1(\sigma) + N_d \, F_\Delta(\sigma) ~,
\end{align}
where $N_i = |N_i|\,e^{i\phi_i}$ are complex.
Because we are insensitive to the overall phase, we fit the relative phase $\phi_d-\phi_c$. To compare the features introduced by each term, we will consider three model cases: C, D, and CD which indicate which coefficients are allowed to be non-zero. Model C, for example, sets $N_d = 0$.

For further comparison, we also consider a twice-subtracted solution, as in conventional KT. This corresponds to a solution of the form \cref{eq:normal_solution} with $m=2$, yielding a decomposition in terms of two basis functions given by \cref{eq:B_solution} with $B = 1$ and $\sigma$, respectively. These basis functions have a different power of $m$ than those in \cref{eq:F_basis_simple} (because the driving term now goes linearly in $\sigma$ rather than being asymptotically constant), and thus the two functions with $B=1$ will differ. We therefore write the conventional KT solution analogous to \cref{eq:F_basis_simple} as:
\begin{equation}
    \label{eq:twice_sub}
    F(\sigma) = N_0 \, F^\prime_1(\sigma) + N_1 \, F^\prime_\sigma(\sigma) ~,
\end{equation}
where the prime serves only to distinguish these isobars from those appearing in \cref{eq:F_basis,eq:F_basis_simple}. Comparing fit results to \cref{eq:twice_sub} assesses to what extent the effects of additional structure in the driving terms, e.g., the logarithmic singularities of \cref{eq:B_deck}, can be absorbed by subtraction polynomials within the usual KT formalism using the same number of free parameters.
\begin{figure}
    \centering
    \includegraphics[width=\linewidth]{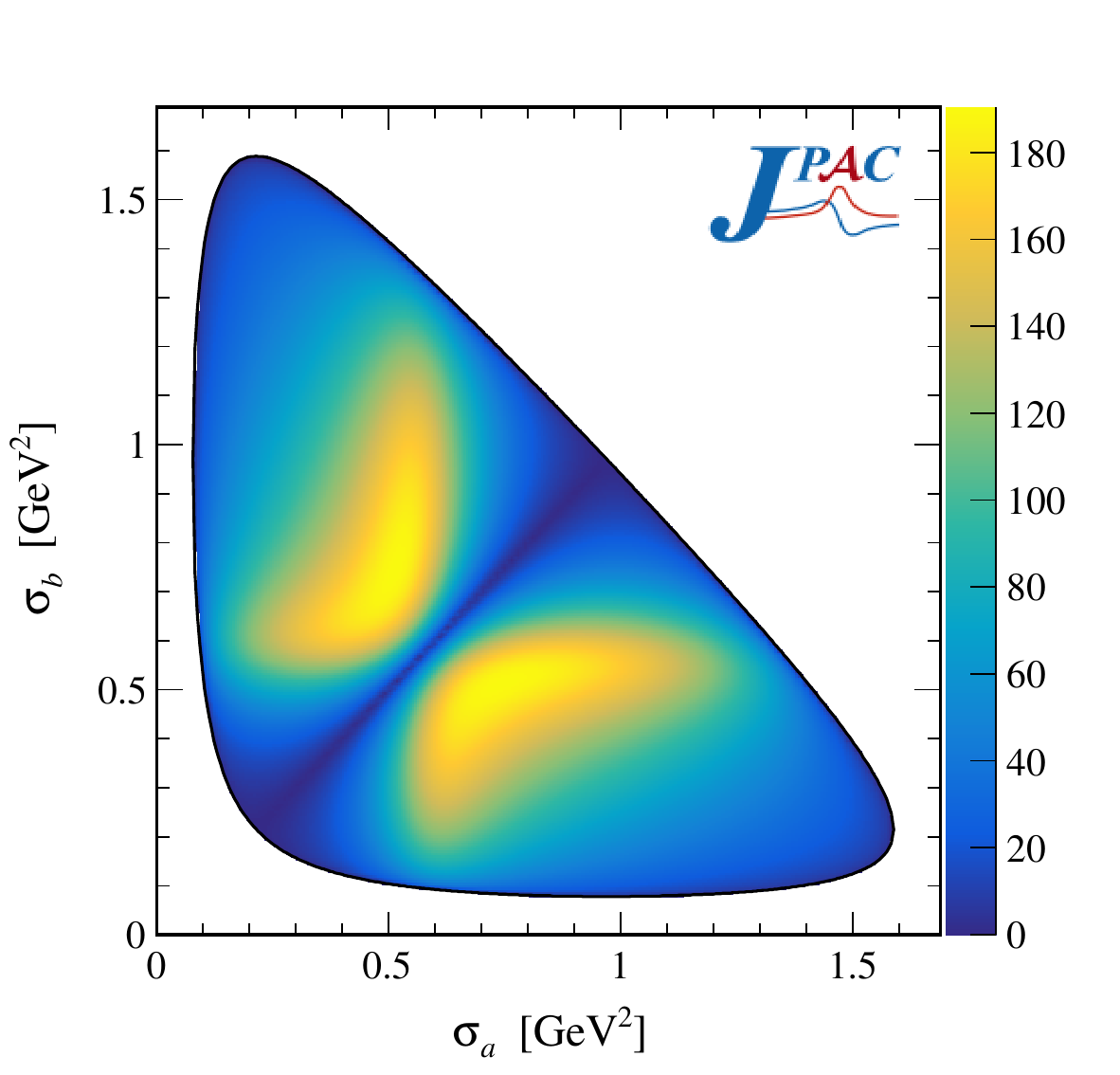}
    \caption{Dalitz plot intensity using the CD isobar at $\mtp = 1.4$ GeV and $t=t_0=-0.12$ GeV$^2$.}
    \label{fig:dalitz}
\end{figure}

For each model, we minimize 
\begin{equation}
    \label{eq:chi2}
    \chi^2 = \sum_{k=0}^{n_{\sigma}-1}\left( \frac{|\mathcal{F}^\text{ex}_{k}| - |\mathcal{F}^\text{th}_k|}{\delta| \mathcal{F}^\text{ex}_{k}|}\right)^2 ~,
\end{equation}
where $k$ runs over the $n_\sigma$ bins of $(\sigma_a, \sigma_b)$ in the Dalitz plot. The moduli of the amplitudes are computed bin-by-bin from \cref{eq:reduced_amplitude} using \cref{eq:freed-amplitude} and \cref{eq:F_basis_simple}, for $|\mathcal{F}^\text{ex}| \pm \delta |\mathcal{F}^\text{ex}|$ and $|\mathcal{F}^\text{th}|$ respectively. Results for \mbox{$\mtp \in \{1.4, 1.6, 1.8\}$ GeV} are tabulated in \cref{tab:single_bin_fits,tab:twice_sub_fits}. We quote only the best-fit value of each parameter since we are only interested in a qualitative assessment of the different models at this stage.
\begin{ruledtabular}
    \begin{table*}[]
        \caption{Summary of best-fit values for the different isobar combinations considering independent, single $\mtp$ bins at fixed $t = t_0 = -0.12$ GeV$^2$. \label{tab:single_bin_fits}}
        \centering
        \begin{tabular}{c|cccc}
                & $|N_c|$ & $|N_d|$ & $\phi_d-\phi_c$ & $\chi^2$/dof \\ \hline
            \multicolumn{5}{c}{$\mtp = 1.4$ GeV}                                   \\ \hline
            C   & 58.1   &         &          & 4.61          \\
            D   &        & 25.2    &          & 4.96          \\
            CD  & 387.5  & 153.6   & $-3.00$  & 3.66          \\\hline
            \multicolumn{5}{c}{$\mtp = 1.6$ GeV}                                   \\ \hline
            C   & 29.7   &         &          & 5.80          \\
            D   &        & 17.2    &          & 6.45          \\
            CD  & 326.8  & 176.6   & $-3.07$  & 3.47          \\\hline
            \multicolumn{5}{c}{$\mtp = 1.8$ GeV}                                   \\ \hline
            C   & 8.38   &         &          & 3.89          \\
            D   &        & 6.25    &          & 4.07          \\
            CD  & 129.0  & 92.0    & $-3.09$  & 3.09          
        \end{tabular}
    \end{table*}
\end{ruledtabular}
\begin{ruledtabular}
    \begin{table}[]
        \caption{Analogous best-fit values using the conventional twice-subtracted KT parameterization in \cref{eq:twice_sub}. \label{tab:twice_sub_fits}
        }
        \centering
        \begin{tabular}{cccc}
            $|N_0|$ & $|N_1|$ & $\phi_1-\phi_0$ & $\chi^2$/dof \\ \hline
            \multicolumn{4}{c}{$\mtp = 1.4$ GeV}               \\ \hline
            1056.8  & 1760.5  & $-3.06$         & 3.94         \\ \hline
            \multicolumn{4}{c}{$\mtp = 1.6$ GeV}               \\ \hline
            562.5   & 929.6  & $-3.10$         & 4.41         \\ \hline
            \multicolumn{4}{c}{$\mtp = 1.8$ GeV}               \\ \hline
            182.3   & 325.3   & $-3.13$         & 3.36         \\ 
        \end{tabular}
    \end{table}
\end{ruledtabular}

In these comparisons, several points become apparent: first, the CD model, plotted in \cref{fig:dalitz}, substantially improves upon both the contact-only and Deck-only models across all cases. This confirms that Deck production contributes non-negligibly to the \compass intensity. We also note that the relative phase between the contact and Deck terms is nearly $-\pi$, so the two terms interfere almost entirely destructively.

\begin{figure*}
    \centering
    \includegraphics[width=0.49\linewidth]{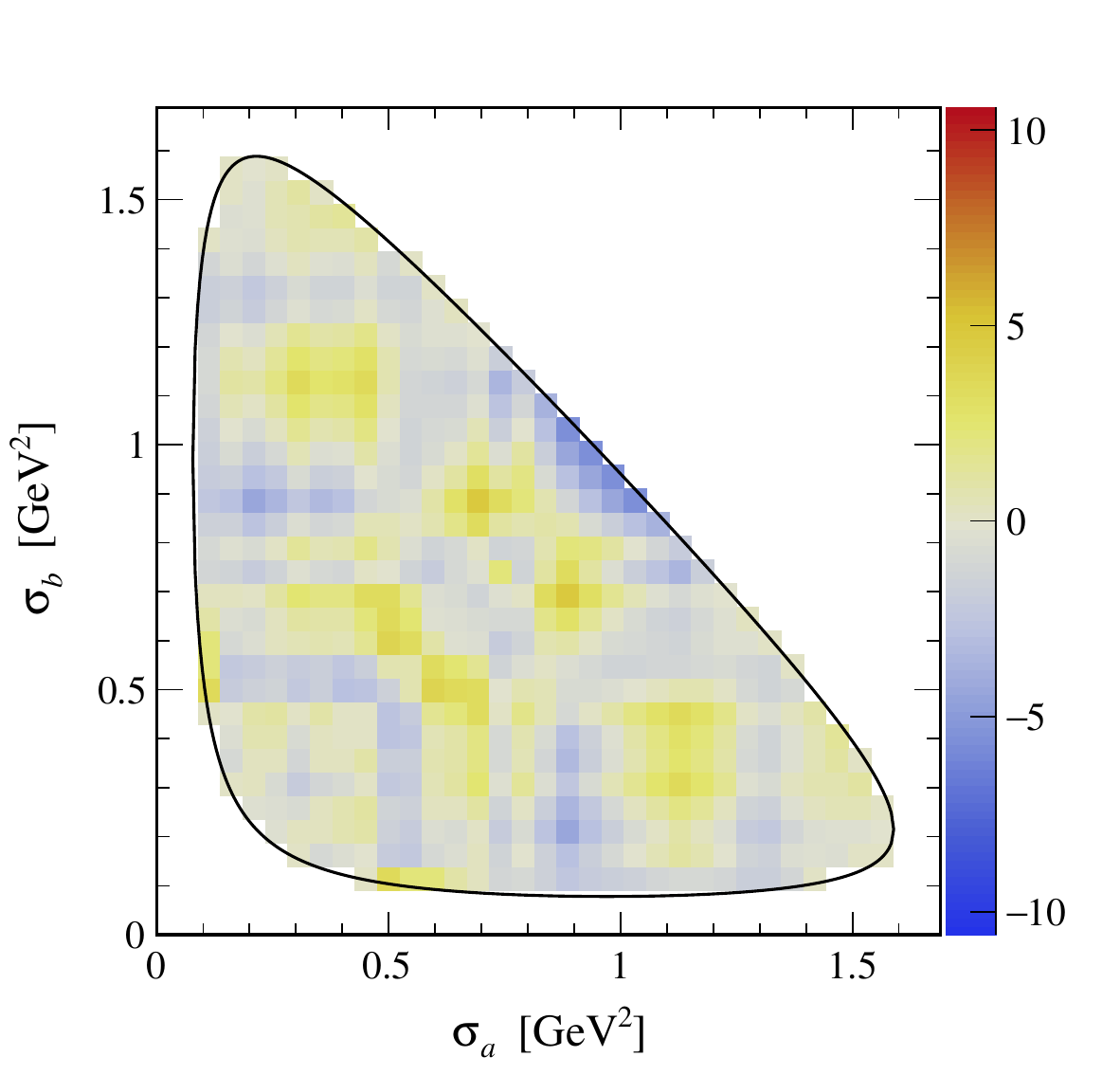}
    \includegraphics[width=0.49\linewidth]{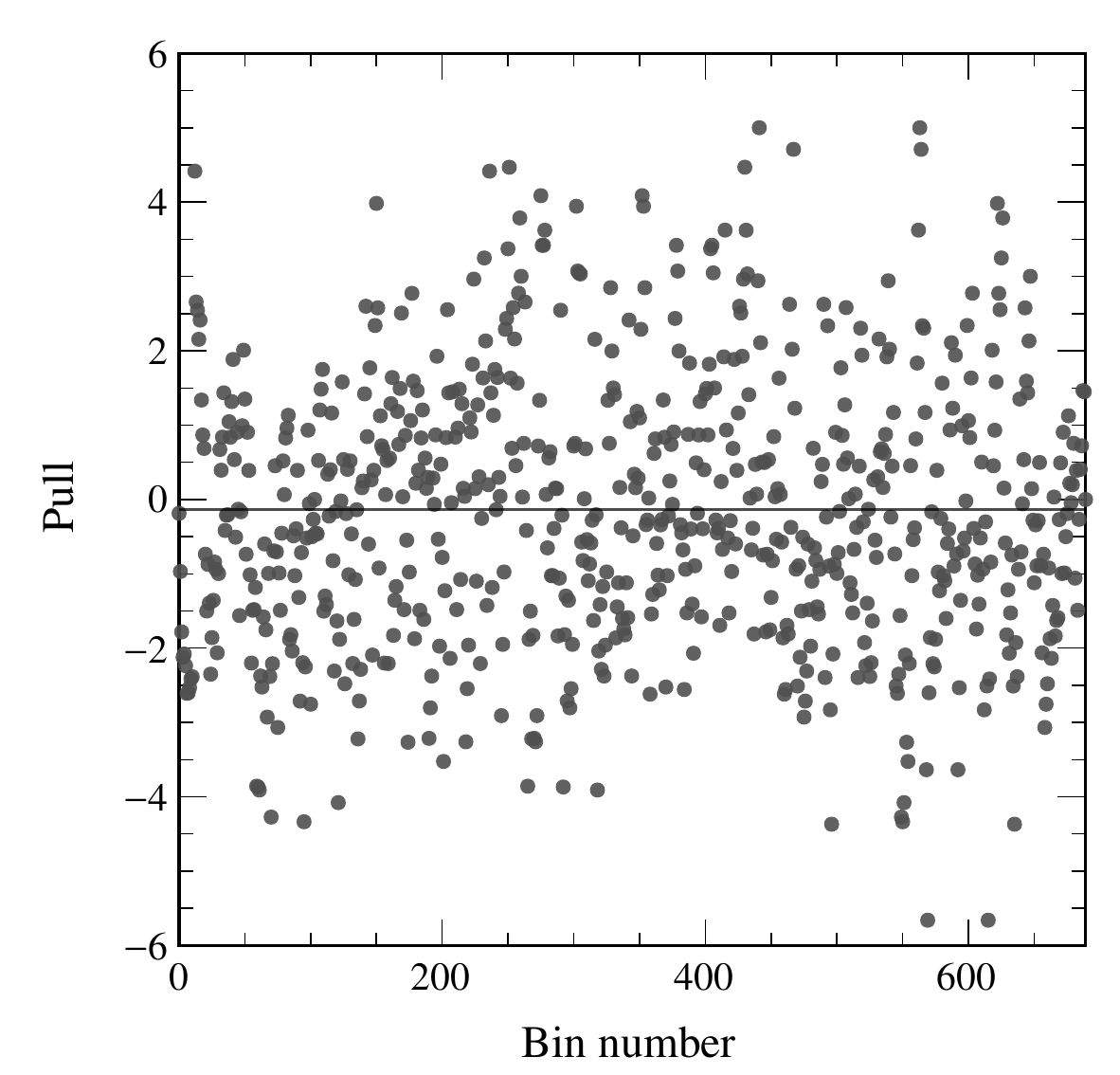}
    \caption{2D (left) and 1D (right) pull distributions corresponding to the $t = t_0$, $\mtp = 1.4$ GeV bin with the CD isobar. We see the large $\chi^2/\text{dof} = 3.66$ is not driven by a systematic deficiency of the model biasing the curves. In the 1D distribution, the horizontal axis refers to the index when scanning bins in the Dalitz plot from left-to-right and bottom-to-top and the horizontal line denotes the average pull.}
    \label{fig:dalitz_pull}
\end{figure*}

Compared with the twice-subtracted KT solution, the CD model fits 10--30\% better in all cases with the same number of free parameters. The largest deviation occurs at the nominal $\pi_1$ mass, supporting the more physics-motivated form of the Deck driving term as capturing more essential dynamics in the data. Regardless, both CD and conventional KT have rather large $\chi^2/\text{dof} \approx 3$--$4$. Even a thrice-subtracted KT solution does not appreciably improve the description beyond 15\% over \cref{eq:twice_sub}. Examining the pull distributions, e.g., of the CD model in \cref{fig:dalitz_pull}, shows no noticeable regions of systematic disagreement between the model and the data. Instead, the residuals show a uniformly large spread, centered at zero, across the entire physical Dalitz region. 

These observations point to an adequate description of the current data set --- at least to the degree that the unknown systematic uncertainties of the freed-isobar analysis can be ignored. 
More specifically, these results highlight the CD model as capturing the essential features present in the data distributions. We also remind that the normalizations of \cref{eq:F_basis_simple} have a straightforward physical interpretation, i.e., as the product of couplings and lineshapes associated with each production term, unlike the coefficients appearing in \cref{eq:twice_sub}. This will allow us to learn substantially more, such as the $t$ and $\mtp$ distributions, when fitting multiple bins simultaneously.

\begin{figure*}
    \centering
    \includegraphics[width=0.49\linewidth]{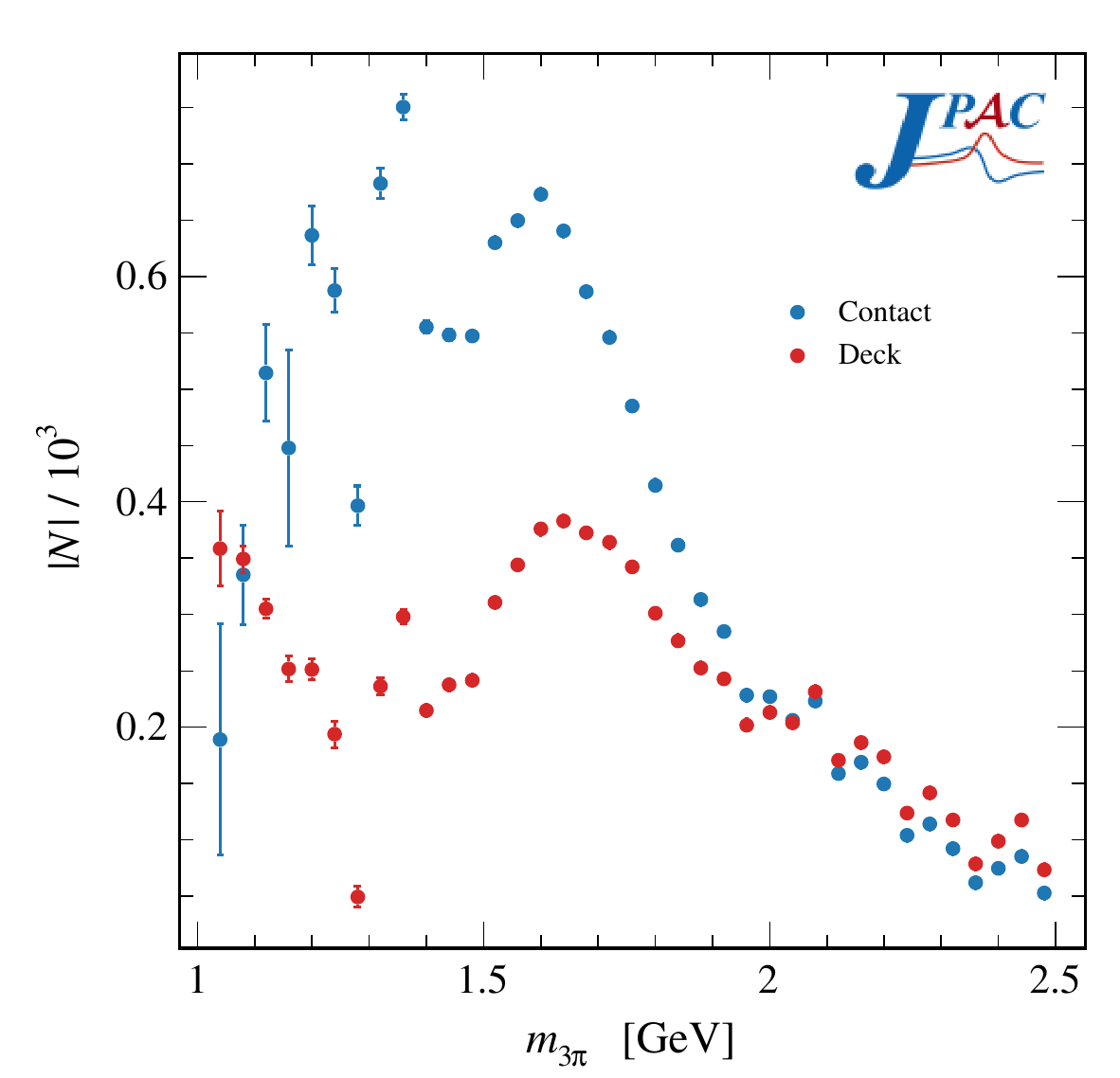}
    \includegraphics[width=0.49\linewidth]{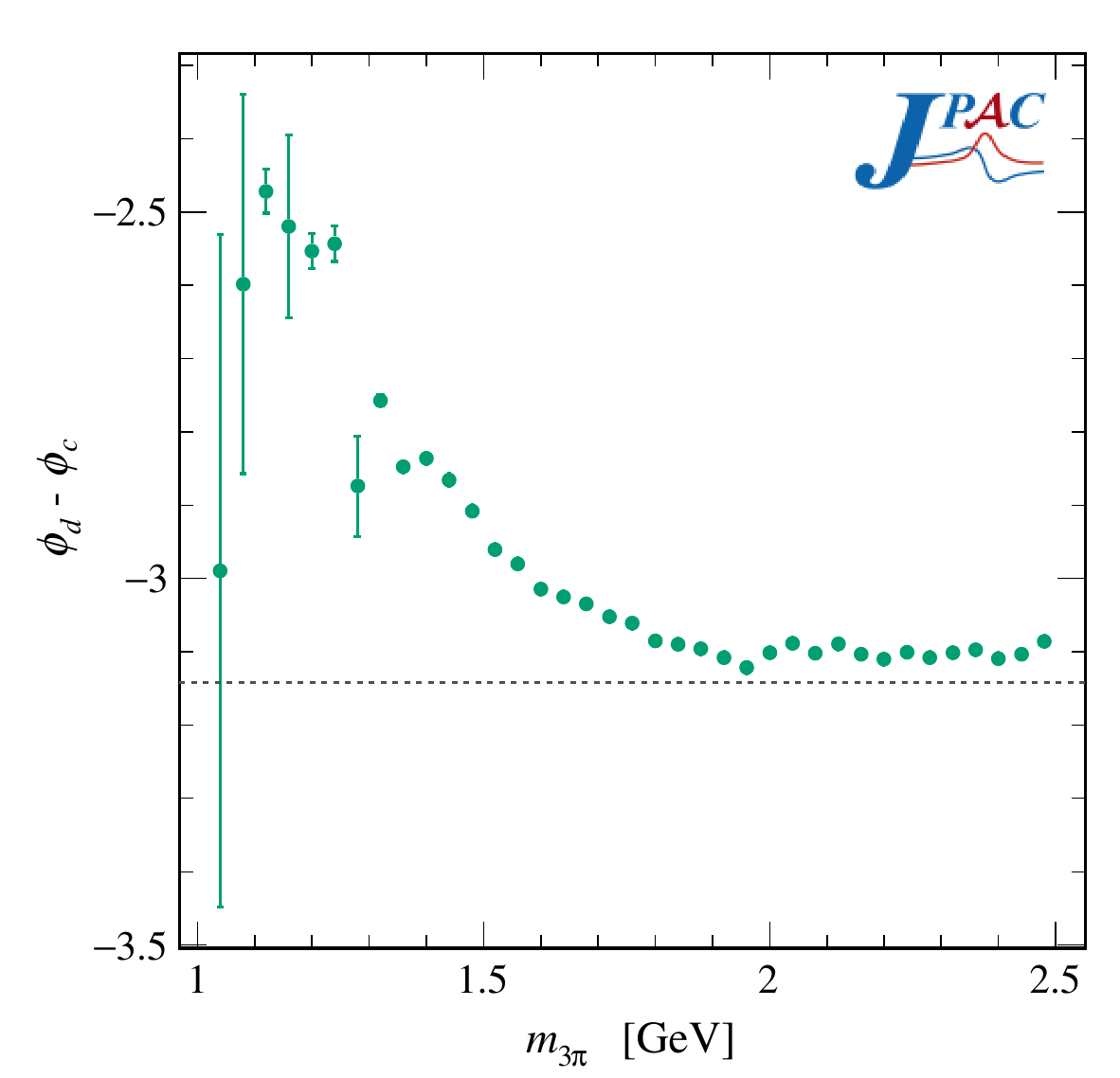}
    \caption{Extracted distributions of the moduli, $|N_c|$ and $|N_d|$ (left) and relative phase $\phi_d - \phi_c$ (right), as a function of $\mtp$ resulting from the global fit to \compass isobar data. All points include bootstrapped error bars even if they appear too small to see.}
    \label{fig:pars}
\end{figure*}

\subsection{Global fit}
\label{subsec:minimal}
We now turn to a more detailed analysis of the \compass data set in its entirety.
The linearity of the basis functions in \cref{eq:F_basis} allows us to precalculate them across the entire grid of production variables spanning the available data. The normalizations, which are expected to be functions of the $[3\pi]_{1^{-+}}$ mass, can then be extracted by associating a floating complex coefficient to each bin, to be determined by fit --- similar to the methodology of the freed-isobar analysis itself. This will allow us to determine the $\mtp$ dependence directly from the data, without bias from an explicit dynamical model. We therefore fit  \cref{eq:F_basis} in a binned form:

\begin{align}
    \label{eq:fit_model}
     &F (t,\mtp^2;\sigma) = \sum_{k=0}^{n_{3\pi}-1} \, \theta(\Delta\mtp^2- 2\,|\mtp^2-m^2_{3\pi,k}|) \, \times                                                    \\
      &\sqrt{-t} \,\bigg[ N_{c,k} \, e^{c(t-t_0)} \, F_1(\sigma) + N_{d,k} \, e^{d(t-t_0)} \, F_\Delta(t,m_{3\pi,k}^2;\sigma) \bigg]~, \nonumber
\end{align}
where the sum runs over each bin centered at $m_{3\pi,k}^2$ with width $\Delta\mtp^2$ (cf. \cref{eq:freed_isobar}). 

Although the $t$-distribution of the data is also binned, the dependence on $t$ enters primarily through the exponential functions, which, for simplicity, we assume share universal, constant slope parameters. We therefore do not need to discretize $c$ and $d$ but do subtract $t_0$ in the exponent so that this factor is unity at the most-forward bin. Because each $\mtp$ bin is almost independent, connected only by the $t$-slopes, every $N_{c,d}$ will be highly correlated with $c$ and $d$ but only weakly correlated with each other. By normalizing the exponent this way, we help decouple this factor from the other parameters, leading to better fit convergence.

All in all, considering $n_{3\pi}$ bins of $\mtp$ requires \mbox{$3n_{3\pi}+2$} real parameters, which we fix via a simultaneous fit of the bins. Because the data sets have different numbers of subchannel bins as a function of $\mtp$, we minimize the average $\chi^2/n_\sigma$ per Dalitz plot:

\begin{equation}
    \bar{\chi}^2 = \frac{1}{4n_{3\pi}}\,\sum_{k=0}^{4 n_{3\pi}} \left[\frac{\chi^2}{n_{\sigma}}\right]_{k}  ~,
\end{equation}
where we also include each of the available $t$ bins for every $\mtp$ value. Each $\chi^2$ is calculated using \cref{eq:chi2} and then divided by the number of subchannel mass bins. This prevents the fit from being biased toward the higher $\mtp$ region, which has larger phase spaces and thus more bins per Dalitz plot.

We fit all $n_{3\pi}$ bins available except for the two bins with smallest total mass, i.e. $\mtp = 0.96$ and $1$ GeV. Initial studies of the $\rho$ peak observed that the nominal $\rho$ mass appears shifted by upwards of 30 MeV in these small $\mtp$ bins~\cite{BartlMasters} and thus difficult to fit with a model with a fixed $\rho$ position given by the $\pi\pi$ phase shift. This shift results in much larger local best-fit $\chi^2$ indicating these two bins are particularly incompatible with the remaining data (fits to these $\mtp$ bins alone have $\chi^2/\text{dof} > 15$) and are thus discarded. 

The resulting bin-by-bin normalization values are plotted in \cref{fig:pars}, where we see a remarkably (albeit not perfectly) smooth $\mtp$ dependence. This smoothness is a direct result of considering all $t$ bins simultaneously, as the universal slopes connect and set a scale for the otherwise completely independent bin-wise normalizations. Because the data suggest strong destructive interference, individual $t$ bins constrain only the difference between the two terms, while $N_c$ and $N_d$ can both be arbitrarily large. Fits without the exponential factor therefore yield scatter plots that are sporadic and featureless across the $\mtp$ range, because adjacent bins are entirely uncorrelated.
The largest deviations from a smooth curve occur toward the ends of distribution and particularly at the low $\mtp$ range. 

\begin{figure}
    \centering
    \includegraphics[width=\linewidth]{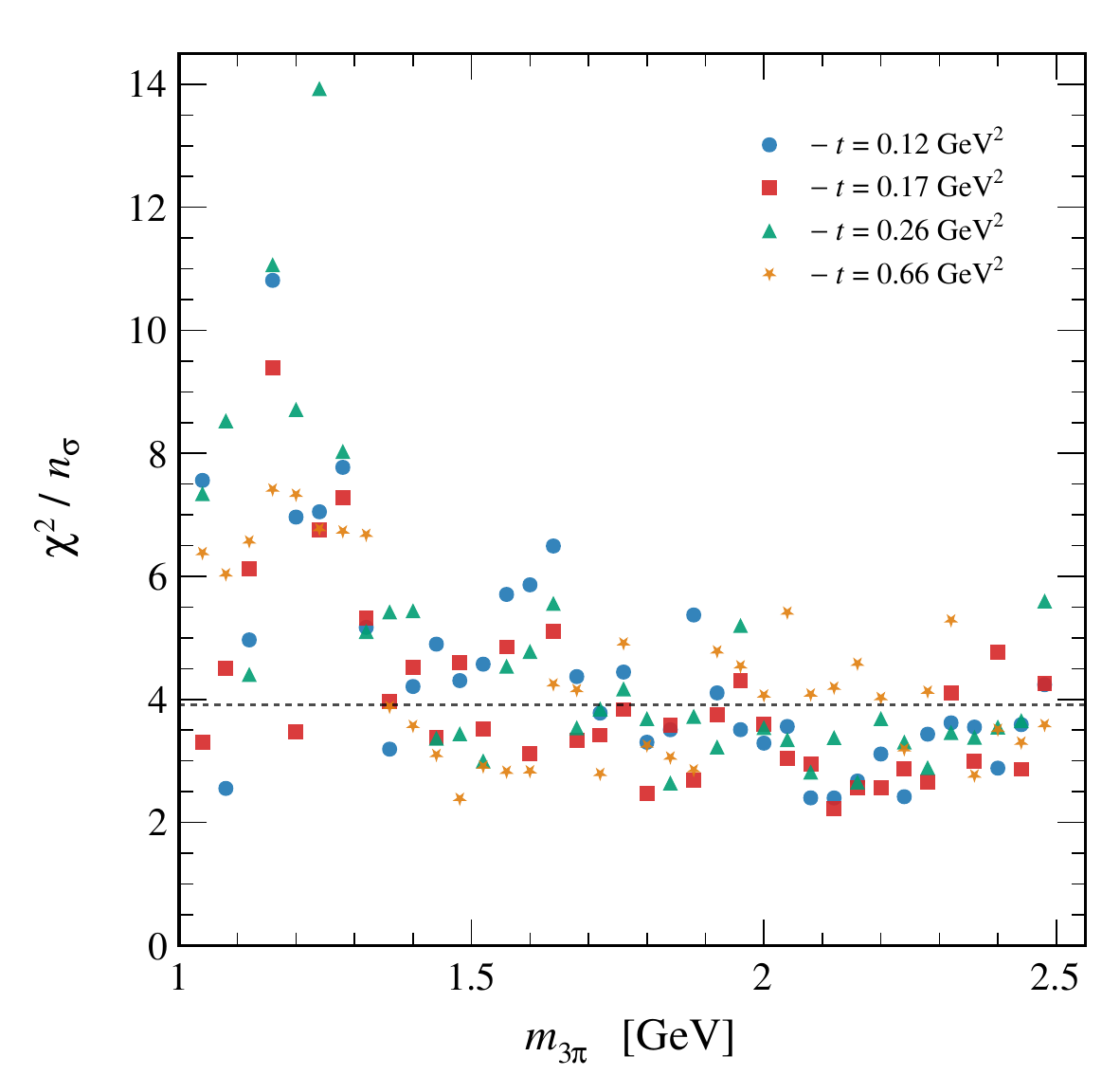}
    \caption{Distribution of $\chi^2/n_\sigma$ for every Dalitz plot bin considered in the global fit. The dashed line denotes the best-fit $\chi^2/\text{dof} = 3.91$.}
    \label{fig:minimal_chi2s}
\end{figure}

The spread of $\chi^2/n_\sigma$ for each of the 144 Dalitz plots is also shown in \cref{fig:minimal_chi2s}. Here, we see a substantial spread of $\chi^2$ values, yielding \mbox{$\bar{\chi}^2 = 4.41$}. We note that the region of $\mtp \leq 1.4$ GeV appears to be slightly less well described than the rest of the bins. This becomes more apparent when calculating the best-fit $\chi^2/\text{dof} = 3.91$, which weights the low-$\mtp$ bins (with smaller $n_\sigma$'s) less heavily, resulting in \mbox{$\chi^2/\text{dof} < \bar{\chi}^2$} by 10\%.\footnote{Fitting only bins with $\mtp > 1.4$ GeV, for instance, yields $\bar{\chi}^2 =3.70$ and $\chi^2/\text{dof} = 3.64$ with nearly identical parameter distributions to those in \cref{fig:pars}. This indicates that the larger $\bar{\chi}^2$ of the full fit does not significantly affect the shape of the higher-energy tail.} 

The quality of the description in the small $\mtp$ region is likely affected by two effects present in the \compass data. First, the measured integrated intensity as a function of $\mtp$ shows prominent fluctuations among bins below the nominal $\pi_1$ mass (cf. Fig. 8 of Ref.~\cite{COMPASS:2021ogp}). As we should anticipate a mostly smooth distribution, the large deviations, several standard deviations between adjacent $\mtp$ bins, point to prominent systematic uncertainties in this region which have not been estimated. Second, Ref.~\cite{COMPASS:2015gxz} previously noted that this $\mtp$ range has a different characteristic $t$-distribution, preventing a good description in terms of the sum of two exponentials with constant $t$-slopes. This may indicate contributions from other production diagrams not considered, which introduce additional $t$-dependence beyond that in our minimal model.

 In order to gauge the effect of additional $t$-dependence, we explore fits with an additional contact term beyond the minimal model in \cref{app:second_contact}. The presence of a third $t$-slope improves the overall fit quality slightly but the $\chi^2/n_\sigma$ distribution still reveal a systematically worse description of the small $\mtp$ region, pointing to some other systematic effect present in these bins.

The uncertainties of each parameter are estimated using a bootstrap analysis~\cite{JPAC:2021rxu}. Because the quoted experimental uncertainties are so small (and the corresponding residuals so large), we generate bootstrap pseudo-data sets by resampling within the local distance to the best-fit model curve (instead of the raw experimental error). This procedure is described in more detail in \cref{app:bootstap} and is expected to yield a more conservative, and therefore \textit{slightly} more realistic, estimate of uncertainties. Despite this rescaling, the resulting bootstrap errors remain unrealistically small and will require further consideration, which we defer to a subsequent analysis.

Examining the distributions of the normalizations themselves, the most glaring feature are the apparent peaks in both normalizations, which are highly suggestive of the resonant $\pi_1$. 
While it may be tempting to fit these distributions with Breit-Wigners, a rigorous identification with the $\pi_1$ pole requires model parameterizations of $N_c(\mtp^2)$ that can be analytically continued to unphysical Riemann sheets below the three-body cut. The parametric distributions observed in \cref{fig:pars} may thus serve to constrain such a model to extract the $\pi_1$ pole parameters. As construction and analytic continuation of three-body amplitudes can be quite delicate (see e.g. Ref.~\cite{Dawid:2023jrj}), this is left for a dedicated follow-up work. 

    \begin{figure}
        \centering
        \includegraphics[width=\linewidth]{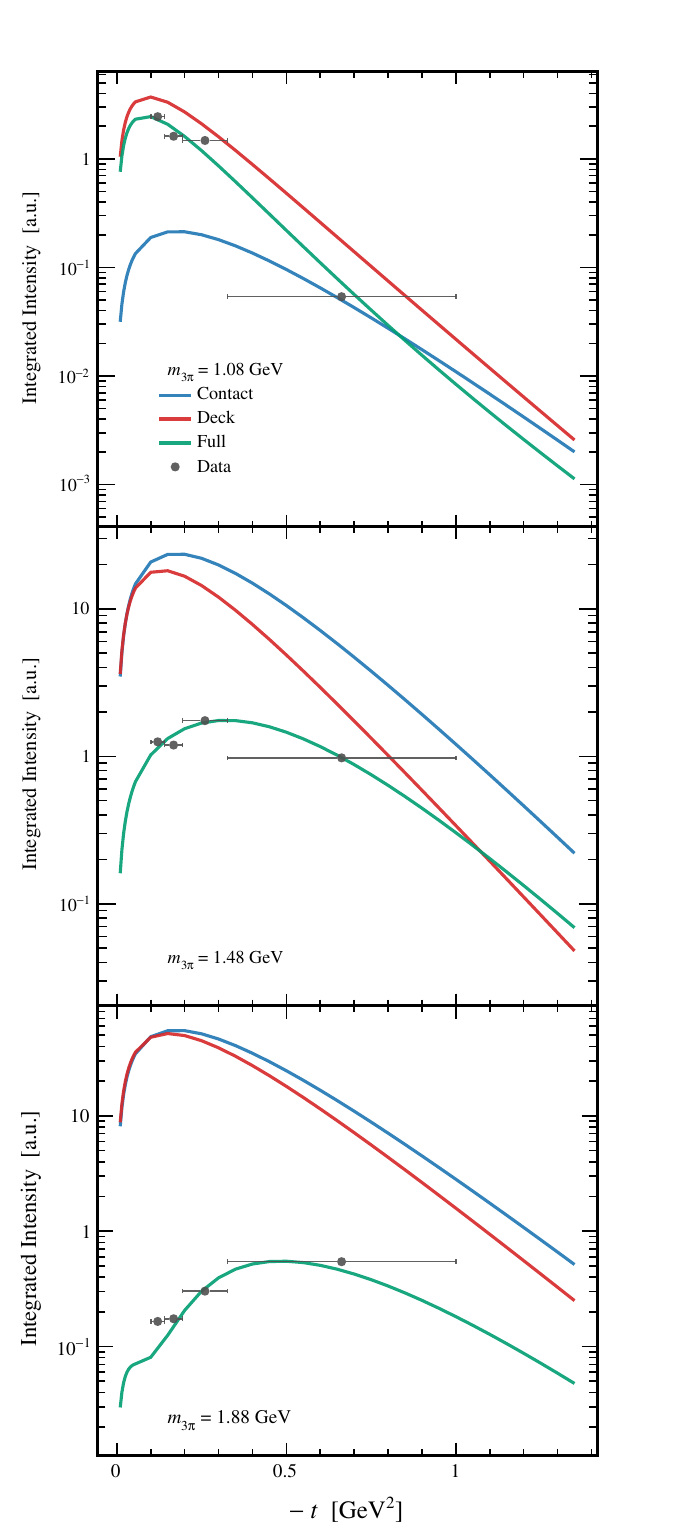}
        \caption{Integrated width as a function of $-t$ at various characteristic values of $\mtp$ for the minimal fit results. In addition to the best-fit results (blue) we plot the contributions to the width from individual components. Bootstrapped uncertainties are very small and thus we do not plot error bands. }
        \label{fig:intensity_t}
    \end{figure}

We note the nontrivial dependence on $\mtp$ that emerges in the Deck coefficient, including an apparent zero around 1.3 GeV. The error bars of these parameters are also generally smaller error bars than those of the contact term, even in the small total mass region. As discussed in \cref{subsec:basis}, the Deck coefficient is nearly entirely determined by the existence of deformations of the $\rho$ away from the lineshape observed in $\pi\pi$ scattering. and thus the distribution in $|N_d|$ emerges from the nontrivial $\mtp$ dependence of the rescattering corrections captured by the freed-isobar analysis. 

\begin{figure*}
    \centering
    \includegraphics[width=0.49\linewidth]{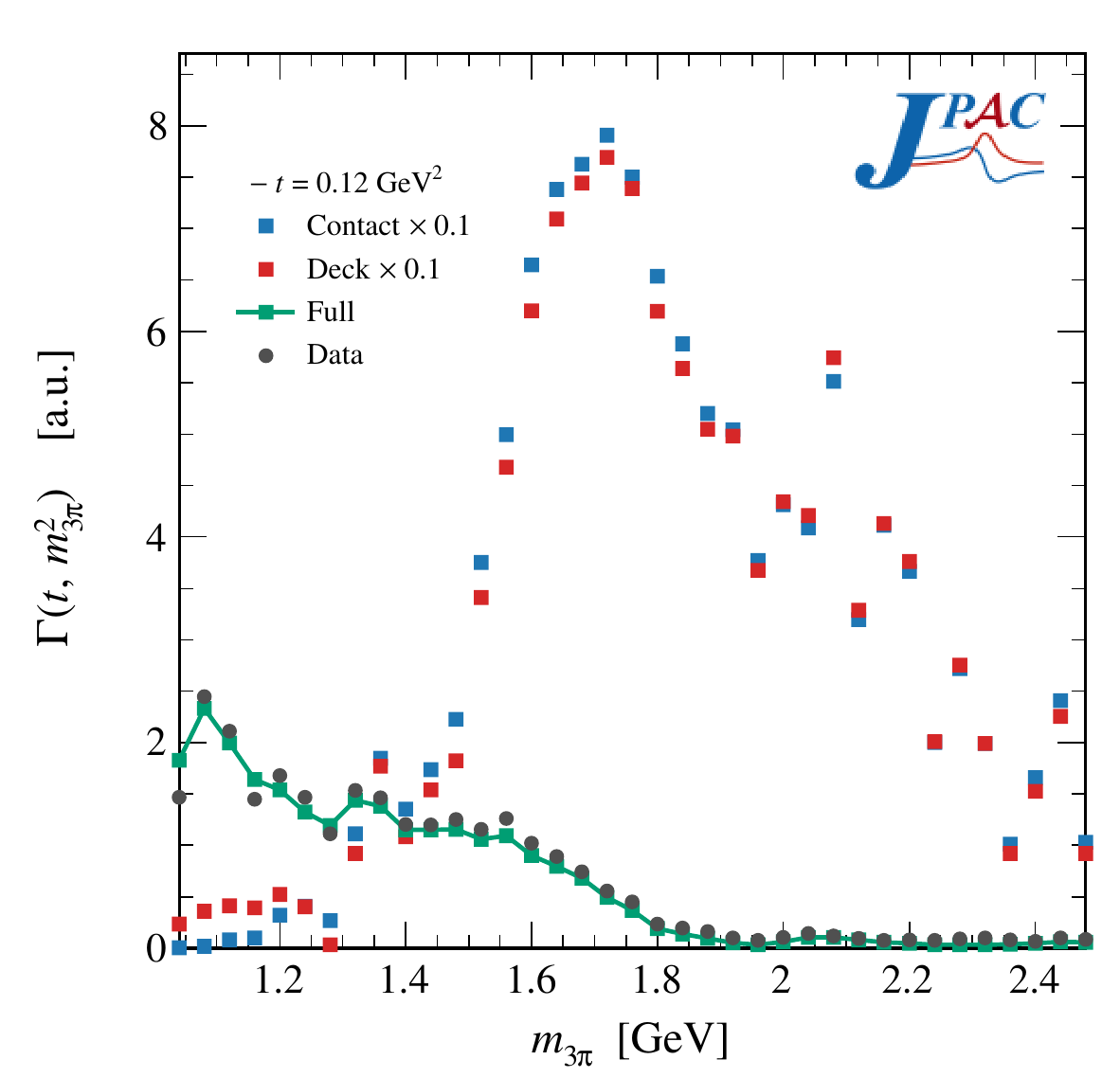}
    \includegraphics[width=0.49\linewidth]{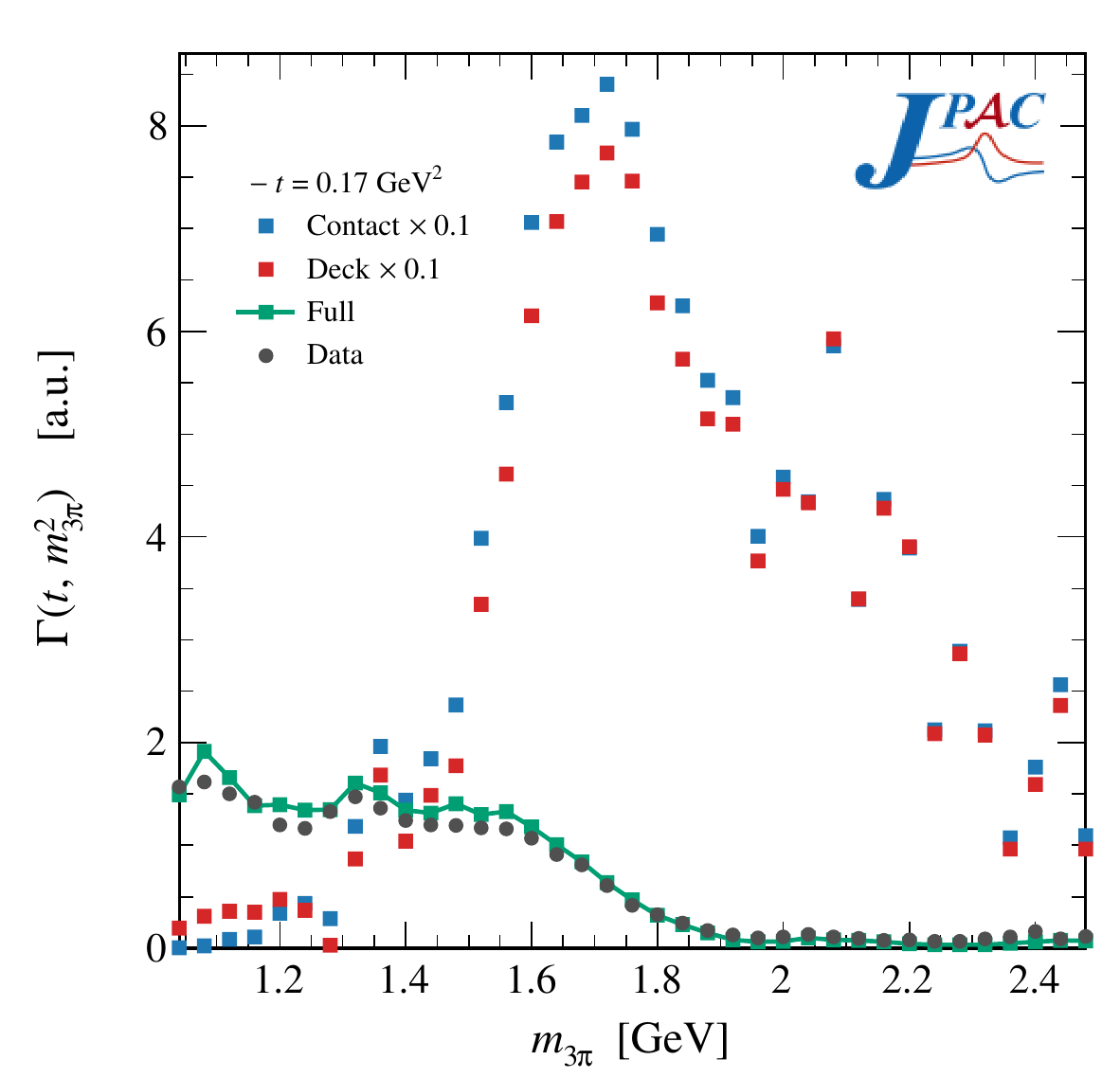}\\
    \includegraphics[width=0.49\linewidth]{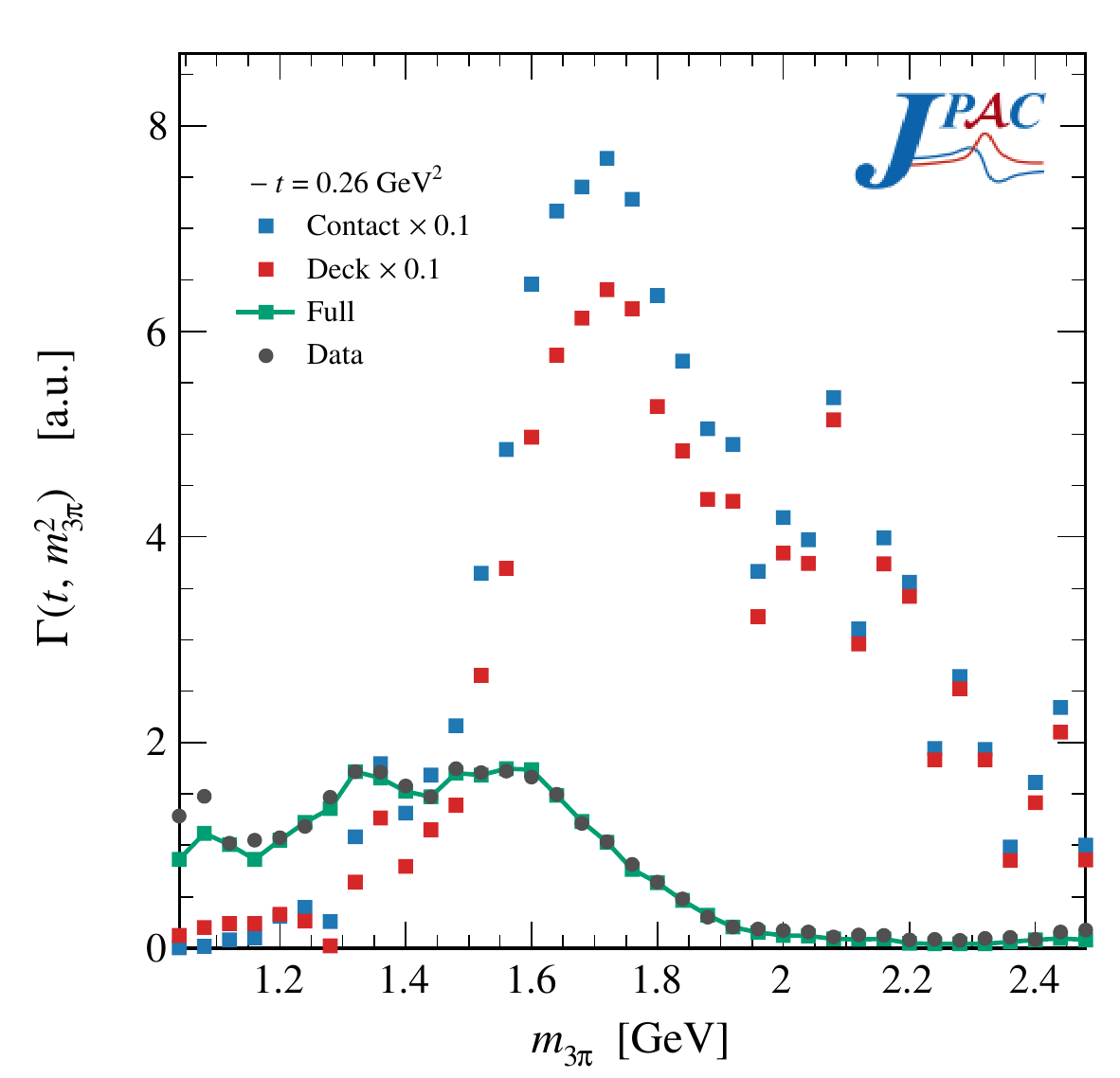}
    \includegraphics[width=0.49\linewidth]{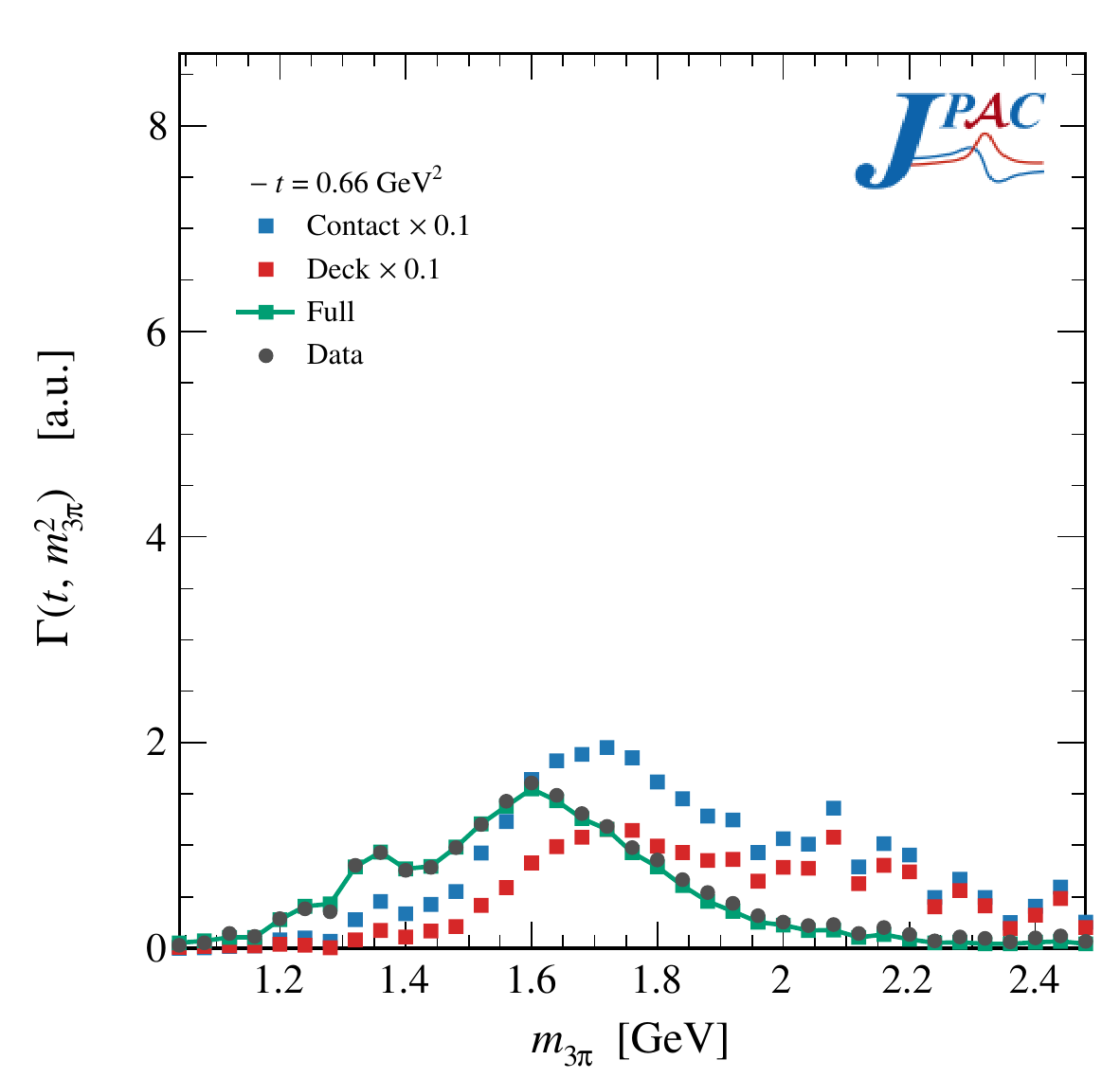}
    \caption{Integrated intensity as a function of $\mtp$ compared to the \compass intensity. Unlike the $t$-dependence in \cref{fig:intensity_t}, the $\mtp$-dependence is binned and thus we plot the bin centers connected by linear interpolations. As there is no explicit constraint of continuity, the lineshape is purely the result of the data. Bootstrapped uncertainties are very small and thus we do not plot error bands. }
    \label{fig:intensities_m3pi}
\end{figure*}

Examining the phase difference in \cref{fig:pars}, we observe a very smooth and relatively slow-varying phase motion, even around the nominal $\pi_1$ mass. If the peaks observed in the normalizations are indeed caused by the $\pi_1$, then it must couple to both production mechanisms in order for the characteristic resonant phase motion to cancel in the difference. 

The phase difference stays near $-\pi$ for most of the energy range with a monotonic increase at the low $\mtp$ end, suggesting a characteristically different interference pattern between the two terms for $\mtp \leq 1.4$ GeV.
Plotting the theoretical integrated intensity,
    \begin{equation}
        \Gamma(t, \mtp^2) = \int d\sigma \, \int d\sigma^\prime \, |\mathcal{F}(t,\mtp^2;\sigma,\sigma^\prime)|^2 ~,
    \end{equation}
as a function of $t$ in \cref{fig:intensity_t} reveals that this phase motion coincides with the region where the contact term is up to an order of magnitude smaller than the Deck contribution. 
As $t$ increases, the intensity distribution is flatter with the two production terms of similar size with strong destructive interference, as previously observed in \cref{subsec:single_fits}.

The best-fit $t$-slopes are given by:
\begin{subequations}
    \label{eq:minimal_slopes}
\begin{align}
    c &= 2.860 \pm 0.004 \text{ GeV}^{-2} \\
    d &= 2.741 \pm 0.005 \text{ GeV}^{-2} ~,
\end{align}
\end{subequations}
which, as they are the only parameters constrained by global features of the data, have very small deviations when bootstrapped. We remind that $c$ and $d$ include the absorbed Regge factor of the $\pom$ that may be subtracted to extract the purely exponential $t$-slope of \cref{eq:amp_contact}. Using $\alpha^\prime_\pom = 0.25 \text{ GeV}^{-2}$~\cite{Donnachie:1992ny} and $s = (19 \text{ GeV})^2$~\cite{COMPASS:2021ogp}, we obtain central values of
\begin{equation}
    c_0 = 1.39 \, \text{GeV}^{-2}
    \quad \text{and} \quad
    d_0 =  1.27 \, \text{GeV}^{-2} ~.
\end{equation}
These generally within expectations for the diffractive peaks in other Pomeron-driven reactions, such as high-energy heavy vector meson photoproduction~\cite{HillerBlin:2016odx}. It is important to note that although $d \lesssim c$, the Deck contribution can still be seen (e.g. in \cref{fig:intensity_t}) to fall off faster than the contact due to the extra $t$-dependence built into $\Delta$ in \cref{eq:Delta}. Even so, the fall-off of the Deck term is still comparable to the contact term and thus unlike the $t$-dependence of the non-resonant background models in \compass analyses~\cite{COMPASS:2018uzl,COMPASS:2021ogp}.

The interplay between the two terms can be illustrated when examining the integrated intensity as a function of $\mtp$ in each of the $t$-bins in which data is available, which we plot in \cref{fig:intensities_m3pi}. Here we see the contributions from the two terms are very similar peaks with the combined intensity driven primarily by large destructive interference. In the most forward-$t$ bin the distributions are nearly identical as needed to produce the peak-less distribution of the data. As $t$ increases, the difference between Deck and contact curves grows revealing the resonance-like peak in the data with largest-$t$.

We reiterate that although the integrated intensities show large destructive interference, this is not the result of a fine-tuned difference between the two production terms. This interference instead emerges from fitting the four-dimensional Dalitz plot data with the Deck constrained by the subtle modifications of the $\rho$ lineshape. Indeed, the apparent smoothness of the resulting parameter and intensity distributions across $\mtp$ bins, despite no by-hand constraint of continuity in the fitting procedure (beyond universal $t$-slopes), as well as Gaussian-like bootstrap distributions without nearby local minima point to this being a robust feature of our model.

\section{Summary and Outlook}
\label{sec:conclusions}

In this work, we analyzed the \compass collaboration's free-isobar data set relevant to the spin-exotic wave with $J^{PC}=1^{-+}$. The experimental analysis assumes that the $\pi_1 \to 3\pi$ decay proceeds primarily via $\rho\pi$ in $P$- wave, i.e., the so-called isobar approximation, and is thus decomposable into functions that parameterize individual $\pi\pi$ subchannels. This is the same dynamical assumption that enters the Khuri-Treiman formalism and makes the analysis readily comparable with dispersive models that are consistent with low-energy analyticity, unitarity, and crossing symmetry.

Because the $\pi_1$ is so wide, the effects of left-hand cut singularities from production processes must be taken into account to properly identify the exotic resonant signal. We outlined how the KT formalism can be extended to incorporate physics-motivated production amplitudes that replace the typical subtraction polynomials. With the \compass hadro-production in mind, we explore solutions to the KT equations seeded by Born terms that are consistent with high-energy exchange and Regge phenomenology.
These iterative solutions incorporate the ladder of $\rho\pi$ rescattering to all orders, as required by unitarity, and yield the characteristic modifications to the $\rho$ lineshape expected from different production amplitudes.

The unitarized isobars were used to perform a global fit to the \compass data set, considering the full $3\pi$ Dalitz plots across production momentum transfer and three-body invariant mass simultaneously. This allows us to constrain the unknown lineshapes that modify each production term as a function of $\mtp$ directly from the data, in a minimally model-dependent way. The result is a unitarized isobar model that describes the entire four-dimensional kinematic dependence of both production and decay variables. Notably, the $3\pi$ mass dependence revealed a surprisingly smooth distribution peaking at $1.6$ GeV, suggestive of a possible spin-exotic resonance.

As discussed, to properly identify this lineshape in the presence of such a resonance, one must be able to analytically continue the model into the complex plane and search for poles below the three-body cut. Such a continuation of the KT framework has been formally explored in the past~\cite{Aitchison:1966lpz}, but, to our knowledge, has never been generalized beyond the $S$-wave nor implemented to extract a three-body resonance pole from data (analyses with similar approaches, however, include Refs.~\cite{Jackura:2016llm,Mikhasenko:2017jtg}). This analysis is the first step toward this goal, with the required continuation left as the focus of future work.

The unitarization of three-body production amplitudes within the isobar model presented in this analysis can also be readily adapted to other reactions of interest. The Deck mechanism, specifically, contributes to all partial waves explored by the freed-isobar analysis, and thus the driving terms presented here can be adapted to the analysis of any $3\pi$ decay by simply considering the appropriate analogue of \cref{eq:L_contact} for the projecting Lagrangian. This would allow the exploration of other members of the predicted hybrid multiplet, such as those in the $2^{-+}$ sector~\cite{Klempt:2007cp,Dudek:2011bn}. The three-body dynamics relevant to non-exotic decays are also of interest. The $a_1(1260)$, in particular, has recently been the subject of several studies~\cite{Feng:2024wyg,Sadasivan:2021emk,Mai:2021nul,Molina:2021awn}, making a direct comparison with the \compass freed $1^{++}\ [\pi\pi]_{1^{--}}\pi ~S$- and $1^{++}\ [\pi\pi]_{0^{++}}\pi~P$-waves of timely interest.

The same can also apply to production effects relevant to $3\pi$ states in photo- or electro-production processes studied at Jefferson Lab~\cite{CLAS:2008zko,Tsaris:2016oty,Bookwalter:2012ixa}. In these cases, the beam particle is a photon, and thus the exchange of a reggeized charged $\pi$ or $\rho$ is expected to be the dominant production mechanism. These mechanisms have been argued to be enhanced relative to Pomeron exchange due to the hybrid nature of the $\pi_1$~\cite{Szczepaniak:2001qz,Afanasev:1999rb,Close:2003af}. The possibility of considering the $\pi_1^-\Delta^{++}$ final state may also enhance production cross sections, making photoproduction a promising avenue for spin-exotic searches in various decay modes~\cite{GlueX:2024erj,Albrecht:2024qdh} with the goal of understanding the properties of the elusive $\pi_1$.

The KT formalism has also been generalized to a number of hadronic decays to non-symmetric final states, including $K\pi\pi$~\cite{Niecknig:2017ylb,Kou:2023kvp,Niecknig:2015ija}, $\eta\pi\pi$~\cite{Akdag:2023oob,Isken:2017dkw,Akdag:2021efj}, and, more recently, $D\bar{D}K$~\cite{Hu:2025ppi}. The methodology presented here may also be used to extend these analyses to production processes. This may become particularly important for investigations of three-body systems relevant to heavy exotic states, e.g., $J/\psi \pi \pi$ for both the $X(3872)$ and $Y(4260)$, at future collider facilities such as the EIC~\cite{AbdulKhalek:2021gbh} and JLab22~\cite{Accardi:2023chb}.

Code to reproduce all the results of this work are publicly available at Ref.~\cite{iteratekt,*bootstrapPi1}.

\acknowledgments
We thank Bastian Kubis for useful discussions and encouraging comments on early versions of the manuscript. We also thank the high-performance computing team of the Laboratorio de Modelos y Datos (LAMOD) at ICN-UNAM for their guidance with the TOCHTLI cluster which was used extensively in this work. 
This material is based upon work supported by the U.S. Department of Energy, Office of Science, Office of Nuclear Physics under Contract No. 89243126CSC000213, by U.S.\ Department of Energy Grant No.~\mbox{DE-FG02-87ER40365}, and it contributes to the aims of the U.S. Department of Energy \mbox{ExoHad} Topical Collaboration, contract \mbox{DE-SC0023598}.
The work of S.G-S., G.M. and V.M. is supported by the Ministerio de Ciencia, Innovaci\'on y Universidades Grant No. PID2023-147112NB-C21, as well as through the award ``Unit of Excellence María
de Maeztu 2025-2029'' to the Institute of Cosmos Sciences,
Grant No. CEX2024-001451-M.
S.G-S. also acknowledges additional support from the Generalitat de Catalunya (AGAUR) through grant
2021SGR01095. S. G-S. is a Serra Húnter Fellow. V.M. is a Serra Húnter Professor. G.M. is a Beatriu de Pin\'os Fellow. We acknowledge financial support under the program ``Progetti di Rilevante Interesse Nazionale'' (PRIN 2022), published on 2.2.2022 by the Italian Ministry of University and Research (MUR), Project Title ``The X(3872) files'' -- CUP J53C24002600006 -- Grant Assignment Decree No. 20429 adopted on 6.11.2024 by the Italian Ministry of University and Research (MUR).

\appendix
\section{Derivation of isobar decomposition}
\label{app:isobar_derivation}
The necessary structure of the isobar decomposition for an arbitrary decay $X_J\to3\pi$ can be derived entirely from the crossing matrix (i.e.\ by requiring that the amplitude is crossing symmetric) and has been worked out in Ref.~\cite{Albaladejo:2019huw}. Here we briefly sketch the derivation for the $\pi_1$ case and refer to the reference above for more detailed discussion.

Because all particles involved are isovectors, the decay $\pi_1^i\to\pi^a\pi^b\pi^c$ for arbitrary isospin projections is related to that of $\pi_1^i \pi^{\bar{c}}\to \pi^a\pi^b$ and thus to a single scalar (reduced) amplitude by (we briefly show all three variables in \cref{eq:sigmas_def} to better highlight symmetries):
\begin{align}
    \label{eq:F_iabc}
     & \hat{\mathcal{F}}_{i\to abc}(\sigma_a,\sigma_b,\sigma_c) =
    \delta^{i\bar{c}}\delta^{ab} \, \hat{\mathcal{F}}(\sigma_a,\sigma_b,\sigma_c)
    \\
     & \quad+\delta^{ib}\delta^{a\bar{c}} \, \hat{\mathcal{F}}(\sigma_c,\sigma_a,\sigma_b) + \delta^{ia}\delta^{b\bar{c}} \, \hat{\mathcal{F}}(\sigma_b,\sigma_c,\sigma_a) \nonumber ~.
\end{align}
Alternatively, \cref{eq:F_iabc} can be written in terms of isospin-projectors and amplitudes of definite isospin, $\hat{\mathcal{F}}^I$, in the $c$-frame (i.e.\ the center-of-mass frame for the $i\bar{c}\to ab$ reaction). These isospin amplitudes are related to the scalar amplitude by:
\begin{equation}
    \label{eq:scalar_isospin_relation}
    \begin{bmatrix}
        \hat{\mathcal{F}}(\sigma_a, \sigma_b, \sigma_c) \\
        \hat{\mathcal{F}}(\sigma_b, \sigma_c, \sigma_a) \\
        \hat{\mathcal{F}}(\sigma_c, \sigma_a, \sigma_b)
    \end{bmatrix}
    = \frac{1}{6}
    \begin{bmatrix}
        2 & 0  & -2 \\
        0 & 3  & 3  \\
        0 & -3 & 3
    \end{bmatrix}
    \begin{bmatrix}
        \hat{\mathcal{F}}^0(\sigma_a, \sigma_b, \sigma_c) \\
        \hat{\mathcal{F}}^1(\sigma_a, \sigma_b, \sigma_c) \\
        \hat{\mathcal{F}}^2(\sigma_a, \sigma_b, \sigma_c)
    \end{bmatrix}
    ~.
\end{equation}

Each amplitude on the right-hand side can be expanded into isospin partial waves as:
\begin{equation}
    \label{eq:PWE}
    \hat{\mathcal{F}}^I(\sigma_a,\sigma_b,\sigma_c) = \sum_{j=1}^\infty \kappa_c^{j-1} \, P_j^\prime(z_c) \, \hat{f}_j^I(\sigma_c)~,
\end{equation}
where a $j=0$ term is prohibited due to the transversity of the $\pi_1$ which also yields angular structure in terms of the first derivative of the Legendre polynomials. The $c$-frame scattering angle is related to the crossed channel decay variables by:
\begin{subequations}
    \label{eq:z_and_kappa}
    \begin{equation}
        z_c = \frac{\sigma_b - \sigma_a}{\kappa_c} ~,
    \end{equation}
    with
    \begin{align}
        \kappa^2_c& = \kappa^2(\sigma_c) = (1-\sth/\sigma_c)(p_\text{th} - \sigma_c) (r_\text{th} - \sigma_c)~,
    \end{align}
\end{subequations}
denoting the Kacser function~\cite{Kacser:1963zz} in terms of the $2\pi$ threshold $\sth = 4\mu^2$ and the ``regular" and ``pseudo"-thresholds of the scattering reaction: $r_\text{th} = (\mtp+\mu)^2$ and $p_\text{th} = (\mtp-\mu)^2$ respectively. Prefactors of $\kappa$ are the angular momentum barrier factors and are included in \cref{eq:PWE} such that it is free from kinematic singularities.

The isobar model assumption then amounts to truncating \cref{eq:PWE} to a finite $j_\text{max}$, replacing $\hat{f}_j^I(\sigma)$ with isobars $F_j^I(\sigma)$ and, to reinstate crossing symmetry, adding analogous truncated expansions corresponding to the projections of the $a$- and $b$-channels. These crossed channel terms must enter with the appropriate Wigner matrices to rotate both the helicity and isospin projections back onto the $c$-channel~\cite{Trueman:1964zzb}. Luckily, the crossing structure of natural-parity vector decays is trivial and the general isobar decomposition reads:
\begin{align}
    \label{Eq:FI}
    \hat{\mathcal{F}}^I(\sigma_a,\sigma_b,\sigma_c) = 2 \sum_{j=1}^{j_\text{max}} \bigg\{ \kappa_c^{j-1} \, P_j^\prime(z_c) \,F_j^I(\sigma_c) &
    \nonumber                                                                                                                                                               \\
    - \sum_{I^\prime} C_{II^\prime} \, \bigg[\kappa_b^{j-1} \, P^\prime_j(z_b) \, F_j^{I^\prime}(\sigma_b)                                    &
    \\
    + (-1)^{I+I^\prime}\, \kappa_a^{j-1} \, P^\prime_j(z_a) \,F_j^{I^\prime}(\sigma_a)                                                        & \bigg] \bigg\} \nonumber ~,
\end{align}
where the isospin crossing matrix is
\begin{equation}
    C_{II^\prime} = \begin{bmatrix}
        \frac{1}{3} & 1             & \frac{5}{3}  \\
        \frac{1}{3} & \frac{1}{2}   & -\frac{5}{6} \\
        \frac{1}{3} & - \frac{1}{2} & \frac{1}{6}
    \end{bmatrix} ~,
\end{equation}
and the crossed channel $\kappa$ and $z$ given by \cref{eq:z_and_kappa} with the appropriate interchange of $c\leftrightarrow b$ or $c\leftrightarrow a$. A factor of 2 is also included in the normalization for convenience to remove overall factors in the final expression.

As a final step, at small values of $\sigma$ (e.g.\ within the physical decay region), the barrier factors will suppress contributions from higher spin and we truncate to $j_\text{max} =1$. Bose symmetry will thus enforce that the only relevant isobar will be $F_1^1(\sigma)$ in each channel. Inserting \cref{Eq:FI} with this truncation into \cref{eq:scalar_isospin_relation}, yields the final decomposition of \cref{eq:isobar_decomposition}.

\section{$\Delta(t,\mtp^2;\sigma)$ and its analytic continuation}
\label{app:continuation_of_triangle}
The discontinuity of the Deck triangle diagram in \cref{eq:Delta} is best written in terms of $s$-channel center-of-mass frame variables of the $\pi\pom\to\pi\rho$ reaction. To this end, we write the (moduli of) incoming and outgoing momenta as:
\begin{equation}
    \label{eq:prod_momenta}
    |\bm{k}|= \frac{\sqrt{\lambda(\mtp^2,t,\mu^2)}}{2\,\mtp}~, \quad
    |\bm{q}| = \frac{\sqrt{\lambda(\mtp^2,\sigma,\mu^2)}}{2\,\mtp}~,
\end{equation}
in terms of the usual K\"{a}ll\'{e}n triangle function~\cite{Byckling:1971vca}. We note that the ``initial state" Pomeron is spacelike and thus $t\leq0$, but, because both $|\bm{k}|$ and $|\bm{q}|$ have straightforward analytic continuations, this poses no problems. We may also define the  kinematic phase space factor of the $\rho\pi$ system in $P$-wave:
\begin{equation}
    \label{eq:rhoj}
    \rho_{\pi\rho P} = \frac{2\,
    |\bm{q}|^{3}}{\mtp}~,
\end{equation}
which we emphasize is a function of (and can be continued to arbitrary complex values of) both the total center-of-mass energy and floating $\rho$ mass, i.e.\ $\mtp$ and $\sigma$ respectively. 

From \cref{fig:deck_diagram}, the exchanged pion enters as a pole in the momentum transfer between the beam pion and $\rho$, which can be written in terms of \cref{eq:prod_momenta} and a new scattering angle as:
\begin{align}
    \label{eq:tau}
    \tau( & z) =                                                                                               \\
          & \mu^2 + \sigma - \frac{(\mtp^2+\mu^2-t)(\mtp^2-\mu^2+\sigma)}{2\,\mtp^2} + 2|\bm{q}||\bm{k}| \, z \nonumber ~.
\end{align}
Projecting this pole onto the $P$-wave yields \cref{eq:Delta} in terms of the Legendre function of the second kind, $Q_0(\zeta)$, with argument~\cite{Jackura:2023qtp}
\begin{equation}
    \label{eq:zeta}
    \zeta = \frac{1}{2|\bm{q}| |\bm{k}|} \left[\mu^2 - \tau(0) \right] ~.
\end{equation}
The Legendre functions of the second kind have a straightforward analytic continuation to arbitrary values of \cref{eq:zeta}~\cite{abramowitz+stegun}. We note that $Q_0$ (and therefore $\Delta$) develops an imaginary part at $\sigma < 0$ corresponding to the left-hand cut which emerges when projecting the crossed channel exchanges onto the $s$-channel. 

In addition, for certain kinematics, the $Q_0$ may also develop a complex cut on for $\Re\sigma >0$. Such a right-hand cut is an kinematical artifact of the partial-wave projection and does not correspond to the opening of a multi-body threshold. As a result, it may always be avoided by a suitable analytical continuation. The branch-point corresponding to $|\zeta|=1$ in the right-half complex $\sigma$-plane, however, will manifest as a pinch-singularity which must be treated with care, see e.g. discussion in Ref.~\cite{Dawid:2023jrj}. Luckily, in the kinematics considered in this work, these complications are sufficiently far from the physical region (and path of integration required by solutions to KT) and thus do not need to be treated explicitly. 

As discussed in \cref{subsec:deck}, $\Delta$ corresponds to the discontinuity of the Feynman loop \cref{eq:T_loop}. In fact, it is analytic for real $\mtp$ (and physical $2\mu \leq \sqrt{\sigma} \leq \mtp - \mu$) and, because $t\leq 0$, the only intermediate state that contributes to the loop is that of $\mtp$ and thus it can be entirely reconstructed by dispersing (with the phase space):
\begin{align}
    \hat{\mathcal{T}}(t, & \mtp^2;\sigma) =                                                                                                                                \\
                   & \hat{\mathcal{T}}(t,0;\sigma)  + \frac{\mtp^2}{\pi}\int_{(\sqrt\sigma+\mu)^2}^\infty \frac{dx}{x} \, \frac{\rho_{\pi\rho P} \, \Delta(t,x;\sigma)}{x-\mtp^2} \nonumber ~.
\end{align}
This implies that $\Im \hat{\mathcal{T}}(t,\mtp^2; \sigma)= \rho_{\pi\rho P} \, \Delta(t,\mtp^2; \sigma)$ within the Dalitz region. These two will deviate for unphysical $\sigma$ where the discontinuity itself becomes complex.

\section{Other production diagrams}
\label{app:not_considered}
In this appendix, we discuss other production diagrams which can be considered in addition to the contact and Deck production processes in \cref{subsec:contact,subsec:deck} but are ignored in our analysis. These effectively fall into three categories: Deck-like processes with heavier exchanges, nucleon spin-flip diagrams, and non-Pomeron exchanges. 

The first class of production diagrams include Deck-like mechanisms involving the exchange of heavier particles. These arise from the production via $\rho/f_2$ exchange, inclusion of a form factor, or reggeization effects of the pion pole at large $\mtp^2$~\cite{JointPhysicsAnalysisCenter:2024kck}.
\begin{figure}[h]
    \centering
        \begin{tikzpicture}[scale=5]
        \begin{feynman}
        \vertex (b)   at (-0.4, +0.25) {};
        \vertex (pom) at (-0.4, -0.25) {};
        \vertex (p2) [square dot] at (+0, 0.) {};
        \vertex (p3) [square dot] at (+0.4, 0.) {};
        \vertex (pi1) at (0.7, 0) {};
        \diagram*
        {
            (b)  -- [plain]  (p2),
            (pom)  -- [zigzag] (p2),
            (p2) -- [plain, half right] (p3) -- [triple] (pi1),
            (p2) -- [double, half left] (p3)
        };
        \draw [dashed,color={jpac_red},line width = 0.4mm] ($(0.2,0.315)$) -- ($(0.2,-0.315)$);
        \end{feynman}
    \end{tikzpicture} 
    \caption{Bubble diagram emerging from heavier Deck-like diagrams. The heavier exchange propagator can be shrunk to contact $\pi\pi\rho \pom$ interaction resulting in an effective bubble diagram.}
    \label{fig:deck_bubble}
\end{figure}
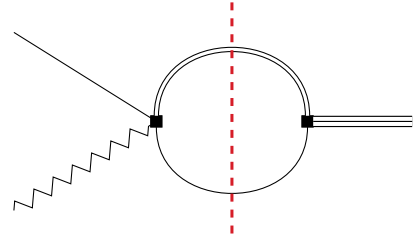

Naively replacing $\mu \to \Lambda$ in \cref{eq:deck_pwp} causes the amplitude to vanish asymptotically as $\Lambda \to \infty$ (at fixed $\mtp$).
This is because increasing the exchanged mass moves the logarithmic singularity contained in $Q_0(\zeta)$ farther from the Dalitz region~\cite{Aitchison:1965ghs}. We may thus reasonably expect a hierarchy in which the lightest exchange (i.e., the pion exchange in \cref{subsec:deck}) is the most influential for the amplitude lineshape, while the effects of heavier exchanges are smoother and therefore can be absorbed into the normalizations of the contact or Deck terms.

To investigate the effect of finite $\Lambda$, we may simply expand the vertical propagator in the Deck loop integrand and ignore momenta with $\ell^2 > \Lambda^2$:
\begin{equation}
    \label{eq:prop_expansion}
    \frac{1}{\Lambda^2-\ell^2} = \frac{1}{\Lambda^2} \left[1+ \mathcal{O}\left(\frac{\ell^2}{\Lambda^2}\right)\right]~.
\end{equation}
This effectively shrinks the triangle diagram in \cref{fig:deck_triangle} to a bubble, as shown in \cref{fig:deck_bubble}, and thus instead of \cref{eq:mT} we arrive at:
\begin{equation}
    \label{eq:mB}
    \mathcal{T}(t,\mtp^2; \sigma) \approx  \left[ \epsilon_{\mu\nu\alpha\beta}\, \pi_1^\mu\, k^\nu q^\alpha \,\pom^\beta\right] \times \Lambda^{-2} \, \hat{\mathcal{B}}(\mtp^2;\sigma) ~,
\end{equation}
with $\hat{\mathcal{B}}$ is a scalar, logarithmically-divergent two-point integral.\footnote{Again, in the notation of \texttt{LoopTools}, $\hat{\mathcal{B}}(\mtp^2;\sigma) = (4/\pi) \, B_{00}(\mtp^2, \sigma, \mu^2)$.} Proceeding as in \cref{subsec:deck}, we cut the loop and note that 
    \begin{align}
        \frac{1}{2i}\left[\hat{\mathcal{B}}(\mtp^2+i\epsilon;\sigma) - \hat{\mathcal{B}}(\mtp^2-i\epsilon;\sigma)\right] = \rho_{\pi\rho P} ~.
    \end{align}
This means that after removing the phase space factor, we are left with a constant and thus these diagrams are indistinguishable from the contact term of \cref{eq:L_contact} to first-order.

The second class of diagrams beyond those considered in the main text include nucleon spin-flip interactions at the bottom vertex. The effect of such interactions arises from \cref{eq:t_factors} where we may consider $|\lambda - \lambda^\prime| = 1$. This would permit us to add a second copy of either $B_\text{Contact}$ or $B_\text{Deck}$ with an additional factor of $\sqrt{-t}$ and a different normalization. In principle, such a term would be required to describe the large-$t$ kinematics where spin-flip contributions are expected to dominate, but are unnecessary for the current data set with $t < 1$ GeV$^2$.  

Finally, Regge exchanges other than the Pomeron are also possible in the bottom vertex (i.e., exchanged with the target), such as those involving the $\rho$ or $f_2$ trajectories. These meson exchanges will all have Regge intercepts less than unity and thus will be asymptotically suppressed relative to the Pomeron as $s\to \infty$~\cite{Winney:2025tla,Collins:1977jy}.
At finite kinematics, however, these diagrams may still have a non-negligible effect --- especially in high-statistics analyses. For instance, $f_2$ exchange with the nucleon has recently been shown to be necessary to accurately describe the momentum-transfer distribution of $\eta^{(\prime)}\pi$ pairs~\cite{Bibrzycki:2021rwh,Pekeler:2025ehx,Pekeler:2025,COMPASS:2026mcg}, contradicting the typical assumption that the full range of \compass kinematics requires only Pomeron exchange.

In our current application, both Deck and contact production can proceed via a subleading meson exchange.
The primary effect of considering such a diagram enters through the amplitude's total $s$-dependence, e.g., through the $s^{\alpha(t)}$ factor in \cref{eq:amp_contact}. Since we are not sensitive to this dependence, swapping the Pomeron for any other reggeon simply adds functionally identical copies of the production term with different parameters. For example, contact-like production mediated by $\rho$ exchange is included by adding a second \cref{eq:B_contact} with different parameters. If the parameters of both $\rho$ and $\pom$ diagrams are \textit{a priori} unknown and to be determined by fit, it will not be possible to unambiguously identify which term corresponds to which exchange. Therefore, we will generally continue to assume Pomeron dominance so that we may focus on comparing different top vertex dynamics.

\section{The $\sigma \to \infty$ limit and double Regge asymptotics}
\label{app:asymptotic}

In order to determine the number of subtractions needed in \cref{eq:B_solution}, one typically considers the properties of the isobars in the $\sigma \to \infty$ limit. Specifically, as the power $m$ governs the asymptotic behavior of the rescattering term, it is customarily assumed to match the leading power in the homogeneous solution in \cref{eq:fb_homogenous}. If $B(t,\mtp^2;\sigma\to\infty) \propto \sigma^i$, then, we choose the number of subtractions to be $m = i+1$ so that the inhomogeneous term is asymptotically of the same order. When multiplied by the Omn\`{e}s function, if the phase shift asymptotically approaches $\delta(\sigma\to\infty) = k\pi$ for integer $k$,\footnote{As our model is only constrained by elastic unitarity, the extrapolation of the $\delta(\sigma)$ beyond a few GeV (and therefore the choice of $k$) is largely arbitrary. We follow Ref.~\cite{Danilkin:2014cra} and match the CFD fit of Ref.~\cite{Garcia-Martin:2011iqs} to the form:
$$
    \delta(s>s_m) = \pi - \frac{a}{b + (s/s_m)^{3/2}}~,
$$
where $a$ and $b$ are chosen to smoothly match $\delta(s_m)$ and $\delta^\prime(s_m)$ at $s_m = 1.3$ GeV$^2$.} the full isobar will go as $F_B(t,\mtp^2;\sigma\to\infty) \propto \sigma^{i-k}$. Although this fixed-power scaling is unphysical for any $i$ and $k$~\cite{Gribov:2009cfk}, their values are still chosen to at least minimize violations of high-energy expectations, such as the Froissart bound~\cite{Froissart:1961ux}.

\begin{figure}[b]
    \centering
    \begin{tikzpicture}[scale=5]
    \begin{feynman}
    \vertex (b)    at (-0.45, 0.25) {};
    \vertex (et)   [dot] at (0,  0.25) {};
    \vertex (eb)   [dot] at (0, -0.025) {};
    \vertex (r)    at (+0.45, 0.25) {};
    \vertex (p3)   at (+0.45, -0.025) {};
    \vertex (p4)   at (+0.45, -0.125) {};
    \vertex (bpom1) [dot] at (0,    -0.125) {};
    \vertex (bpom2) at (0,    -0.4);
    \vertex (pr1)  at (-0.45, -0.4) {};
    \vertex (pr2)  at (+0.45, -0.4) {};
    \vertex (tt1) at (-0.08, -0.28) {$\mathbb{P}$};
    \vertex (tt2) at (-0.08, 0.11) {$\mathbb{P}$};

    \diagram*
    {
        (b)    -- [plain]  (et),
        (et)   -- [zigzag]  (eb),
        (p4) -- [plain] (bpom1)  -- [plain] (eb),
        (eb)   -- [plain]  (p3),
        (et)   -- [plain] (r),
        (bpom1) -- [zigzag] (bpom2),
        (pr1)  -- [fermion] (bpom2) -- [fermion] (pr2)
    };
    \vertex (m3pi) at (+0.6,  0.125) {\color{red} $\sigma$};
    \draw [<->,jpac_red, thick] (r)   to [bend left=40] (p3);
    \vertex (s)    at (-0.65, -0.05) {\color{jpac_red} $s$};
    \draw [<->,jpac_red, thick] (b)   to [bend right=40] (pr1);
    \vertex (m3pi) at (+0.6,  -0.3) {\color{jpac_red} $s_{\pi p}$};
    \draw [<->,jpac_red, thick] (p4)   to [bend left=40] (pr2);
    %
    \end{feynman}
\end{tikzpicture}
    \caption{The ``fast-$\pi$" diagram of Deck-like production in the double Regge limit. The invariant mass of the centrally-produced $\pi\pi$ system is assumed to be small compared to $\sigma$ and $s_{\pi p}$ such that the middle pion exchange does not reggeize.}
    \label{fig:2pi_prod}
\end{figure}
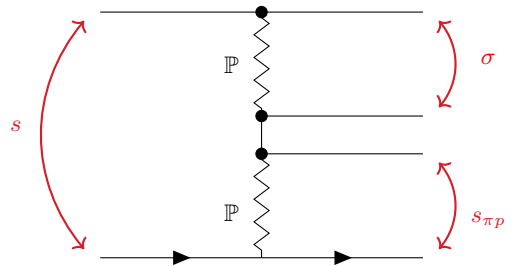

Because the $\rho\pi$ system is embedded in a production amplitude with Regge kinematics, one should, in principle, consider the effect of bottom $\pom$ exchange when taking $\sigma \to \infty$.
Specifically, if $s > \mtp^2> \sigma$ and $s_{\pi p}$ are all large, the reaction will resemble central $\pi\pi$ production, with both the beam and target diffracting, i.e., exchanging a Pomeron, as in \cref{fig:2pi_prod}. Such ``double Regge" processes are being investigated in $\eta^{(\prime)}\pi$ production relevant for the $\pi_1$ at \compass~\cite{COMPASS:2014vkj,Bibrzycki:2021rwh} and at GlueX~\cite{Montana:2025asi,RebeccaPhD}. 

Avoiding unphysical overlapping singularities in the $3\pi$ and $\pi\pi p$ channels imposes very nontrivial constraints on the possible $\sigma$ dependence required by analyticity in the double Regge limit~\cite{Shimada:1978sx}. Taking $\tau = t \approx 0$ (cf., \cref{fig:deck_diagram}), the leading $\sigma$ behavior will be the same for diagrams in which the final state $\pi$ or $\pi\pi$ system are the ``fast particle" (i.e., travel in the beam direction) and yield
    \begin{equation}
        \label{eq:double_regge}
        \mathcal{F}(t,\mtp^2; \sigma \to \infty, \sigma^\prime) \propto \sigma^{\alpha_\pom(0) - \alpha_\pom(0)}  = \sigma^0~.
    \end{equation}
This bound is more constraining than the naive single Regge limit where one ignores that the bottom Pomeron exchange is already reggeized, i.e., \hbox{$ \sigma^{\alpha_\pom(0)} \approx \sigma^1$}. In order to satisfy the upper bound set by \cref{eq:double_regge}, we should expect our isobars to be asymptotically bounded by $\sigma^{-1}$ in order to counteract the Kibble factor in \cref{eq:reduced_amplitude}. With a standard parameterization of the $P$-wave $\pi\pi$ phase shift~\cite{Garcia-Martin:2011iqs} extrapolated to $\delta(\sigma\to\infty) = \pi$, this limit is saturated by the contact and Deck driving terms of \cref{subsec:contact,subsec:deck} respectively with $m=1$. 

It is important to note that unlike the contact term, the Deck contribution can introduce arbitrary polynomials of $\sigma$ by considering non-minimal vertices with derivatives in the Feynman loop of \cref{fig:deck_triangle}. Any such coupling will, however, exceed the bound imposed by \cref{eq:double_regge} and thus violate high-energy unitarity and analyticity. 

\section{Fitting a third $t$-slope}
\label{app:second_contact}

The global fit results of \cref{sec:fits} suggest a possibility of production effects beyond those present in the nominal contact-plus-Deck model. Specifically, the region of small $\mtp$ is seen to be systematically less well described as the higher mass tail and coincides with the observation of a steeper characteristic $t$-slope in Ref.~\cite{COMPASS:2015gxz}. This may point to contributions of subleading nucleon spin-flip or non-Pomeron exchange diagrams (c.f. \cref{app:not_considered}).

To gauge whether this is the case, here, we consider fits with a non-minimal $t$-dependence by introducing an additional contact term. We make the replacement:
    \begin{equation}
        N_c(\mtp^2) \, e^{ct} \to N_c(\mtp^2) \, e^{ct} + N^\prime_c(\mtp^2) \, e^{c^\prime t} ~,
    \end{equation}
in terms of a third $t$-slope and set of normalizations. We assume this new term to have the same phase as the original contact term (i.e. $\phi_{c}^\prime = 0$), so as to not over-proliferate the number of free parameters but still allow interference with the Deck term. 

\begin{figure*}
    \centering
    \includegraphics[width=0.48\linewidth]{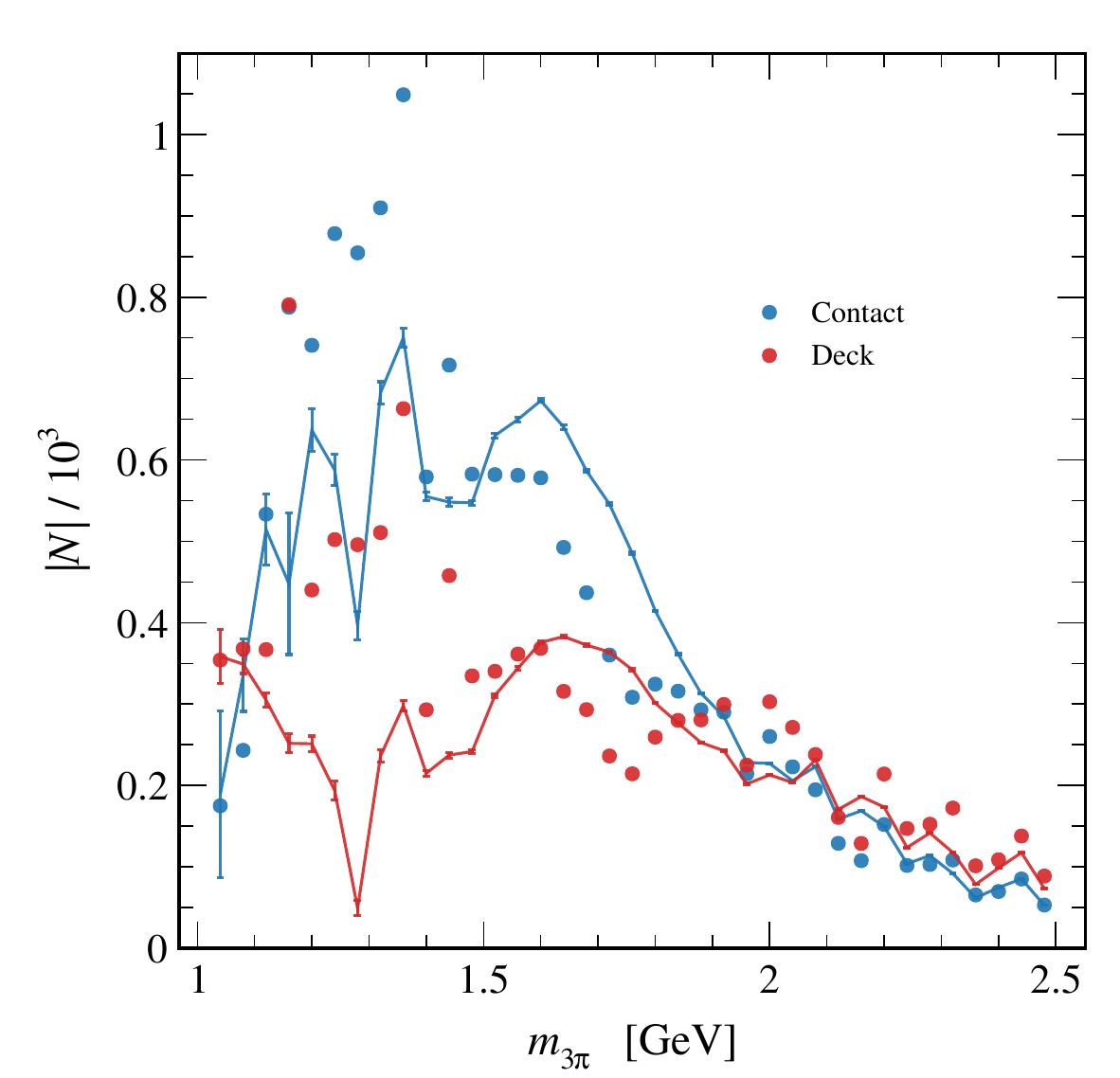}
    \includegraphics[width=0.48\linewidth]{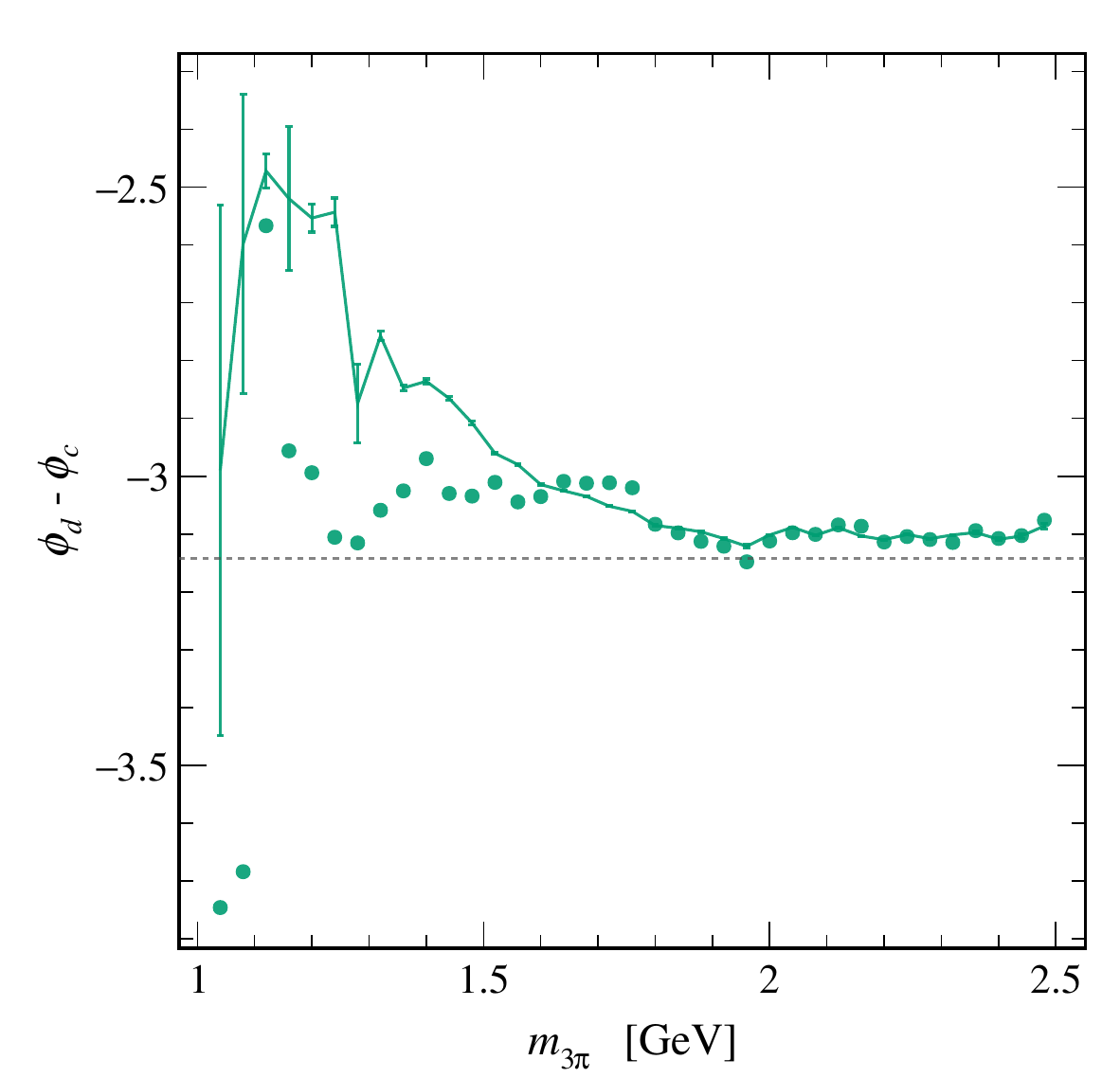}
    \caption{ $|N_c|$ and $|N_d|$ distributions (left) and phase difference (right) resulting from the extended fit. For comparison, we superimpose linear interpolations of the distributions from the minimal model (i.e. \cref{fig:pars}) as solid curves.}
    \label{fig:nonminimal_pars}
\end{figure*}

\begin{figure}
    \centering
    \includegraphics[width=\linewidth]{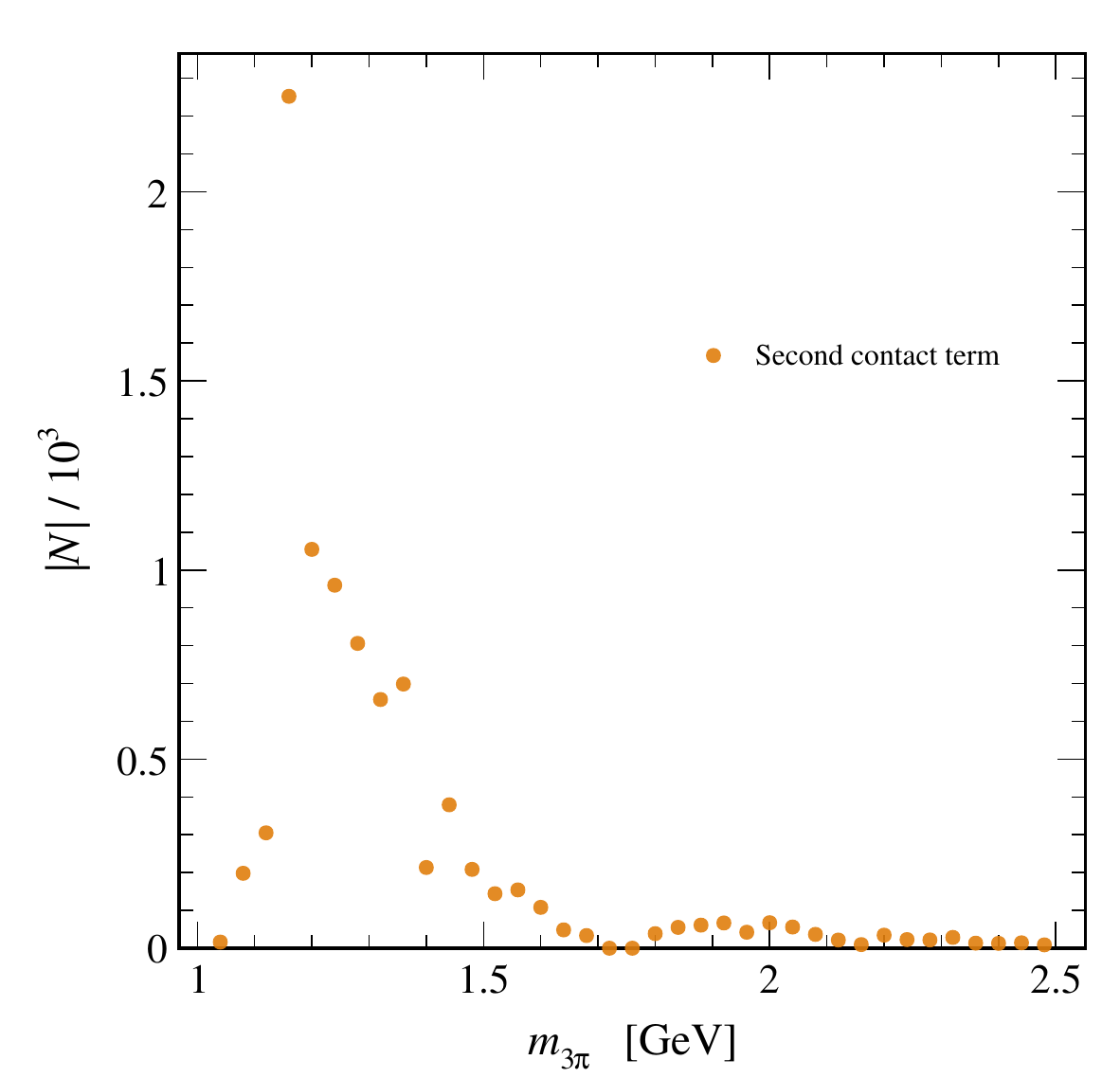}
    \caption{Bin-wise distributions of the second contact term in the extended fit.}
    \label{fig:nonminimal_Ncp}
\end{figure}

\begin{figure}
    \centering
    \includegraphics[width=\linewidth]{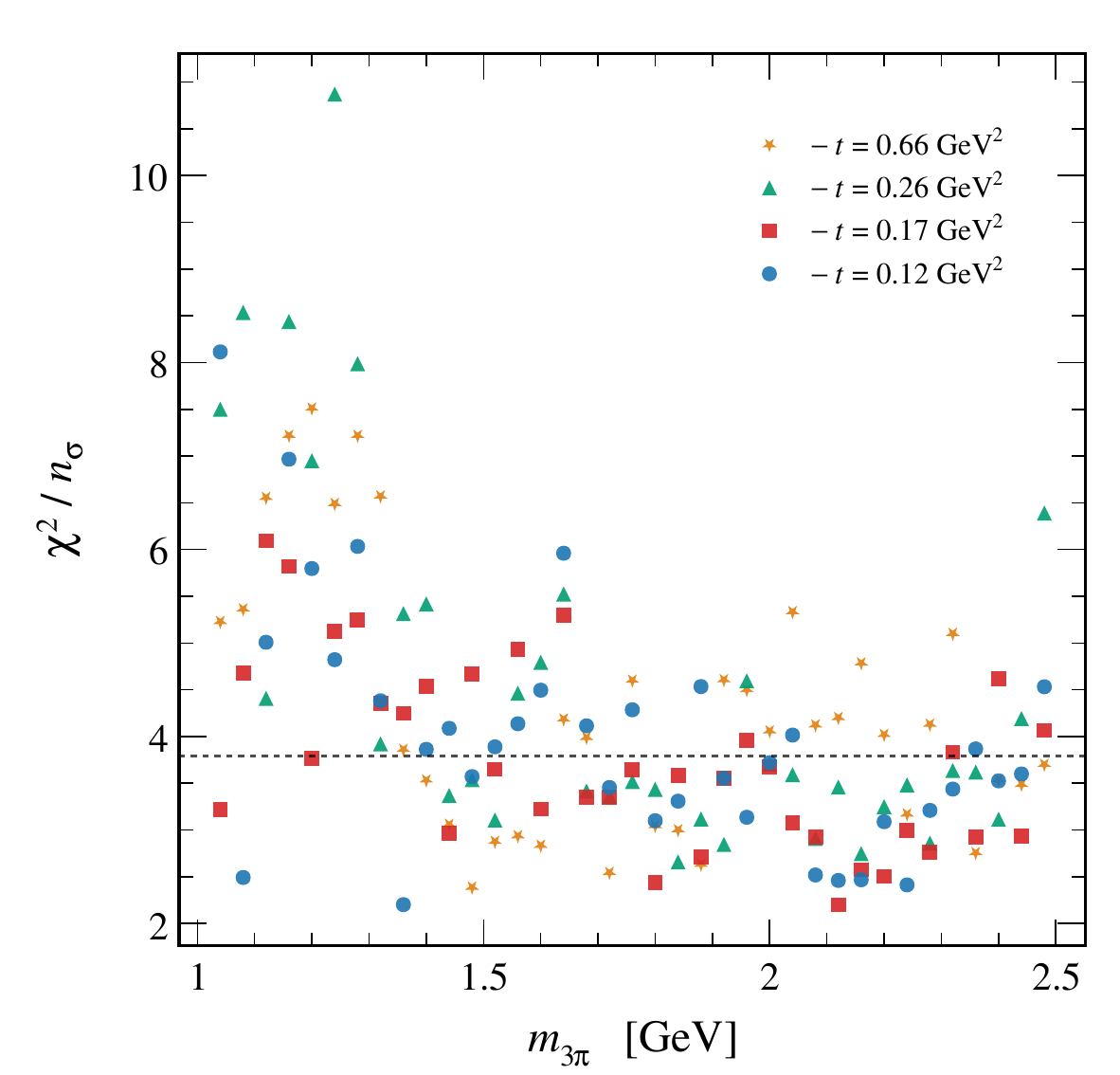}
    \caption{Distribution of normalized $\chi^2$'s resulting from the extended model. The dashed line denotes the $\chi^2/\text{dof} = 3.80$.} 
    \label{fig:nonminimal_chi2}
\end{figure}

The resulting bin-wise parameters and $\chi^2$ distributions are plotted in \cref{fig:nonminimal_pars,fig:nonminimal_Ncp,fig:nonminimal_chi2}. The best-fit values of $t$-slopes for the original contact and Deck terms are
    \begin{align}
        c = 2.78 \, \text{GeV}^{-2}
        \quad \text{and} \quad
        d = 2.91 \, \text{GeV}^{-2} ~,
    \end{align}
which are roughly comparable with the values seen in \cref{eq:minimal_slopes}, while the new contact term has a steeper $t$-dependence with $c^\prime = 5.66 \, \text{GeV}^{-2}$. Examining the $N^\prime_c$ distribution, this term indeed contributes primarily in the $\mtp < 1.5$ GeV region, as expected, while the $N_c$ and $N_d$ distributions now have a milder $\mtp$ dependence with a less obvious peak at $1.6$ GeV when compared to \cref{fig:pars}. The phase difference of the new fit also now hovers around $-\pi$ throughout the entire energy range indicating all normalizations are approximately real with purely destructive interference. We note that, because of this interference, the lack of as obvious of a peak in the normalizations does not indicate the absence of a resonance, since the final lineshape in the small mass region is determined by possible cancellations between individual amplitudes. This result also validates the increase of $\Im N_d$ observed in the minimal fit as driven by the different interference pattern needed to account for the steeper $t$-dependence at small $3\pi$ masses. 

The quantitative improvement of the fit quality, however, is relatively little. The extended fit results in a $\bar{\chi}^2 = 4.15$ and a $\chi^2/\text{dof} = 3.79$ which is less than a 10\% difference over the quoted results of \cref{sec:fits}. 
Examining the bin-by-bin distribution of $\chi^2$'s in \cref{fig:nonminimal_chi2}, we see a very similar distribution as \cref{fig:minimal_chi2s} with systematically larger $\chi^2/n_\sigma$ in the small $\mtp^2$ bins. This suggests that, while the inclusion of additional $t$-dependence allows small (likely over-fitted) improvements throughout the entire energy range, it does not address the primary deficiency of the current description in the low mass region. As such, we will continue to quote the minimal contact-plus-Deck model as our primary result with the caveat that the extracted parameter distributions are still subject to deviations based on the model considered. Ultimately how much these changes affect an extraction of the $\pi_1$ pole will need to be taken into account in the uncertainty analysis of any future work. 


\section{Bootstrap analysis}
\label{app:bootstap}

The uncertainties in the fit parameters are determined as the standard deviation of parameter distributions resulting from $10^4$ bootstrap samples. Because the quoted experimental uncertainties are so small, doing a traditional bootstrap analysis, i.e.\ assuming each parameter is uncorrelated and generating pseudo-data of each data point within its Gaussian error, produces tiny, sub-1\% uncertainties in all parameters. This is clearly unrealistic, especially considering the size of the pull and $\chi^2$ distributions in \cref{fig:dalitz_pull,fig:minimal_chi2s}. 

Instead, for each data point $\mathcal{F}_k^\text{ex}$ corresponding to the four-dimensional intensity, we calculate:
    \begin{equation}
        \label{eq:pull}
        S_k = \left| \frac{|\mathcal{F}_k^\text{ex}| - |\mathcal{F}_k^\text{BF}|}{\delta|\mathcal{F}_k^\text{ex}|} \right| ~,
    \end{equation}
where $|\mathcal{F}_k^\text{BF}|$ is the best-fit model curve evaluated at the same kinematic point. Clearly \cref{eq:pull} corresponds to absolute value of the pull calculated from the the best-fit curve with $\chi^2 = \sum_k S_k^2$. If $S_k \geq 1$, bootstrap pseudo-data are resampled within a Gaussian of width $S_k \, \sigma_k$ otherwise the original experimental uncertainty $\sigma_k$ is used. The bootstrapped data then is fit with \cref{eq:chi2} as before with the original errors. This procedure scales the spread of data points by their local distance to the best-fit curve and thus is expected to capture a slightly more realistic estimation of parameter uncertainties. 
    \begin{figure}
        \centering
        \includegraphics[width=\linewidth]{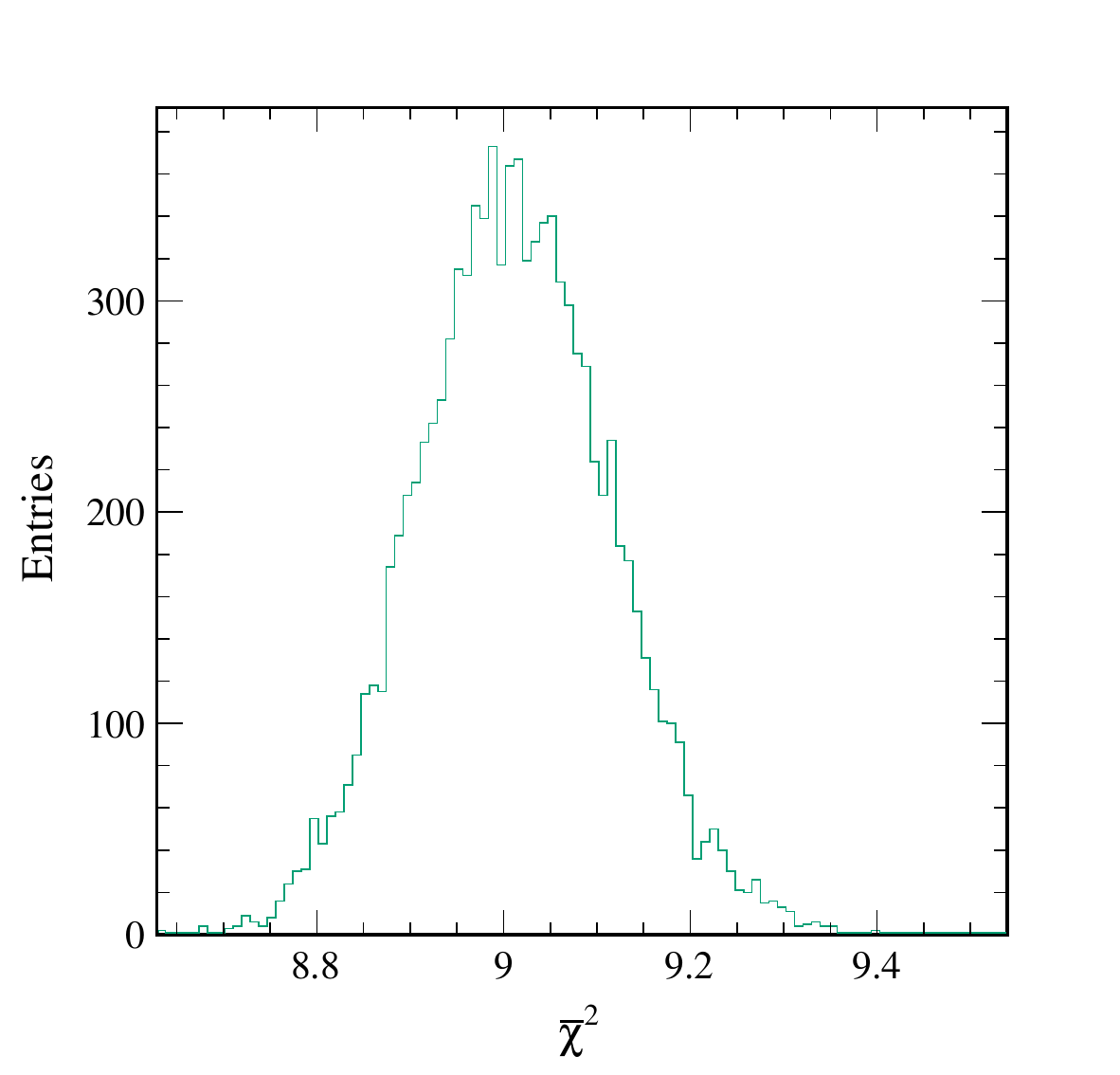}
        \caption{Bootstrapped distributions of the fit function $\bar{\chi}^2$ for the global fits.}
        \label{fig:chi2_bootstrap}
    \end{figure}

\begin{figure*}
        \centering
        \includegraphics[width=0.49\linewidth]{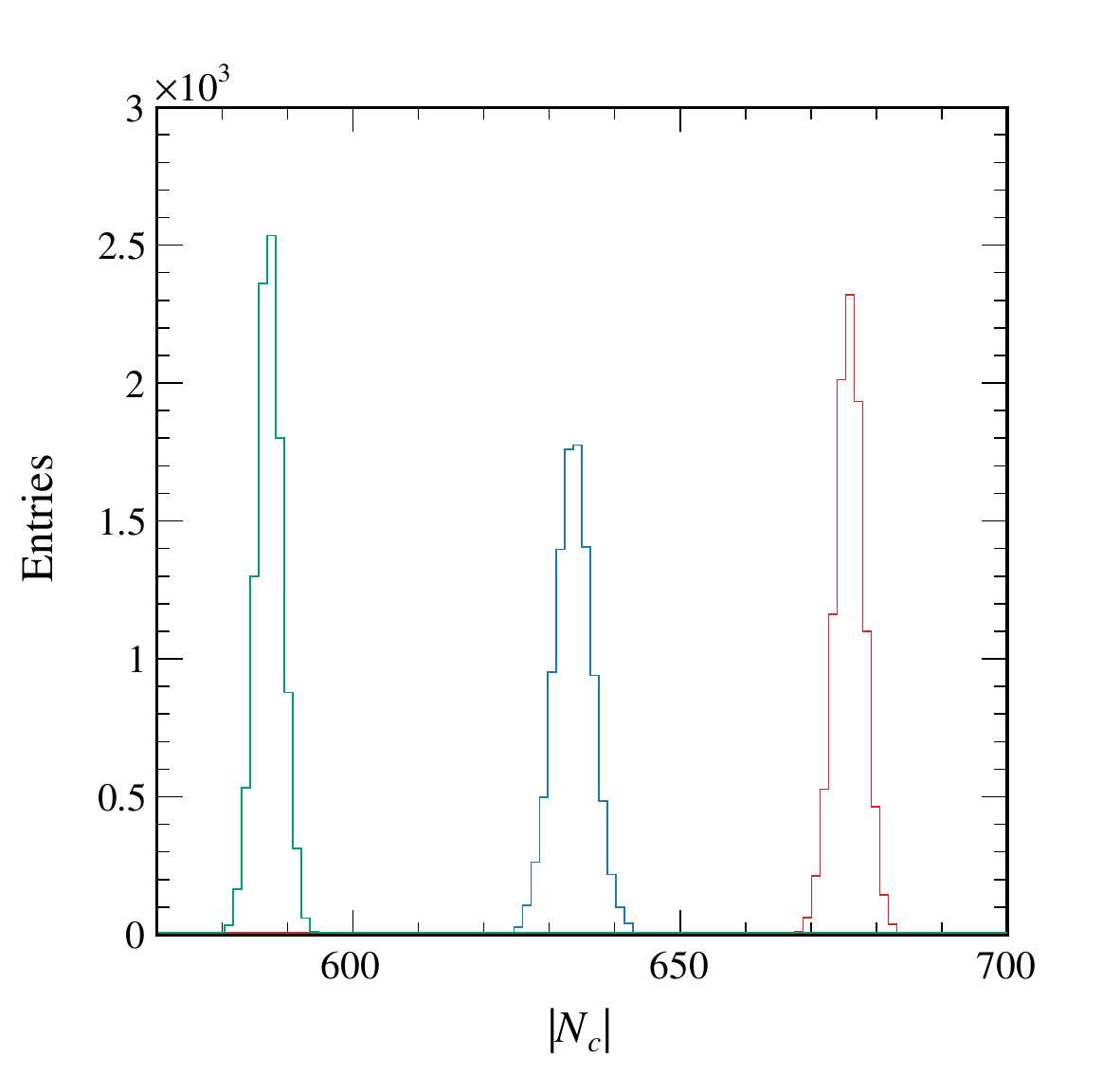}
        \includegraphics[width=0.49\linewidth]{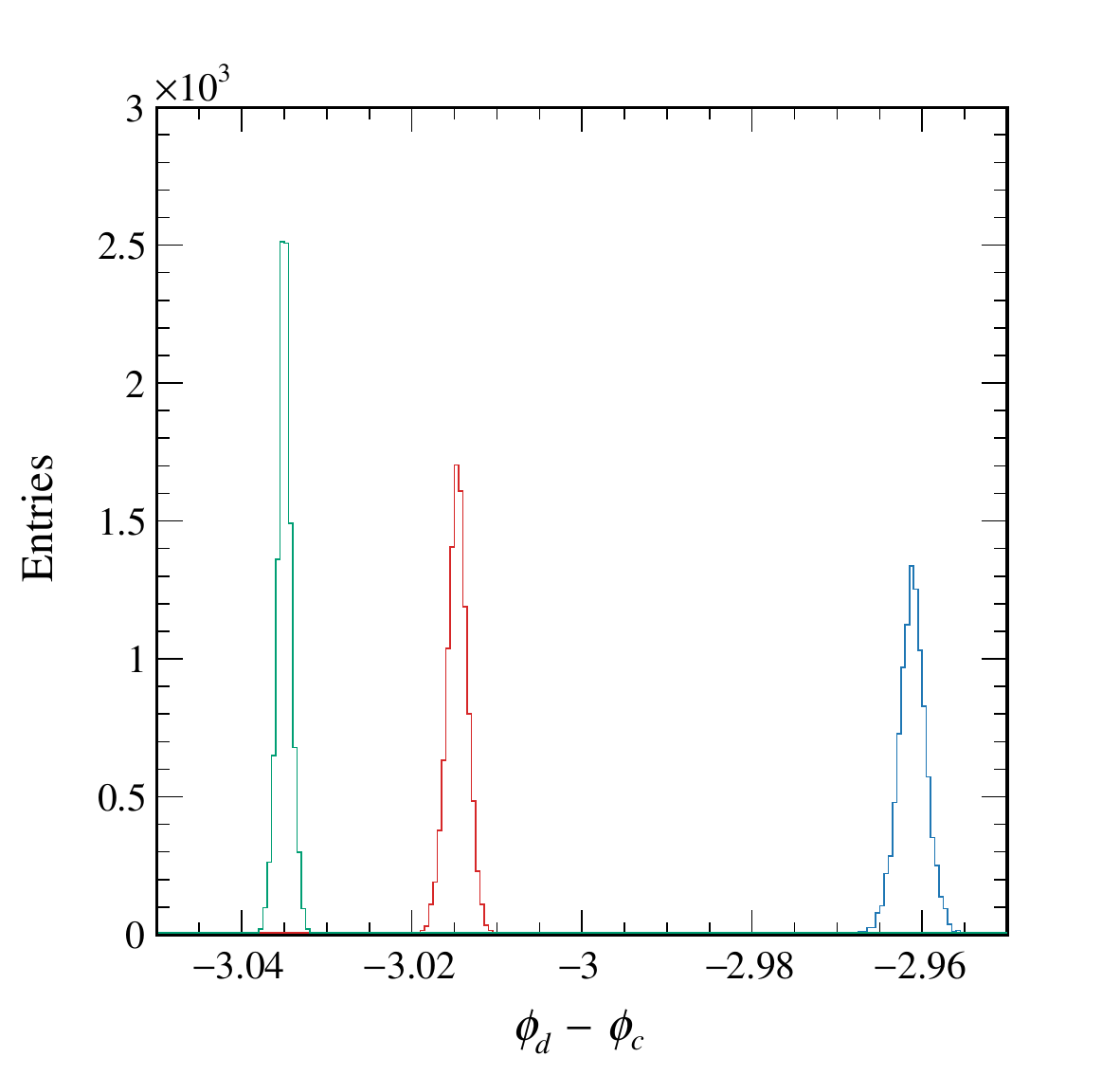}
        \caption{Bootstrapped distributions of the $|N_c|$ (left) and the phase difference (right) at fixed values of $\mtp = 1.52$ (blue), 1.6 (red), and 1.68 GeV (green).}
        \label{fig:Nc_bootstrap}
    \end{figure*}

\begin{figure}
        \centering
        \includegraphics[width=\linewidth]{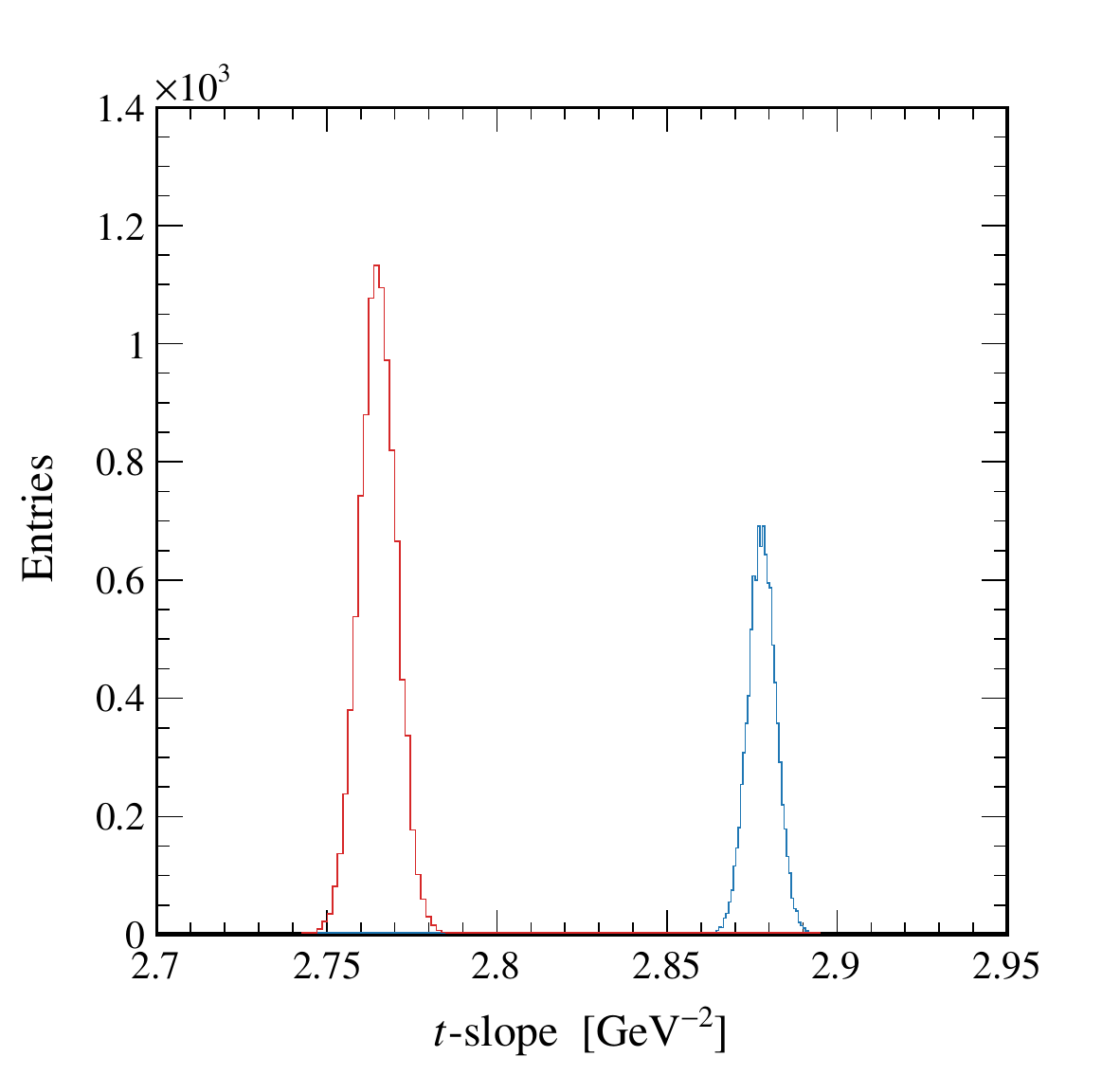}
        \caption{Bootstrapped distributions of the $t$-slope for the contact $c$ (blue) and Deck $d$ (red) terms.}
        \label{fig:tslope_bootstrap}
    \end{figure}

The resulting $\bar{\chi}^2$ distribution of the global fit is plotted in \cref{fig:chi2_bootstrap}, where we indeed see a Gaussian-like distribution with a mean approximately at twice the best-fit value. Example parameter distributions, e.g.\ for $|N_c|$, are also plotted in \cref{fig:Nc_bootstrap} and those of the $t$-slopes in \cref{fig:tslope_bootstrap}. Despite such a multi-dimensional fit, all normalization and phase parameters appear fairly Gaussian-like, without nearby local minima, and therefore allow straightforward calculations of the $1\sigma$ uncertainty. This likely reflects the relatively uncorrelated nature of the bin-wise normalizations of the different terms. 


We note that the bin-by-bin correlations of the COMPASS isobars do not explicitly factor into \cref{eq:pull}. These correlations are used in the original calculation of the data uncertainty (as detailed in \cref{sec:fits}) but are much smaller than the effect of rescaling by an average of $\bar{S}_k =\sqrt{\bar{\chi}^2} \approx 2$,\footnote{ To check consistency, we conducted the original fit including correlations with all errors inflated by a global factor of ${\sqrt{\bar{\chi}^2}}$, which indeed yields $\chi^2/\text{dof} \approx 1$.}
and thus ignored in the bootstrap resampling. 
The correlations will still enter implicitly as systematically shifted points are assumed to have a similarly correlated distance to the smooth best-fit curve. 

In the end, despite the rescaling, the estimated uncertainties remain very small and possibly still a severe underestimation of the true values. We remind that this smallness is related to the lack of systematic error estimation in the experimental freed-isobar analysis at present. As such, our parameter uncertainties are likely a reliable estimation of the relative error but will require further scaling in their absolute size in any subsequent analysis.

\bibliographystyle{apsrev4-2.bst}
\bibliography{refs}

@article{Stamen:2022eda,
    author = "Stamen, Dominik and Isken, Tobias and Kubis, Bastian and Mikhasenko, Mikhail and Niehus, Malwin",
    title = "{Analysis of rescattering effects in $3\pi $ final states}",
    eprint = "2212.11767",
    archivePrefix = "arXiv",
    primaryClass = "hep-ph",
    doi = "10.1140/epjc/s10052-023-11665-x",
    journal = "Eur. Phys. J. C",
    volume = "83",
    number = "6",
    pages = "510",
    year = "2023",
    note = "[Erratum: Eur.Phys.J.C 83, 586 (2023)]"
}

@article{Ascoli:1974hi,
    author = "Ascoli, G. and Cutler, R. and Jones, L. M. and Kruse, U. and Roberts, T. and Weinstein, B. and Wyld, H. W.",
    title = "{Deck-model calculation of pi- p ---\ensuremath{>} pi- pi+ pi- p}",
    doi = "10.1103/PhysRevD.9.1963",
    journal = "Phys. Rev. D",
    volume = "9",
    pages = "1963--1979",
    year = "1974"
}

@article{Ascoli:1973htr,
    author = "Ascoli, G. and Jones, L. M. and Weinstein, B. and Wyld, H. W.",
    title = "{Partial-wave analysis of the deck amplitude for pi n ---\ensuremath{>} pi pi pi n}",
    doi = "10.1103/PhysRevD.8.3894",
    journal = "Phys. Rev. D",
    volume = "8",
    pages = "3894--3919",
    year = "1973"
}

@article{COMPASS:2021ogp,
    author = "Alexeev, M. G. and others",
    collaboration = "COMPASS",
    title = "{Exotic meson $\pi_1(1600)$ with $J^{PC} = 1^{-+}$ and its decay into $\rho(770)\pi$}",
    eprint = "2108.01744",
    archivePrefix = "arXiv",
    primaryClass = "hep-ex",
    reportNumber = "CERN-EP-2021\textendash{}162",
    doi = "10.1103/PhysRevD.105.012005",
    journal = "Phys. Rev. D",
    volume = "105",
    number = "1",
    pages = "012005",
    year = "2022"
}

@article{Albaladejo:2019huw,
    author = "Albaladejo, M. and Winney, D. and Danilkin, I. V. and Fern\'andez-Ram\'\i{}rez, C. and Mathieu, V. and Mikhasenko, M. and Pilloni, A. and Silva-Castro, J. A. and Szczepaniak, A. P.",
    collaboration = "JPAC",
    title = "{Khuri-Treiman equations for $3\pi$ decays of particles with spin}",
    eprint = "1910.03107",
    archivePrefix = "arXiv",
    primaryClass = "hep-ph",
    reportNumber = "JLAB-THY-19-3066",
    doi = "10.1103/PhysRevD.101.054018",
    journal = "Phys. Rev. D",
    volume = "101",
    number = "5",
    pages = "054018",
    year = "2020"
}

@article{JPAC:2023nhq,
    author = "Albaladejo, M. and others",
    collaboration = "JPAC",
    title = "{Khuri-Treiman analysis of J/\ensuremath{\psi}\textrightarrow{}\ensuremath{\pi}+\ensuremath{\pi}-\ensuremath{\pi}0}",
    eprint = "2304.09736",
    archivePrefix = "arXiv",
    primaryClass = "hep-ph",
    reportNumber = "JLAB-THY-23-3796",
    doi = "10.1103/PhysRevD.108.014035",
    journal = "Phys. Rev. D",
    volume = "108",
    number = "1",
    pages = "014035",
    year = "2023"
}

@article{JPAC:2020umo,
    author = "Albaladejo, M. and Danilkin, I. and Gonzalez-Solis, S. and Winney, D. and Fernandez-Ramirez, C. and Blin, A. N. Hiller and Mathieu, V. and Mikhasenko, M. and Pilloni, A. and Szczepaniak, A.",
    collaboration = "JPAC",
    title = "{$\omega \rightarrow 3\pi $ and $\omega \pi ^{0}$ transition form factor revisited}",
    eprint = "2006.01058",
    archivePrefix = "arXiv",
    primaryClass = "hep-ph",
    reportNumber = "JLAB-THY-20-3200",
    doi = "10.1140/epjc/s10052-020-08576-6",
    journal = "Eur. Phys. J. C",
    volume = "80",
    number = "12",
    pages = "1107",
    year = "2020"
}

@article{Bibrzycki:2021rwh,
    author = "Bibrzycki, L. and Fernandez-Ramirez, C. and Mathieu, V. and Mikhasenko, M. and Albaladejo, M. and Blin, A. N. Hiller and Pilloni, A. and Szczepaniak, A. P.",
    collaboration = "JPAC",
    title = "{$\pi^-p\to\eta^{(\prime)}\, \pi^- p$ in the double-Regge region~
}",
    eprint = "2104.10646",
    archivePrefix = "arXiv",
    primaryClass = "hep-ph",
    reportNumber = "JLAB-THY-21-3354",
    doi = "10.1140/epjc/s10052-021-09594-8",
    journal = "Eur. Phys. J. C",
    volume = "81",
    pages = "647",
    year = "2021",
    note = "[Erratum: Eur.Phys.J.C 81, 915 (2021)]"
}

@article{Jackura:2023qtp,
    author = "Jackura, Andrew W. and Brice{\~n}o, Ra{\'u}l A.",
    title = "{Partial-wave projection of the one-particle exchange in three-body scattering amplitudes}",
    eprint = "2312.00625",
    archivePrefix = "arXiv",
    primaryClass = "hep-ph",
    doi = "10.1103/PhysRevD.109.096030",
    journal = "Phys. Rev. D",
    volume = "109",
    number = "9",
    pages = "096030",
    year = "2024"
}

@article{Shimada:1978sx,
    author = "Shimada, T. and Martin, Alan D. and Irving, A. C.",
    title = "{DOUBLE REGGE EXCHANGE PHENOMENOLOGY}",
    reportNumber = "Print-78-0676 (DURHAM)",
    doi = "10.1016/0550-3213(78)90209-2",
    journal = "Nucl. Phys. B",
    volume = "142",
    pages = "344--364",
    year = "1978"
}

@article{Niecknig:2012sj,
    author = "Niecknig, Franz and Kubis, Bastian and Schneider, Sebastian P.",
    title = "{Dispersive analysis of $\omega -> 3\pi$ and $\phi -> 3\pi$ decays}",
    eprint = "1203.2501",
    archivePrefix = "arXiv",
    primaryClass = "hep-ph",
    doi = "10.1140/epjc/s10052-012-2014-1",
    journal = "Eur. Phys. J. C",
    volume = "72",
    pages = "2014",
    year = "2012"
}

@article{Stamen:2024gfz,
    author = "Stamen, D. and Winney, D. and Rodas, A. and Fern{\'a}ndez-Ram{\'\i}rez, C. and Mathieu, V. and Monta{\~n}a, G. and Pilloni, A. and Szczepaniak, A. P.",
    title = "{Toward a unified description of hadron scattering at all energies}",
    eprint = "2409.09172",
    archivePrefix = "arXiv",
    primaryClass = "hep-ph",
    reportNumber = "JLAB-THY-24-4182",
    doi = "10.1103/PhysRevD.110.114023",
    journal = "Phys. Rev. D",
    volume = "110",
    number = "11",
    pages = "114023",
    year = "2024"
}

@article{Deck:1964hm,
  title = {Kinematical Interpretation of the First {$\pi$} - {$\rho$} Resonance},
  author = {Deck, Robert T.},
  year = 1964,
  journal = {Phys. Rev. Lett.},
  volume = {13},
  pages = {169--173},
  doi = {10.1103/PhysRevLett.13.169}
}

@article{Kibble:1960zz,
    author = "Kibble, T. W. B.",
    title = "{Kinematics of General Scattering Processes and the Mandelstam Representation}",
    doi = "10.1103/PhysRev.117.1159",
    journal = "Phys. Rev.",
    volume = "117",
    pages = "1159--1162",
    year = "1960"
}

@article{COMPASS:2018uzl,
    author = "Aghasyan, M. and others",
    collaboration = "COMPASS",
    title = "{Light isovector resonances in $\pi^- p \to \pi^-\pi^-\pi^+ p$ at 190 GeV/${\it c}$}",
    eprint = "1802.05913",
    archivePrefix = "arXiv",
    primaryClass = "hep-ex",
    reportNumber = "CERN-EP-2018-021",
    doi = "10.1103/PhysRevD.98.092003",
    journal = "Phys. Rev. D",
    volume = "98",
    number = "9",
    pages = "092003",
    year = "2018"
}

@book{Collins:1977jy,
    author = "Collins, P. D. B.",
    title = "{An Introduction to Regge Theory and High Energy Physics}",
    doi = "10.1017/9781009403269",
    isbn = "978-1-009-40326-9, 978-1-009-40329-0, 978-1-009-40328-3, 978-0-521-11035-8",
    publisher = "Cambridge University Press",
    year = "1977"
}

@article{Hahn:1998yk,
    author = "Hahn, T. and Perez-Victoria, M.",
    title = "{Automatized one loop calculations in four-dimensions and D-dimensions}",
    eprint = "hep-ph/9807565",
    archivePrefix = "arXiv",
    reportNumber = "UG-FT-87-98, KA-TP-7-1998",
    doi = "10.1016/S0010-4655(98)00173-8",
    journal = "Comput. Phys. Commun.",
    volume = "118",
    pages = "153--165",
    year = "1999"
}

@book{Gribov:2003nw,
    author = "Gribov, V. N.",
    title = "{The theory of complex angular momenta: Gribov lectures on theoretical physics}",
    doi = "10.1017/CBO9780511534959",
    isbn = "978-0-521-03703-7, 978-0-521-81834-6, 978-0-511-05504-1",
    publisher = "Cambridge University Press",
    series = "Cambridge Monographs on Mathematical Physics",
    month = "6",
    year = "2007"
}

@article{JointPhysicsAnalysisCenter:2024kck,
    author = "Montana, Gloria and others",
    collaboration = "Joint Physics Analysis Center",
    title = "{Revisiting gauge invariance and Reggeization of pion exchange}",
    eprint = "2407.19577",
    archivePrefix = "arXiv",
    primaryClass = "hep-ph",
    reportNumber = "JLAB-THY-24-4129",
    doi = "10.1103/PhysRevD.110.114012",
    journal = "Phys. Rev. D",
    volume = "110",
    number = "11",
    pages = "114012",
    year = "2024"
}

@book{Byckling:1971vca,
    author = "Byckling, Eero and Kajantie, K.",
    title = "{Particle Kinematics}",
    publisher = "University of Jyvaskyla",
    address = "Jyvaskyla, Finland",
    year = "1971"
}

@article{Aitchison:1965ghs,
    author = "Aitchison, I. J. R. and Kacser, C.",
    title = "{Singularities and discontinuities of the triangle graph, as a function of an internal mass}",
    doi = "10.1007/bf02721045",
    journal = "Nuovo Cim. A",
    volume = "40",
    number = "2",
    pages = "576--588",
    year = "1965"
}

@article{Kacser:1963zz,
    author = "Kacser, C.",
    title = "{Analytic Structure of Partial-Wave Amplitudes for Production and Decay Processes}",
    reportNumber = "TID-19083",
    doi = "10.1103/PhysRev.132.2712",
    journal = "Phys. Rev.",
    volume = "132",
    number = "6",
    pages = "2712",
    year = "1963"
}

@article{Trueman:1964zzb,
    author = "Trueman, T. L. and Wick, G. C.",
    title = "{Crossing relations for helicity amplitudes}",
    doi = "10.1016/0003-4916(64)90254-4",
    journal = "Annals Phys.",
    volume = "26",
    pages = "322--335",
    year = "1964"
}

@Book{		  abramowitz+stegun,
  address	= {New York},
  author	= {Abramowitz, Milton and Stegun, Irene A.},
  description	= {BibTeX - Wikipedia, the free encyclopedia},
  edition	= {ninth Dover printing, tenth GPO printing},
  publisher	= {Dover},
  title		= {Handbook of Mathematical Functions with Formulas, Graphs,
		  and Mathematical Tables},
  year		= 1964
}

@phdthesis{Akdag:2023oob,
    author = "Akdag, Mehmet Hakan",
    title = "{C and CP Violation in Light-Meson Decays}",
    school = "U. Bonn (main)",
    year = "2023"
}

@article{Omnes:1958hv,
    author = "Omnes, R.",
    title = "{On the Solution of certain singular integral equations of quantum field theory}",
    doi = "10.1007/BF02747746",
    journal = "Nuovo Cim.",
    volume = "8",
    pages = "316--326",
    year = "1958"
}

@book{Muskhelishvili:1958,
    title = {Singular Integral Equations},
    author = {Muskhelishvili, N. I.},
    year = {1958},
    publisher = {Wolters-Noordhoff Publishing},
    doi = {10.1007/978-94-009-9994-7},
}

@article{Mikhasenko:2019vhk,
    author = "Mikhasenko, M. and Wunderlich, Y. and Jackura, A. and Mathieu, V. and Pilloni, A. and Ketzer, B. and Szczepaniak, A. P.",
    title = "{Three-body scattering: Ladders and Resonances}",
    eprint = "1904.11894",
    archivePrefix = "arXiv",
    primaryClass = "hep-ph",
    reportNumber = "JLAB-THY-19-2924",
    doi = "10.1007/JHEP08(2019)080",
    journal = "JHEP",
    volume = "08",
    pages = "080",
    year = "2019"
}

@article{Pasquier:1968zz,
    author = "Pasquier, R. and Pasquier, J. Y.",
    title = "{Khuri-Treiman-Type Equations for Three-Body Decay and Production Processes}",
    doi = "10.1103/PhysRev.170.1294",
    journal = "Phys. Rev.",
    volume = "170",
    pages = "1294--1309",
    year = "1968"
}

@article{Pasquier:1969dt,
    author = "Pasquier, R. and Pasquier, J. Y.",
    title = "{Khuri-treiman-type equations for three-body decay and production processes. 2.}",
    doi = "10.1103/PhysRev.177.2482",
    journal = "Phys. Rev.",
    volume = "177",
    pages = "2482--2493",
    year = "1969"
}

@article{Aitchison:1965zz,
  title = {Dispersion {{Theory Model}} of {{Three-Body Production}} and {{Decay Processes}}},
  author = {Aitchison, I. J. R.},
  year = 1965,
  journal = {Phys. Rev.},
  volume = {137},
  pages = {B1070-B1084},
  doi = {10.1103/PhysRev.137.B1070},
}

@article{Aitchison:1966lpz,
    author = "Aitchison, I. J. R. and Pasquier, R.",
    title = "{Three-Body Unitarity and Khuri-Treiman Amplitudes}",
    doi = "10.1103/PhysRev.152.1274",
    journal = "Phys. Rev.",
    volume = "152",
    number = "4",
    pages = "1274",
    year = "1966"
}

@article{COMPASS:2014vkj,
    author = "Adolph, C. and others",
    collaboration = "COMPASS",
    title = "{Odd and even partial waves of $\eta\pi^-$ and $\eta'\pi^-$ in $\pi^-p\to\eta^{(\prime)}\pi^-p$ at $191\,\textrm{GeV}/c$}",
    eprint = "1408.4286",
    archivePrefix = "arXiv",
    primaryClass = "hep-ex",
    reportNumber = "CERN-PH-EP-2014-204",
    doi = "10.1016/j.physletb.2014.11.058",
    journal = "Phys. Lett. B",
    volume = "740",
    pages = "303--311",
    year = "2015",
    note = "[Erratum: Phys.Lett.B 811, 135913 (2020)]"
}

@book{Gribov:2009cfk,
    author = "Gribov, Vladimir",
    title = "{Strong Interactions of Hadrons at High Energies : Gribov Lectures on Theoretical Physics}",
    doi = "10.1017/9781009290227",
    isbn = "978-1-009-29022-7, 978-1-009-29027-2, 978-1-009-29024-1",
    publisher = "Oxford University Press",
    year = "2009"
}

@article{Froissart:1961ux,
    author = "Froissart, Marcel",
    title = "{Asymptotic behavior and subtractions in the Mandelstam representation}",
    doi = "10.1103/PhysRev.123.1053",
    journal = "Phys. Rev.",
    volume = "123",
    pages = "1053--1057",
    year = "1961"
}

@article{Krinner:2017dba,
    author = "Krinner, F. and Greenwald, D. and Ryabchikov, D. and Grube, B. and Paul, S.",
    title = "{Ambiguities in model-independent partial-wave analysis}",
    eprint = "1710.09849",
    archivePrefix = "arXiv",
    primaryClass = "hep-ph",
    doi = "10.1103/PhysRevD.97.114008",
    journal = "Phys. Rev. D",
    volume = "97",
    number = "11",
    pages = "114008",
    year = "2018"
}

@phdthesis{Krinner:2018bwg,
    author = "Krinner, Fabian",
    title = "{Freed-Isobar Partial-Wave Analysis}",
    school = "Munich, Tech. U.",
    year = "2018"
}

@article{Stern:1993rg,
    author = "Stern, J. and Sazdjian, H. and Fuchs, N. H.",
    title = "{What pi - pi scattering tells us about chiral perturbation theory}",
    eprint = "hep-ph/9301244",
    archivePrefix = "arXiv",
    reportNumber = "PURD-TH-93-01, IPNO-TH-92-106",
    doi = "10.1103/PhysRevD.47.3814",
    journal = "Phys. Rev. D",
    volume = "47",
    pages = "3814--3838",
    year = "1993"
}

@article{Garcia-Martin:2011iqs,
    author = "Garcia-Martin, R. and Kaminski, R. and Pelaez, J. R. and Ruiz de Elvira, J. and Yndurain, F. J.",
    title = "{The Pion-pion scattering amplitude. IV: Improved analysis with once subtracted Roy-like equations up to 1100 MeV}",
    eprint = "1102.2183",
    archivePrefix = "arXiv",
    primaryClass = "hep-ph",
    doi = "10.1103/PhysRevD.83.074004",
    journal = "Phys. Rev. D",
    volume = "83",
    pages = "074004",
    year = "2011"
}

@article{Passarino:1978jh,
    author = "Passarino, G. and Veltman, M. J. G.",
    title = "{One Loop Corrections for e+ e- Annihilation Into mu+ mu- in the Weinberg Model}",
    reportNumber = "Print-79-0284 (UTRECHT)",
    doi = "10.1016/0550-3213(79)90234-7",
    journal = "Nucl. Phys. B",
    volume = "160",
    pages = "151--207",
    year = "1979"
}

@article{E852:1997gvf,
    author = "Thompson, D. R. and others",
    collaboration = "E852",
    title = "{Evidence for exotic meson production in the reaction pi- p ---{\ensuremath{>}} eta pi- p at 18-GeV/c}",
    eprint = "hep-ex/9705011",
    archivePrefix = "arXiv",
    doi = "10.1103/PhysRevLett.79.1630",
    journal = "Phys. Rev. Lett.",
    volume = "79",
    pages = "1630--1633",
    year = "1997"
}

@article{E852:1999xev,
    author = "Chung, S. U. and others",
    collaboration = "E852",
    title = "{Evidence for exotic J(PC) = 1-+ meson production in the reaction pi- p ---{\ensuremath{>}} eta pi- p at 18-GeV/c}",
    eprint = "hep-ex/9902003",
    archivePrefix = "arXiv",
    doi = "10.1103/PhysRevD.60.092001",
    journal = "Phys. Rev. D",
    volume = "60",
    pages = "092001",
    year = "1999"
}

@article{E862:2006cfp,
    author = "Adams, G. S. and others",
    collaboration = "E862",
    title = "{Confirmation of a pi(1)0 Exotic Meson in the eta pi0 System}",
    eprint = "hep-ex/0612062",
    archivePrefix = "arXiv",
    doi = "10.1016/j.physletb.2007.07.068",
    journal = "Phys. Lett. B",
    volume = "657",
    pages = "27--31",
    year = "2007"
}

@article{E852:2001ikk,
    author = "Ivanov, E. I. and others",
    collaboration = "E852",
    title = "{Observation of exotic meson production in the reaction pi- p ---{\ensuremath{>}} eta-prime pi- p at 18-GeV / c}",
    eprint = "hep-ex/0101058",
    archivePrefix = "arXiv",
    reportNumber = "JLAB-PHY-01-96",
    doi = "10.1103/PhysRevLett.86.3977",
    journal = "Phys. Rev. Lett.",
    volume = "86",
    pages = "3977--3980",
    year = "2001"
}

@article{Khokhlov:2000tk,
    author = "Khokhlov, Yu. A.",
    editor = "Faldt, G. and Hoistad, B. and Kullander, S.",
    collaboration = "VES",
    title = "{Study of X(1600) 1-+ hybrid}",
    doi = "10.1016/S0375-9474(99)00663-6",
    journal = "Nucl. Phys. A",
    volume = "663",
    pages = "596--599",
    year = "2000"
}

@article{CrystalBarrel:1998cfz,
    author = "Abele, A. and others",
    collaboration = "Crystal Barrel",
    title = "{Exotic eta pi state in anti-p d annihilation at rest into pi- pi0 eta p(spectator)}",
    doi = "10.1016/S0370-2693(98)00123-3",
    journal = "Phys. Lett. B",
    volume = "423",
    pages = "175--184",
    year = "1998"
}

@article{Horn:1977rq,
    author = "Horn, D. and Mandula, J.",
    title = "{A Model of Mesons with Constituent Gluons}",
    reportNumber = "CALT-68-575",
    doi = "10.1103/PhysRevD.17.898",
    journal = "Phys. Rev. D",
    volume = "17",
    pages = "898",
    year = "1978"
}

@article{Isgur:1984bm,
    author = "Isgur, Nathan and Paton, Jack E.",
    title = "{A Flux Tube Model for Hadrons in QCD}",
    reportNumber = "Print-84-0830 (TORONTO)",
    doi = "10.1103/PhysRevD.31.2910",
    journal = "Phys. Rev. D",
    volume = "31",
    pages = "2910",
    year = "1985"
}

@article{Chanowitz:1982qj,
    author = "Chanowitz, Michael S. and Sharpe, Stephen R.",
    title = "{Hybrids: Mixed States of Quarks and Gluons}",
    reportNumber = "LBL-14865",
    doi = "10.1016/0550-3213(83)90635-1",
    journal = "Nucl. Phys. B",
    volume = "222",
    pages = "211--244",
    year = "1983",
    note = "[Erratum: Nucl.Phys.B 228, 588--588 (1983)]"
}

@article{Barnes:1982tx,
    author = "Barnes, Ted and Close, F. E. and de Viron, F.",
    title = "{Q anti-Q G Hermaphrodite Mesons in the MIT Bag Model}",
    reportNumber = "RL-82-088",
    doi = "10.1016/0550-3213(83)90004-4",
    journal = "Nucl. Phys. B",
    volume = "224",
    pages = "241",
    year = "1983"
}

@article{Dudek:2013yja,
    author = "Dudek, Jozef J. and Edwards, Robert G. and Guo, Peng and Thomas, Christopher E.",
    collaboration = "Hadron Spectrum",
    title = "{Toward the excited isoscalar meson spectrum from lattice QCD}",
    eprint = "1309.2608",
    archivePrefix = "arXiv",
    primaryClass = "hep-lat",
    reportNumber = "JLAB-THY-13-1786, TCDMATH-13-11",
    doi = "10.1103/PhysRevD.88.094505",
    journal = "Phys. Rev. D",
    volume = "88",
    number = "9",
    pages = "094505",
    year = "2013"
}

@article{Woss:2020ayi,
    author = "Woss, Antoni J. and Dudek, Jozef J. and Edwards, Robert G. and Thomas, Christopher E. and Wilson, David J.",
    collaboration = "Hadron Spectrum",
    title = "{Decays of an exotic $1{-+}$ hybrid meson resonance in QCD}",
    eprint = "2009.10034",
    archivePrefix = "arXiv",
    primaryClass = "hep-lat",
    reportNumber = "JLAB-THY-20-3249",
    doi = "10.1103/PhysRevD.103.054502",
    journal = "Phys. Rev. D",
    volume = "103",
    number = "5",
    pages = "054502",
    year = "2021"
}

@article{Meyer:2015eta,
    author = "Meyer, C. A. and Swanson, E. S.",
    title = "{Hybrid Mesons}",
    eprint = "1502.07276",
    archivePrefix = "arXiv",
    primaryClass = "hep-ph",
    doi = "10.1016/j.ppnp.2015.03.001",
    journal = "Prog. Part. Nucl. Phys.",
    volume = "82",
    pages = "21--58",
    year = "2015"
}

@article{Meyer:2010ku,
    author = "Meyer, C. A. and Van Haarlem, Y.",
    title = "{The Status of Exotic-quantum-number Mesons}",
    eprint = "1004.5516",
    archivePrefix = "arXiv",
    primaryClass = "nucl-ex",
    doi = "10.1103/PhysRevC.82.025208",
    journal = "Phys. Rev. C",
    volume = "82",
    pages = "025208",
    year = "2010"
}

@article{JPAC:2018zyd,
    author = "Rodas, A. and others",
    collaboration = "JPAC",
    title = "{Determination of the pole position of the lightest hybrid meson candidate}",
    eprint = "1810.04171",
    archivePrefix = "arXiv",
    primaryClass = "hep-ph",
    reportNumber = "JLAB-THY-18-2839",
    doi = "10.1103/PhysRevLett.122.042002",
    journal = "Phys. Rev. Lett.",
    volume = "122",
    number = "4",
    pages = "042002",
    year = "2019"
}

@article{Kopf:2020yoa,
    author = {Kopf, B. and Albrecht, M. and Koch, H. and K{\"u}{\ss}ner, M. and Pychy, J. and Qin, X. and Wiedner, U.},
    title = "{Investigation of the lightest hybrid meson candidate with a coupled-channel analysis of ${{\bar{p}}p}$-, $\pi ^- p$- and ${\pi \pi }$-Data}",
    eprint = "2008.11566",
    archivePrefix = "arXiv",
    primaryClass = "hep-ph",
    doi = "10.1140/epjc/s10052-021-09821-2",
    journal = "Eur. Phys. J. C",
    volume = "81",
    number = "12",
    pages = "1056",
    year = "2021"
}

@article{CrystalBarrel:2019zqh,
    author = "Albrecht, M. and others",
    collaboration = "Crystal Barrel",
    title = "{Coupled channel analysis of ${\bar{p}p}\,\rightarrow \,\pi ^0\pi ^0\eta $, ${\pi ^0\eta \eta }$ and ${K^+K^-\pi ^0}$ at 900 MeV/c and of ${\pi \pi }$-scattering data}",
    eprint = "1909.07091",
    archivePrefix = "arXiv",
    primaryClass = "hep-ex",
    doi = "10.1140/epjc/s10052-020-7930-x",
    journal = "Eur. Phys. J. C",
    volume = "80",
    number = "5",
    pages = "453",
    year = "2020"
}

@article{GlueX:2024erj,
    author = "Afzal, F. and others",
    collaboration = "GlueX",
    title = "{Upper Limit on the Photoproduction Cross Section of the Spin-Exotic {\ensuremath{\pi}}1(1600)}",
    eprint = "2407.03316",
    archivePrefix = "arXiv",
    primaryClass = "nucl-ex",
    doi = "10.1103/PhysRevLett.133.261903",
    journal = "Phys. Rev. Lett.",
    volume = "133",
    number = "26",
    pages = "261903",
    year = "2024"
}

@article{Mai:2018djl,
    author = "Mai, Maxim and Doring, Michael",
    title = "{Finite-Volume Spectrum of $\pi^+\pi^+$ and $\pi^+\pi^+\pi^+$ Systems}",
    eprint = "1807.04746",
    archivePrefix = "arXiv",
    primaryClass = "hep-lat",
    reportNumber = "JLAB-THY-18-2767",
    doi = "10.1103/PhysRevLett.122.062503",
    journal = "Phys. Rev. Lett.",
    volume = "122",
    number = "6",
    pages = "062503",
    year = "2019"
}

@article{Briceno:2025yuq,
    author = "Brice{\~n}o, Ra{\'u}l A. and Hansen, Maxwell T. and Jackura, Andrew W. and Edwards, Robert G. and Thomas, Christopher E.",
    title = "{Isotensor $πππ$ scattering with a $ρ$ resonant subsystem from QCD}",
    eprint = "2510.24894",
    archivePrefix = "arXiv",
    primaryClass = "hep-lat",
    reportNumber = "JLAB-THY-25-4592",
    journal = "",
    month = "10",
    year = "2025"
}

@article{Hansen:2020otl,
    author = "Hansen, Maxwell T. and Brice{\~n}o, Raul A. and Edwards, Robert G. and Thomas, Christopher E. and Wilson, David J.",
    collaboration = "Hadron Spectrum",
    title = "{Energy-Dependent $\pi^+ \pi^+ \pi^+$  Scattering Amplitude from QCD}",
    eprint = "2009.04931",
    archivePrefix = "arXiv",
    primaryClass = "hep-lat",
    reportNumber = "CERN-TH-2020-147, JLAB-THY-20-3242",
    doi = "10.1103/PhysRevLett.126.012001",
    journal = "Phys. Rev. Lett.",
    volume = "126",
    pages = "012001",
    year = "2021"
}

@article{Blanton:2019vdk,
    author = "Blanton, Tyler D. and Romero-L{\'o}pez, Fernando and Sharpe, Stephen R.",
    title = "{$I=3$ Three-Pion Scattering Amplitude from Lattice QCD}",
    eprint = "1909.02973",
    archivePrefix = "arXiv",
    primaryClass = "hep-lat",
    doi = "10.1103/PhysRevLett.124.032001",
    journal = "Phys. Rev. Lett.",
    volume = "124",
    number = "3",
    pages = "032001",
    year = "2020"
}

@article{Yan:2024gwp,
    author = "Yan, Haobo and Mai, Maxim and Garofalo, Marco and Mei{\ss}ner, Ulf-G. and Liu, Chuan and Liu, Liuming and Urbach, Carsten",
    title = "{{\ensuremath{\omega}} Meson from Lattice QCD}",
    eprint = "2407.16659",
    archivePrefix = "arXiv",
    primaryClass = "hep-lat",
    doi = "10.1103/PhysRevLett.133.211906",
    journal = "Phys. Rev. Lett.",
    volume = "133",
    number = "21",
    pages = "211906",
    year = "2024"
}

@article{Fischer:2020jzp,
    author = "Fischer, Matthias and Kostrzewa, Bartosz and Liu, Liuming and Romero-L{\'o}pez, Fernando and Ueding, Martin and Urbach, Carsten",
    title = "{Scattering of two and three physical pions at maximal isospin from lattice QCD}",
    eprint = "2008.03035",
    archivePrefix = "arXiv",
    primaryClass = "hep-lat",
    doi = "10.1140/epjc/s10052-021-09206-5",
    journal = "Eur. Phys. J. C",
    volume = "81",
    number = "5",
    pages = "436",
    year = "2021"
}

@article{Mai:2021nul,
    author = {Mai, Maxim and Alexandru, Andrei and Brett, Ruair{\'\i} and Culver, Chris and D{\"o}ring, Michael and Lee, Frank X. and Sadasivan, Daniel},
    collaboration = "GWQCD",
    title = "{Three-Body Dynamics of the a1(1260) Resonance from Lattice QCD}",
    eprint = "2107.03973",
    archivePrefix = "arXiv",
    primaryClass = "hep-lat",
    doi = "10.1103/PhysRevLett.127.222001",
    journal = "Phys. Rev. Lett.",
    volume = "127",
    number = "22",
    pages = "222001",
    year = "2021"
}

@article{Dawid:2025doq,
    author = {Dawid, Sebastian M. and Draper, Zachary T. and Hanlon, Andrew D. and H{\"o}rz, Ben and Morningstar, Colin and Romero-L{\'o}pez, Fernando and Sharpe, Stephen R. and Skinner, Sarah},
    title = "{Two- and three-meson scattering amplitudes with physical quark masses from lattice QCD}",
    eprint = "2502.17976",
    archivePrefix = "arXiv",
    primaryClass = "hep-lat",
    reportNumber = "MIT-CTP/5846",
    doi = "10.1103/bx16-lp3r",
    journal = "Phys. Rev. D",
    volume = "112",
    number = "1",
    pages = "014505",
    year = "2025"
}

@article{Yan:2025mdm,
    author = {Yan, Haobo and Mai, Maxim and Garofalo, Marco and Feng, Yuchuan and D{\"o}ring, Michael and Liu, Chuan and Liu, Liuming and Mei{\ss}ner, Ulf-G. and Urbach, Carsten},
    title = "{Emergence of the $\pi(1300)$ Resonance from Lattice QCD}",
    eprint = "2510.09476",
    archivePrefix = "arXiv",
    primaryClass = "hep-lat",
    journal = "",
    month = "10",
    year = "2025"
}

@article{COMPASS:2015gxz,
    author = "Adolph, C and others",
    collaboration = "COMPASS",
    title = "{Resonance Production and $\pi\pi$ S-wave in $\pi^- + p \rightarrow \pi^- \pi^- \pi^+ + p_{recoil}$ at 190 GeV$/c$}",
    eprint = "1509.00992",
    archivePrefix = "arXiv",
    primaryClass = "hep-ex",
    reportNumber = "CERN-PH-EP-2015-233",
    doi = "10.1103/PhysRevD.95.032004",
    journal = "Phys. Rev. D",
    volume = "95",
    number = "3",
    pages = "032004",
    year = "2017"
}

@article{Ketzer:2019wmd,
    author = "Ketzer, Bernhard and Grube, Boris and Ryabchikov, Dmitry",
    title = "{Light-Meson Spectroscopy with COMPASS}",
    eprint = "1909.06366",
    archivePrefix = "arXiv",
    primaryClass = "hep-ex",
    doi = "10.1016/j.ppnp.2020.103755",
    journal = "Prog. Part. Nucl. Phys.",
    volume = "113",
    pages = "103755",
    year = "2020",
    note = "[Erratum: Prog.Part.Nucl.Phys. 128, 104000 (2023)]"
}

@article{Khuri:1960zz,
    author = "Khuri, N. N. and Treiman, S. B.",
    title = "{Pion-Pion Scattering and K + /- --{\ensuremath{>}} 3pi Decay}",
    doi = "10.1103/PhysRev.119.1115",
    journal = "Phys. Rev.",
    volume = "119",
    pages = "1115--1121",
    year = "1960"
}

@article{Guo:2016wsi,
    author = "Guo, P. and Danilkin, I. V. and Fern{\'a}ndez-Ram{\'\i}rez, C. and Mathieu, V. and Szczepaniak, A. P.",
    title = "{Three-body final state interaction in {\ensuremath{\eta}} {\textrightarrow} 3 {\ensuremath{\pi}} updated}",
    eprint = "1608.01447",
    archivePrefix = "arXiv",
    primaryClass = "hep-ph",
    reportNumber = "JLAB-THY-16-2302",
    doi = "10.1016/j.physletb.2017.05.092",
    journal = "Phys. Lett. B",
    volume = "771",
    pages = "497--502",
    year = "2017"
}

@article{Danilkin:2014cra,
    author = "Danilkin, I. V. and Fern{\'a}ndez-Ram{\'\i}rez, C. and Guo, P. and Mathieu, V. and Schott, D. and Shi, M. and Szczepaniak, A. P.",
    title = "{Dispersive analysis of {\ensuremath{\omega}}/{\ensuremath{\phi}}{\textrightarrow}3{\ensuremath{\pi}},{\ensuremath{\pi}}{\ensuremath{\gamma}}*}",
    eprint = "1409.7708",
    archivePrefix = "arXiv",
    primaryClass = "hep-ph",
    reportNumber = "JLAB-THY-14-1960",
    doi = "10.1103/PhysRevD.91.094029",
    journal = "Phys. Rev. D",
    volume = "91",
    number = "9",
    pages = "094029",
    year = "2015"
}

@article{Garcia-Lorenzo:2025uzc,
    author = "Garcia-Lorenzo, A. and Albaladejo, M. and Gonzlez-Solis, S. and Hammoud, N. and Mathieu, V. and Montana, G. and Pilloni, A. and Winney, D. and Szczepaniak, A. P.",
    collaboration = "JPAC",
    title = "{$\phi \to 3\pi$ and $\phi\pi^{0}$ transition form factor from Khuri-Treiman equations}",
    eprint = "2505.15309",
    archivePrefix = "arXiv",
    primaryClass = "hep-ph",
    reportNumber = "JLAB-THY-25-4248",
    month = "5",
    journal = "",
    year = "2025"
}

@article{Bernard:2024ioq,
    author = "Bernard, V{\'e}ronique and Descotes-Genon, S{\'e}bastien and Knecht, Marc and Moussallam, Bachir",
    title = "{A dispersive study of final-state interactions in $K\rightarrow \pi \pi \pi $ amplitudes}",
    eprint = "2403.17570",
    archivePrefix = "arXiv",
    primaryClass = "hep-ph",
    doi = "10.1140/epjc/s10052-024-13084-y",
    journal = "Eur. Phys. J. C",
    volume = "84",
    number = "7",
    pages = "744",
    year = "2024"
}

@article{Colangelo:2018jxw,
    author = "Colangelo, Gilberto and Lanz, Stefan and Leutwyler, Heinrich and Passemar, Emilie",
    title = "{Dispersive analysis of $\eta \rightarrow 3 \pi $}",
    eprint = "1807.11937",
    archivePrefix = "arXiv",
    primaryClass = "hep-ph",
    reportNumber = "JLAB-THY-18-2776",
    doi = "10.1140/epjc/s10052-018-6377-9",
    journal = "Eur. Phys. J. C",
    volume = "78",
    number = "11",
    pages = "947",
    year = "2018"
}

@article{Cao:2025ncx,
    author = "Cao, Xiong-Hui and Guo, Feng-Kun and Hanhart, Christoph and Kubis, Bastian",
    title = "{Dispersive analysis of the $J/ \psi \to \pi^0 \gamma^*$ transition form factor with $\rho$-$\omega$ mixing effects}",
    eprint = "2512.00501",
    archivePrefix = "arXiv",
    primaryClass = "hep-ph",
    month = "11",
    journal = "",
    year = "2025"
}

@article{Aitchison:1976nk,
    author = "Aitchison, I. J. R.",
    title = "{Relativistic Three Pion Dynamics Generated by Two-Body Unitarity and Analyticity}",
    reportNumber = "OXFORD-TP-58/76",
    doi = "10.1088/0305-4616/3/2/006",
    journal = "J. Phys. G",
    volume = "3",
    pages = "121",
    year = "1977"
}

@article{Akdag:2023pwx,
    author = "Akdag, Hakan and Kubis, Bastian and Wirzba, Andreas",
    title = "{Correlations of
$C$ and $CP$ violation in $\eta\to \pi^0\ell^+\ell^-$ and $\eta'\to \eta\ell^+\ell^-$}",
    eprint = "2307.02533",
    archivePrefix = "arXiv",
    primaryClass = "hep-ph",
    doi = "10.1007/JHEP03(2024)059",
    journal = "JHEP",
    volume = "03",
    pages = "059",
    year = "2024"
}

@article{Feng:2024wyg,
    author = {Feng, Yuchuan and Gil, Fernando and D{\"o}ring, Michael and Molina, Raquel and Mai, Maxim and Shastry, Vanamali and Szczepaniak, Adam},
    title = "{A unitary coupled-channel three-body amplitude with pions and kaons}",
    eprint = "2407.08721",
    archivePrefix = "arXiv",
    primaryClass = "nucl-th",
    reportNumber = "JLAB-THY-24-4107",
    doi = "10.1103/PhysRevD.110.094002",
    journal = "Phys. Rev. D",
    volume = "110",
    pages = "094002",
    year = "2024"
}

@article{Sadasivan:2021emk,
    author = {Sadasivan, Daniel and Alexandru, Andrei and Akdag, Hakan and Amorim, Felipe and Brett, Ruair{\'\i} and Culver, Chris and D{\"o}ring, Michael and Lee, Frank X. and Mai, Maxim},
    title = "{Pole position of the a1(1260) resonance in a three-body unitary framework}",
    eprint = "2112.03355",
    archivePrefix = "arXiv",
    primaryClass = "hep-ph",
    reportNumber = "JLAB-THY-21-3533",
    doi = "10.1103/PhysRevD.105.054020",
    journal = "Phys. Rev. D",
    volume = "105",
    number = "5",
    pages = "054020",
    year = "2022"
}

@article{Molina:2021awn,
    author = "Molina, R. and Doering, M. and Liang, W. H. and Oset, E.",
    title = "{The $\pi f_0(500)$ decay of the $a_1(1260)$}",
    eprint = "2107.07439",
    archivePrefix = "arXiv",
    primaryClass = "hep-ph",
    reportNumber = "JLAB-THY-21-3464",
    doi = "10.1140/epjc/s10052-021-09574-y",
    journal = "Eur. Phys. J. C",
    volume = "81",
    number = "9",
    pages = "782",
    year = "2021"
}

@article{Culver:2019vvu,
    author = {Culver, Chris and Mai, Maxim and Brett, Ruair{\'\i} and Alexandru, Andrei and D{\"o}ring, Michael},
    title = "{Three pion spectrum in the $I=3$ channel from lattice QCD}",
    eprint = "1911.09047",
    archivePrefix = "arXiv",
    primaryClass = "hep-lat",
    doi = "10.1103/PhysRevD.101.114507",
    journal = "Phys. Rev. D",
    volume = "101",
    number = "11",
    pages = "114507",
    year = "2020"
}

@article{CLAS:2008zko,
    author = "Nozar, M. and others",
    collaboration = "CLAS",
    title = "{Search for the photo-excitation of exotic mesons in the pi+ pi+ pi- system}",
    eprint = "0805.4438",
    archivePrefix = "arXiv",
    primaryClass = "hep-ex",
    reportNumber = "JLAB-PHY-08-820",
    doi = "10.1103/PhysRevLett.102.102002",
    journal = "Phys. Rev. Lett.",
    volume = "102",
    pages = "102002",
    year = "2009"
}

@phdthesis{Tsaris:2016oty,
    author = "Tsaris, Aristeidis",
    title = "{A Study of 3$\pi$ production in $\gamma p \to n\pi^{+}\pi^{+}\pi^{-}$ and $\gamma p \to \Delta^{++}\pi^{+}\pi^{-}\pi^{-}$ with CLAS at Jefferson Lab}",
    reportNumber = "JLAB-PHY-16-2443, DOE/OR/23177-4115",
    doi = "10.2172/1351257",
    school = "Florida State U.",
    year = "2016"
}

@phdthesis{Bookwalter:2012ixa,
    author = "Bookwalter, Craig",
    title = "{A Search for Exotic Mesons in $\gamma p \to \pi^+\pi^+\pi^-n$ with CLAS at Jefferson Lab}",
    school = "Florida State U.",
    year = "2012"
}

@article{Donnachie:1992ny,
    author = "Donnachie, A. and Landshoff, P. V.",
    title = "{Total cross-sections}",
    eprint = "hep-ph/9209205",
    archivePrefix = "arXiv",
    reportNumber = "CERN-TH-6635-92",
    doi = "10.1016/0370-2693(92)90832-O",
    journal = "Phys. Lett. B",
    volume = "296",
    pages = "227--232",
    year = "1992"
}

@article{Mathieu:2018xyc,
    author = "Mathieu, V. and Nys, J. and Fern{\'a}ndez-Ram{\'\i}rez, C. and Jackura, A. and Pilloni, A. and Sherrill, N. and Szczepaniak, A. P. and Fox, G.",
    collaboration = "JPAC",
    title = "{Vector Meson Photoproduction with a Linearly Polarized Beam}",
    eprint = "1802.09403",
    archivePrefix = "arXiv",
    primaryClass = "hep-ph",
    reportNumber = "JLAB-THY-18-2650",
    doi = "10.1103/PhysRevD.97.094003",
    journal = "Phys. Rev. D",
    volume = "97",
    number = "9",
    pages = "094003",
    year = "2018"
}

@article{Jackura:2022gib,
    author = "Jackura, Andrew W.",
    title = "{Three-body scattering and quantization conditions from S-matrix unitarity}",
    eprint = "2208.10587",
    archivePrefix = "arXiv",
    primaryClass = "hep-lat",
    reportNumber = "JLAB-THY-22-3664",
    doi = "10.1103/PhysRevD.108.034505",
    journal = "Phys. Rev. D",
    volume = "108",
    number = "3",
    pages = "034505",
    year = "2023"
}

@article{Jackura:2019bmu,
    author = "Jackura, A. W. and Dawid, S. M. and Fern{\'a}ndez-Ram{\'\i}rez, C. and Mathieu, V. and Mikhasenko, M. and Pilloni, A. and Sharpe, S. R. and Szczepaniak, A. P.",
    title = "{Equivalence of three-particle scattering formalisms}",
    eprint = "1905.12007",
    archivePrefix = "arXiv",
    primaryClass = "hep-ph",
    reportNumber = "JLAB-THY-19-2947",
    doi = "10.1103/PhysRevD.100.034508",
    journal = "Phys. Rev. D",
    volume = "100",
    number = "3",
    pages = "034508",
    year = "2019"
}

@article{Szczepaniak:2001qz,
    author = "Szczepaniak, Adam P. and Swat, Maciej",
    title = "{Role of photoproduction in exotic meson searches}",
    eprint = "hep-ph/0105329",
    archivePrefix = "arXiv",
    reportNumber = "IUNTC-01-01",
    doi = "10.1016/S0370-2693(01)00905-4",
    journal = "Phys. Lett. B",
    volume = "516",
    pages = "72--76",
    year = "2001"
}

@article{Afanasev:1999rb,
    author = "Afanasev, Andrei V. and Szczepaniak, Adam P.",
    title = "{Charge exchange rho0 pi+ photoproduction and implications for searches of exotic meson}",
    eprint = "hep-ph/9910268",
    archivePrefix = "arXiv",
    reportNumber = "JLAB-THY-99-32",
    doi = "10.1103/PhysRevD.61.114008",
    journal = "Phys. Rev. D",
    volume = "61",
    pages = "114008",
    year = "2000"
}

@article{Close:2003af,
    author = "Close, F. E. and Dudek, J. J.",
    title = "{The 'Forbidden' decays of hybrid mesons to pi rho can be large}",
    eprint = "hep-ph/0308099",
    archivePrefix = "arXiv",
    doi = "10.1103/PhysRevD.70.094015",
    journal = "Phys. Rev. D",
    volume = "70",
    pages = "094015",
    year = "2004"
}

@article{Albrecht:2024qdh,
    author = "Albrecht, M.",
    collaboration = "GlueX",
    title = "{Search for exotic hadrons in $\eta^{(')} \pi$ at GlueX}",
    doi = "10.1393/ncc/i2024-24181-1",
    journal = "Nuovo Cim. C",
    volume = "47",
    number = "4",
    pages = "181",
    year = "2024"
}

@article{Niecknig:2017ylb,
    author = "Niecknig, Franz and Kubis, Bastian",
    title = "{Consistent Dalitz plot analysis of Cabibbo-favored $D^+ \to \bar{K} \pi \pi^+$ decays}",
    eprint = "1708.00446",
    archivePrefix = "arXiv",
    primaryClass = "hep-ph",
    doi = "10.1016/j.physletb.2018.03.048",
    journal = "Phys. Lett. B",
    volume = "780",
    pages = "471--478",
    year = "2018"
}

@article{Kou:2023kvp,
    author = "Kou, Emi and Moskalets, Tetiana and Moussallam, Bachir",
    title = "{Isospin symmetry and analyticity in $ D\to \overline{K}\pi \pi $ decays}",
    eprint = "2303.12015",
    archivePrefix = "arXiv",
    primaryClass = "hep-ph",
    doi = "10.1007/JHEP12(2023)177",
    journal = "JHEP",
    volume = "12",
    pages = "177",
    year = "2023"
}

@article{Hu:2025ppi,
    author = "Hu, Xin-Yue and He, Jiahao and Niu, Pengyu and Wang, Qian and Yan, Yupeng",
    title = "{Three-body final state interactions in B+{\textrightarrow}DD{\textasciimacron}K+ decays}",
    eprint = "2509.10039",
    archivePrefix = "arXiv",
    primaryClass = "hep-ph",
    doi = "10.1103/whyt-cx47",
    journal = "Phys. Rev. D",
    volume = "113",
    number = "5",
    pages = "054003",
    year = "2026"
}

@article{Accardi:2023chb,
    author = "Accardi, A. and others",
    title = "{Strong interaction physics at the luminosity frontier with 22 GeV electrons at Jefferson Lab}",
    eprint = "2306.09360",
    archivePrefix = "arXiv",
    primaryClass = "nucl-ex",
    reportNumber = "JLAB-PHY-23-3840, JLAB-THY-23-3848",
    doi = "10.1140/epja/s10050-024-01282-x",
    journal = "Eur. Phys. J. A",
    volume = "60",
    number = "9",
    pages = "173",
    year = "2024"
}

@article{AbdulKhalek:2021gbh,
    author = "Abdul Khalek, R. and others",
    title = "{Science Requirements and Detector Concepts for the Electron-Ion Collider}: {EIC Yellow Report}",
    eprint = "2103.05419",
    archivePrefix = "arXiv",
    primaryClass = "physics.ins-det",
    reportNumber = "BNL-220990-2021-FORE, JLAB-PHY-21-3198, LA-UR-21-20953",
    doi = "10.1016/j.nuclphysa.2022.122447",
    journal = "Nucl. Phys. A",
    volume = "1026",
    pages = "122447",
    year = "2022"
}

@article{Basdevant:1977ya,
    author = "Basdevant, J. L. and Berger, Edmond L.",
    title = "{Unitary Coupled-Channel Analysis of Diffractive Production of the a1 Resonance}",
    reportNumber = "ANL-HEP-PR-77-02",
    doi = "10.1103/PhysRevD.16.657",
    journal = "Phys. Rev. D",
    volume = "16",
    pages = "657",
    year = "1977"
}

@article{Jackura:2016llm,
    author = "Jackura, Andrew and Mikhasenko, Mikhail and Szczepaniak, Adam",
    editor = {Wro{\'n}ska, A. and Magiera, A. and Guaraldo, C. and Str{\"o}her, H.},
    title = "{Amplitude analysis of resonant production in three pions}",
    eprint = "1610.04567",
    archivePrefix = "arXiv",
    primaryClass = "hep-ph",
    reportNumber = "JLAB-THY-16-2292",
    doi = "10.1051/epjconf/201613005008",
    journal = "EPJ Web Conf.",
    volume = "130",
    pages = "05008",
    year = "2016"
}

@article{Mikhasenko:2017jtg,
    author = "Mikhasenko, Mikhail and Jackura, Andrew and Ketzer, Bernhard and Szczepaniak, Adam",
    editor = "Foka, Y. and Brambilla, N. and Kovalenko, V.",
    collaboration = "COMPASS",
    title = "{Unitarity approach to the mass-dependent fit of $3\pi$ resonance production data from the COMPASS experiment}",
    doi = "10.1051/epjconf/201713705017",
    journal = "EPJ Web Conf.",
    volume = "137",
    pages = "05017",
    year = "2017"
}

@article{Dudek:2011bn,
    author = "Dudek, Jozef J.",
    title = "{The lightest hybrid meson supermultiplet in QCD}",
    eprint = "1106.5515",
    archivePrefix = "arXiv",
    primaryClass = "hep-ph",
    reportNumber = "JLAB-THY-11-1387",
    doi = "10.1103/PhysRevD.84.074023",
    journal = "Phys. Rev. D",
    volume = "84",
    pages = "074023",
    year = "2011"
}

@article{Klempt:2007cp,
    author = "Klempt, Eberhard and Zaitsev, Alexander",
    title = "{Glueballs, Hybrids, Multiquarks. Experimental facts versus QCD inspired concepts}",
    eprint = "0708.4016",
    archivePrefix = "arXiv",
    primaryClass = "hep-ph",
    doi = "10.1016/j.physrep.2007.07.006",
    journal = "Phys. Rept.",
    volume = "454",
    pages = "1--202",
    year = "2007"
}

@article{JPAC:2021rxu,
    author = "Albaladejo, Miguel and others",
    collaboration = "JPAC",
    title = "{Novel approaches in hadron spectroscopy}",
    eprint = "2112.13436",
    archivePrefix = "arXiv",
    primaryClass = "hep-ph",
    reportNumber = "LA-UR-21-31664, JLAB-THY-22-3459",
    doi = "10.1016/j.ppnp.2022.103981",
    journal = "Prog. Part. Nucl. Phys.",
    volume = "127",
    pages = "103981",
    year = "2022"
}

@article{H1:2005dtp,
    author = "Aktas, A. and others",
    collaboration = "H1",
    title = "{Elastic J/psi production at HERA}",
    eprint = "hep-ex/0510016",
    archivePrefix = "arXiv",
    reportNumber = "DESY-05-161",
    doi = "10.1140/epjc/s2006-02519-5",
    journal = "Eur. Phys. J. C",
    volume = "46",
    pages = "585--603",
    year = "2006"
}

@article{ZEUS:1995bfs,
    author = "Derrick, M. and others",
    collaboration = "ZEUS",
    title = "{Measurement of elastic $\rho^0$ photoproduction at HERA}",
    eprint = "hep-ex/9507011",
    archivePrefix = "arXiv",
    reportNumber = "DESY-95-143",
    doi = "10.1007/s002880050004",
    journal = "Z. Phys. C",
    volume = "69",
    pages = "39--54",
    year = "1995"
}

@article{Winney:2025tla,
    author = "Winney, Daniel and Szczepaniak, Adam P.",
    title = "{Regge theory in hadron physics}",
    eprint = "2512.21805",
    archivePrefix = "arXiv",
    primaryClass = "hep-ph",
    month = "12",
    year = "2025",
    journal ={}
}

@article{Pekeler:2025ehx,
    author = "Pekeler, Henri",
    collaboration = "COMPASS",
    title = "{Understanding COMPASS data on $\pi^-+p\to\eta^{(\prime)}\pi^-+p$ in the double-Regge region}",
    doi = "10.22323/1.500.0152",
    journal = "PoS",
    volume = "HADRON2025",
    pages = "152",
    year = "2026"
}

@article{Niecknig:2015ija,
    author = "Niecknig, Franz and Kubis, Bastian",
    title = "{Dispersion-theoretical analysis of the D$^{+}$ {\textrightarrow} K$^{−}$ {\ensuremath{\pi}}$^{+}$ {\ensuremath{\pi}}$^{+}$ Dalitz plot}",
    eprint = "1509.03188",
    archivePrefix = "arXiv",
    primaryClass = "hep-ph",
    doi = "10.1007/JHEP10(2015)142",
    journal = "JHEP",
    volume = "10",
    pages = "142",
    year = "2015"
}

@article{Isken:2017dkw,
    author = "Isken, Tobias and Kubis, Bastian and Schneider, Sebastian P. and Stoffer, Peter",
    title = "{Dispersion relations for $\eta '\rightarrow \eta \pi \pi $}",
    eprint = "1705.04339",
    archivePrefix = "arXiv",
    primaryClass = "hep-ph",
    doi = "10.1140/epjc/s10052-017-5024-1",
    journal = "Eur. Phys. J. C",
    volume = "77",
    number = "7",
    pages = "489",
    year = "2017"
}

@article{Akdag:2021efj,
    author = "Akdag, Hakan and Isken, Tobias and Kubis, Bastian",
    title = "{Patterns of C- and CP-violation in hadronic {\ensuremath{\eta}} and {\ensuremath{\eta}}' three-body decays}",
    eprint = "2111.02417",
    archivePrefix = "arXiv",
    primaryClass = "hep-ph",
    doi = "10.1007/JHEP02(2022)137",
    journal = "JHEP",
    volume = "02",
    pages = "137",
    year = "2022",
    note = "[Erratum: JHEP 12, 156 (2022)]"
}

@article{Montana:2025asi,
    author = "Monta{\~n}a, Gloria and others",
    title = "{High-energy {\ensuremath{\eta}}('){\ensuremath{\pi}} photoproduction and the nature of exotic waves}",
    eprint = "2510.14549",
    archivePrefix = "arXiv",
    primaryClass = "hep-ph",
    reportNumber = "JLAB-THY-25-4567, MIT-CTP/5935",
    doi = "10.1016/j.physletb.2025.140101",
    journal = "Phys. Lett. B",
    volume = "872",
    pages = "140101",
    year = "2026"
}

@article{Hoferichter:2014vra,
    author = "Hoferichter, Martin and Kubis, Bastian and Leupold, Stefan and Niecknig, Franz and Schneider, Sebastian P.",
    title = "{Dispersive analysis of the pion transition form factor}",
    eprint = "1410.4691",
    archivePrefix = "arXiv",
    primaryClass = "hep-ph",
    doi = "10.1140/epjc/s10052-014-3180-0",
    journal = "Eur. Phys. J. C",
    volume = "74",
    pages = "3180",
    year = "2014"
}

@phdthesis{RebeccaPhD,
  author  = {Barsotti, Rebecca},
  title   = {Photoproduction of  $\eta \pi^{0}$ by Double-Regge Exchange},
  school  = {Indiana University},
  year    = {2025},
  address = {Bloomington IN, USA}
}

@mastersthesis{BartlMasters,
  author  = "{Bartl, Martin}",
  title   = "{Study of the $\pi^-\pi^+$ subsystem with $J^{PC}=1^{--}$ in the diffractively produced $\pi^-\pi^+\pi^+$ final state at COMPASS}",
  school  = "{Technische Universität München}",
  year    = "{2023}"
}

@article{Aaron:1977wa,
    author = "Aaron, R. and Longacre, R. S. and Sacco, J. E.",
    title = "{Analysis of the A1}",
    reportNumber = "NUB 2340",
    doi = "10.1016/0003-4916(79)90044-7",
    journal = "Annals Phys.",
    volume = "117",
    pages = "56",
    year = "1979"
}

@article{Dawid:2023jrj,
    author = "Dawid, Sebastian M. and Islam, Md Habib E. and Brice{\~n}o, Ra{\'u}l A.",
    title = "{Analytic continuation of the relativistic three-particle scattering amplitudes}",
    eprint = "2303.04394",
    archivePrefix = "arXiv",
    primaryClass = "nucl-th",
    doi = "10.1103/PhysRevD.108.034016",
    journal = "Phys. Rev. D",
    volume = "108",
    number = "3",
    pages = "034016",
    year = "2023"
}

@article{HillerBlin:2016odx,
    author = "Hiller Blin, A. N. and Fern{\'a}ndez-Ram{\'\i}rez, C. and Jackura, A. and Mathieu, V. and Mokeev, V. I. and Pilloni, A. and Szczepaniak, A. P.",
    title = "{Studying the P$_c$(4450) resonance in J/$\psi$ photoproduction off protons}",
    eprint = "1606.08912",
    archivePrefix = "arXiv",
    primaryClass = "hep-ph",
    reportNumber = "JLAB-THY-16-2277",
    doi = "10.1103/PhysRevD.94.034002",
    journal = "Phys. Rev. D",
    volume = "94",
    number = "3",
    pages = "034002",
    year = "2016"
}

@article{Jackura:2018xnx,
    author = "Jackura, A. and Fern{\'a}ndez-Ram{\'\i}rez, C. and Mathieu, V. and Mikhasenko, M. and Nys, J. and Pilloni, A. and Salda{\~n}a, K. and Sherrill, N. and Szczepaniak, A. P.",
    collaboration = "JPAC",
    title = "{Phenomenology of Relativistic $\mathbf{3} \to \mathbf{3}$ Reaction Amplitudes within the Isobar Approximation}",
    eprint = "1809.10523",
    archivePrefix = "arXiv",
    primaryClass = "hep-ph",
    reportNumber = "JLAB-THY-18-2817",
    doi = "10.1140/epjc/s10052-019-6566-1",
    journal = "Eur. Phys. J. C",
    volume = "79",
    number = "1",
    pages = "56",
    year = "2019"
}

@phdthesis{Pekeler:2025,
    author = "Pekeler, Henri",
    title = "{Production of $\eta \pi^-$ and $\eta' \pi^-$ Final States in High Energy $\pi^- p$ Scattering at COMPASS}",
    school = "Bonn Universit{\"a}t",
    month = "10",
    year = "2025",
    url = "https://hdl.handle.net/20.500.11811/14270",
}

@article{Albaladejo:2017hhj,
    author = "Albaladejo, M. and Moussallam, B.",
    title = "{Extended chiral Khuri-Treiman formalism for $\eta\to 3\pi$ and the role of the $a_0(980)$, $f_0(980)$ resonances}",
    eprint = "1702.04931",
    archivePrefix = "arXiv",
    primaryClass = "hep-ph",
    doi = "10.1140/epjc/s10052-017-5052-x",
    journal = "Eur. Phys. J. C",
    volume = "77",
    number = "8",
    pages = "508",
    year = "2017"
}

@article{Lesniak:2003gf,
    author = "Lesniak, L. and Szczepaniak, A. P.",
    title = "{Theoretical model of the phi meson photoproduction amplitudes}",
    eprint = "hep-ph/0304007",
    archivePrefix = "arXiv",
    journal = "Acta Phys. Polon. B",
    volume = "34",
    pages = "3389--3400",
    year = "2003"
}

@article{Close:1999bi,
    author = "Close, Frank E. and Schuler, Gerhard A.",
    title = "{Evidence that the pomeron transforms as a nonconserved vector current}",
    eprint = "hep-ph/9905305",
    archivePrefix = "arXiv",
    reportNumber = "CERN-TH-99-131",
    doi = "10.1016/S0370-2693(99)00875-8",
    journal = "Phys. Lett. B",
    volume = "464",
    pages = "279--285",
    year = "1999"
}

@article{JPAC:2017dbi,
    author = "Jackura, A. and others",
    collaboration = "JPAC, COMPASS",
    title = "{New analysis of $\eta\pi$ tensor resonances measured at the COMPASS experiment}",
    eprint = "1707.02848",
    archivePrefix = "arXiv",
    primaryClass = "hep-ph",
    reportNumber = "CERN-EP-2017-169, JLAB-THY-17-2468",
    doi = "10.1016/j.physletb.2018.01.017",
    journal = "Phys. Lett. B",
    volume = "779",
    pages = "464--472",
    year = "2018"
}

@software{iteratekt,
  author       = {D. Winney},
  title        = {{iterateKT (v1.1)}},
  month        = oct,
  year         = 2026,
  publisher    = {Zenodo},
  version      = {v1.1},
  doi          = {10.5281/zenodo.23426016},
  url          = {https://doi.org/10.5281/zenodo.23426016},
  note = {https://doi.org/10.5281/zenodo.23426016}
}

@software{bootstrapPi1,
  author       = {D. Winney},
  title        = {{bootstrapPi1 (v1.0)}},
  month        = oct,
  year         = 2026,
  publisher    = {Zenodo},
  version      = {v1.0},
  doi          = {10.5281/zenodo.23491459},
  url          = {https://doi.org/10.5281/zenodo.23491460},
  note = {https://doi.org/10.5281/zenodo.23491460}
}

@preprint{COMPASS:2026mcg,
    author = "Alexeev, G. D. and others",
    collaboration = "COMPASS, JPAC",
    title = "{First Observation of an Exotic Reggeon}",
    eprint = "2607.17850",
    archivePrefix = "arXiv",
    primaryClass = "hep-ex",
    reportNumber = "CERN-EP-2026-209, JLAB-THY-26-4814, MIT-CTP/6061",
    month = "7",
    year = "2026"
}

\end{document}